\documentclass{article}
\usepackage[cmex10]{amsmath}
\usepackage{amssymb}
\usepackage{algorithmic}
\usepackage{algorithm}
\usepackage{graphicx}
\usepackage{listings}
\usepackage{tikz}
\usetikzlibrary{matrix}
\usetikzlibrary{shapes}
\usetikzlibrary{trees}
\usetikzlibrary{calc}
\usepackage{arydshln}
\usepackage{color}
\usepackage[tight,small]{subfigure}
\usepackage{listings}
\usepackage{xspace}
\usepackage{ifthen}
\usepackage{geometry}
\usepackage{url}
\lstdefinelanguage{myoctave}[]{Octave}{
numberstyle=\footnotesize,
sensitive=true,%
tabsize=3,
numbers=left,
linewidth=\linewidth,
showspaces=false,
showtabs=false,
xleftmargin=20pt,
xrightmargin=10pt,
mathescape=true,
morekeywords=[1]{p_matrix},
escapeinside={(*@}{@*)}
}%

\newcommand{\mytexttt}[1]{\texttt{\hyphenchar\font45 #1}}

\lstdefinelanguage{pseudocode}{
numberstyle=\footnotesize,
sensitive=true,%
tabsize=3,
numbers=left,
linewidth=\linewidth,
showspaces=false,
showtabs=false,
xleftmargin=20pt,
xrightmargin=10pt,
mathescape=true,
escapeinside={(*@}{@*)}
}%

\providecommand{\tabularnewline}{\\}

\newcommand{\naive}{na\"{i}ve\xspace}

\newcommand\psmatrix{\texttt{p\_matrix}\xspace}

\newcommand\bigo{\mathcal O}

\newcommand{\ME}[3]{\ensuremath{#1_{#2#3}}}
\newcommand{\MEg}[3]{\textcolor{gray}{\ME{#1}{#2}{#3}}}
\newcommand{\MES}[3]{\ensuremath{#1_{#2,#3}}}
\newcommand{\MESg}[3]{\textcolor{gray}{\MES{#1}{#2}{#3}}}
\newcommand{\MSpace}{\ensuremath{S}}
\newcommand{\Div}{\ensuremath{\delta}}

\newcommand{\ESpace}{\ensuremath{s}}
\def\tdots{\operatorname{..}}
\pgfdeclarelayer{background}
\pgfdeclarelayer{foreground}
\pgfsetlayers{background,main,foreground}
\newcommand{\atab}{\hspace{4mm}}

\begin{document}
\title{Accelerating the Execution of Matrix Languages on the Cell Broadband Engine Architecture}
\author{Raymes~Khoury,
        Bernd~Burgstaller
        and~Bernhard~Scholz\footnote{R.~Khoury and B.~Scholz are with the School
of Information Technologies, University of Sydney, Sydney, Australia.
B.~Burgstaller is with the Department of Computer Science, Yonsei University, South Korea.}}
		
\maketitle

\begin{abstract}
Matrix languages, including MATLAB and Octave, are established standards 
for applications in science and engineering. They provide interactive 
programming environments that are easy to use due to their scripting 
languages with matrix data types. Current implementations of matrix 
languages do not fully utilise high-performance, special-purpose chip 
architectures such as the IBM PowerXCell processor (Cell), which is 
currently used in the fastest computer in the world.

We present a new framework that extends Octave to harness the
computational power of the Cell. With this framework the programmer is
relieved of the burden of introducing explicit notions of
parallelism. Instead the programmer uses a new matrix data-type to
execute matrix operations in parallel on the synergistic processing
elements (SPEs) of the Cell.  We employ lazy evaluation semantics for
our new matrix data-type to obtain execution traces of matrix
operations. Traces are converted to data dependence graphs; operations
in the data dependence graph are lowered (split into sub-matrices),
scheduled and executed on the SPEs. Thereby we exploit (1) data
parallelism, (2) instruction level parallelism, (3) pipeline
parallelism and (4) task parallelism of matrix language programs.  We
conducted extensive experiments to show the validity of our
approach. Our Cell-based implementation achieves speedups of up to a
factor of 12 over code run on recent Intel Core2 Quad processors.
\end{abstract}

\section{Introduction} \label{sec:intro}

Matrix languages including MATLAB~\cite{matlab} and Octave~\cite{octave}
are established standards for rapid-proto\-typing in scientific and engineering domains.
One of the main reasons for the widespread adoption of these languages is their ease of use.
They provide interactive execution of code and simple, high-level syntax for matrix 
calculations. Complex scientific and engineering problems are solved 
with a few lines of code because there exists a cornucopia of commercial and open source 
libraries for standard
mathematical problems.

Despite their ease of use, matrix languages traditionally have sequential execution
semantics and utilise a single thread of execution only~\cite{moler1995tit}. 
While the performance growth of
single-core processors is reaching its limits~\cite{sutter2005flo}, scientists and engineers have increasingly
large data-sets which must be processed efficiently~\cite{hoschek2000data}. Thus, the use of matrix languages will
plateau in the near future if not adapted to modern parallel computer architectures.

With the advent of hardware accelerators for high-performance computing such as 
General Purpose Graphics Processors (GPGPUs)~\cite{nvidia2007compute}
and the Cell Broadband Engine~\cite{1130803}, significant performance boosts over single-core 
architectures 
are possible. However, harnessing their computational power is challenging in the context 
of matrix languages. Hardware accelerators for high-performance computing  are attributed to having non-uniform memory accesses and complex parallel
programming patterns. Extending matrix languages to execute on high-performance, accelerator architectures can be achieved by adopting either (1) an explicit parallel programming
model which requires users to manually introduce a notion of parallelism in their matrix language program, or (2) an implicit parallel programming model in which parallelism is
elicited from a matrix language program with little user intervention. The explicit 
model contradicts the initial design goals of matrix languages 
as the users of these languages are most often untrained in concurrent programming.
Hence, it is of paramount importance to the continued success of matrix languages that
an implicit model is adopted which is capable of fully utilising the computational power of modern accelerator architectures for high-performance computing.

A large body of
research~\cite{1386655,1032114,fernandez2006std,Bliss0801,Bliss08012007,sharma2009mlp,derose1996fmi,derose1996mft}
already exists on how matrix languages can be parallelised for
\emph{distributed} parallel architectures that were popular before the turn
of last century. Some of the parallel extensions developed are
reported to offer good performance, however this performance gain was
often paid at the expense of ease of use by the
programmer~\cite{1386655}. Little research has been conducted in how
matrix languages can be parallelised for modern accelerator
architectures which present very different challenges in achieving
good performance.  Current state of the art techniques (e.g.~those
employed in MATLAB) for parallelisation on multicore CPUs involve
using parallel math libraries~\cite{atlas_sc98} to exploit data
parallelism within matrix operations. Several emerging
projects~\cite{jacket1,gpulib} have investigated using
simple bindings to execute MATLAB functions on GPGPUs, again
exploiting data parallelism in matrix operations. However, these
projects offer a \naive approach, neglecting other types of
parallelism that exist in matrix language programs and resulting in
under-utilisation of their target architectures.

In this paper, we introduce a new framework for the automatic parallelisation
of matrix languages which is specifically targeted towards modern hardware
accelerators for high-performance computing. We have implemented this framework 
as an extension to the Octave interpreter running on the Cell Broadband Engine.
The Cell Broadband Engine is a heterogeneous multicore
architecture, which is currently deployed in
the fastest computer in
the world~\cite{barker2008entering}, as well as the Sony Playstation 3 computer games console.
The Cell consists of a PowerPC core that is connected to
several Synergistic Processing Elements~(SPEs) via a high-speed
interconnect double ring bus. The PowerPC unit is an in-order RISC
processor with two hyper-threads, whereas the SPEs are small-sized
vector (i.e.~SIMD) machines with 256kB of local memory that is shared
for data and instructions. Each SPE delivers approximately 25~GFLOPs peak performance
for fused multiply-add operations~\cite{ibmcellhandbook}.

GNU Octave is an open source alternative to MATLAB (a commercial product developed
by MathWorks). Octave mimics MATLAB's syntax and has been used in our work because MATLAB
does not support PowerPC-based architectures such as Cell Broadband Engine. 

Our framework exploits several types of parallelism in an Octave program
to obtain high utilisation of the Cell processor:
\begin{enumerate}
\item \textbf{Data parallelism} is exploited by partitioning large matrices
  into sub-matrices and executing operations on the sub-matrices
  in parallel,
\item \textbf{Instruction level parallelism} is exploited by executing matrix
   operations of an execution trace in parallel if there is no data dependence
   between them,
\item \textbf{Pipeline parallelism} is exploited by overlapping
communication between cores with computation of matrix operations,
\item \textbf{Task parallelism} is exploited by overlapping execution of the
  Octave interpreter, construction of the schedule, and execution
  of the matrix operations on the SPEs.
\end{enumerate}

Lazy evaluation is used to generate execution traces of matrix operations
by deferring their computation until a result is required. 
A key feature of our framework is that partitioned matrix operations
are scheduled among the parallel execution elements of the Cell processor
 in a way that satisfies inter-operation data
dependencies, prior to execution of the trace. Using estimates of the
execution time for each operation, operations can be scheduled  such
that the utilisation of parallel elements is improved and the total execution
time (makespan) is reduced.

We perform an extensive evaluation of our framework using 9 Octave benchmark programs.
The execution time of these benchmarks running on our framework with the Cell processor
is compared with their execution time on a default installation of Octave
on an Intel Core2 Quad processor. Comparisons are also made with several other 
configurations. Furthermore, we perform experiments to evaluate the benefits of
scheduling matrix operations and to determine the extent to which the Cell 
architecture is utilised.

The contributions of this work are as follows:
\begin{enumerate}
\item We introduce a framework for the execution of matrix languages on
modern parallel architectures. Our framework exploits both data parallelism
and instruction level parallelism of matrix
programs. Instruction level parallelism is achieved through lazy evaluation of
matrix operations.
\item We provide a $\thickapprox$7000 line C/C++ implementation of this framework
for the Octave programming language on the Cell Broadband Engine architecture.
\item We provide a novel, efficient technique for partitioning matrix operations
to maximise the parallelism available within an execution trace.
\item We develop accurate time models for estimating the execution times
of matrix operations through multivariate regression analysis.
\item We introduce a new heuristic scheduling for scheduling matrix operations
among parallel processing elements. The algorithm takes
into account the estimated execution times of operations and the
pipelined nature of processing elements.
\item We formulate the scheduling problem as an integer linear program
and compare the obtained optimal solution with the solution produced by the
heuristic scheduling algorithm.
\end{enumerate}

This paper is organised as follows:
In section~\ref{sec:related-work} we survey related work on (1) parallelising
matrix languages and (2) the scheduling problem for precedence constrained tasks on
multiprocessors. In section~\ref{sec:overview} we give an overview 
of our framework and describe each of the major components. In section~\ref{sec:motivating} 
we describe a motivating example that illustrates how an Octave program is executed 
in parallel with our framework. 
In section~\ref{sec:lowering} the lowering process is explained which decomposes 
operations on large matrices into operations on smaller matrices. In 
section~\ref{sec:scheduling} we describe the scheduling problem and
provide a heuristic algorithm as well as an integer linear programming
formulation which yields the optimal solution. In section~\ref{sec:cell} we 
give an overview of the Cell Broadband Engine architecture and we discuss the 
implementation of our framework on this architecture.
In section~\ref{sec:impl} we describe details of the development, testing and
optimisation of our framework.
In section~\ref{sec:experiment} 
we present the experimental results and discuss the observed performance of our
framework. We summarise our work and draw our conclusions in section~\ref{sec:conclusion}. 

\section{Related Work}\label{sec:related-work}

\subsection{Parallel MATLAB}

There are a variety of extensions for MATLAB designed to utilise parallel
computers. The methods of achieving parallelism, the target
architecture and the extent to which the parallelisation process is
automated vary from extension to extension.  In our work we present a new
parallel system for a matrix language with two identifying goals:
\begin{enumerate}
\item It allows automatic parallelisation of code with no intervention from the programmer.

\item It is designed specifically to achieve high performance on modern hardware accelerator architectures.
\end{enumerate}
Choy and Edelman provide a survey~\cite{1386655} of 27~projects that
extend matrix languages with parallel features. The survey classifies
the projects in four main categories:
\begin{itemize}
\item \textbf{Embarrassingly parallel:} These projects make use of
  multiple MATLAB processes running simultaneously. There is
  only communication involved when a new process is spawned or a
  process has completed its task. These parallel extensions for
  MATLAB are limited to applications that can adopt an embarrassingly
  parallel programming scheme.
\item \textbf{Message passing:} These projects provide message passing
  functionality between MATLAB processes. The complexity of these
extensions varies from simple wrappers for MPI functions~\cite{MaS98}
to higher level abstractions. The programmer has to express the parallelism explicitly.
\item \textbf{MATLAB compilers:} These projects compile MATLAB scripts
  into an executable form, either directly or through the use of an
  intermediate language such as Fortran or C. Some of these projects
  link their executables with parallel math libraries while others
  generate code that utilises MPI.
\item \textbf{Backend support:} These projects use a single MATLAB
  process as a front-end, which generates jobs that are submitted to a computation
  engine and executed in parallel, often using numerical libraries
  like ScaLAPACK.
\end{itemize}
We now examine the key parallel MATLAB systems in each of these categories. 
We have chosen to include systems that we
believe are representative of the current state of research including
several which are more recent than the survey provided by Choy and
Edelman.

\subsubsection{Message Passing and Embarrassingly Parallel Extensions}
MatlabMPI~\cite{1032114} is an implementation of the Message Passing
Interface (MPI) for MATLAB, developed at
MIT. MatlabMPI works by spawning several MATLAB processes that
communicate via a shared file system. A sender process
writes a variable to a data file on the file system and touches
(creates) a lock file when the send is complete.  A receiver polls
for the existence of the lock file and when it exists,
reads in the data file and does any necessary cleanup.  The system
consists of 350 lines of pure MATLAB code which makes it very
portable. There are several similar projects in existence which aim to
provide pure message passing functionality in MATLAB including the MPI
Toolbox from the University of Granada~\cite{fernandez2006std}. They
provide flexibility and control in parallelising a program. The
performance of MatlabMPI is also reported to approach that of
equivalent C MPI code for large messages. However, these systems do
not reduce the complexity of concurrent programming for a user and
can perform very poorly for certain workloads.

In recognition of the need to reduce this complexity for the typical
user, Bliss and Kepner developed pMATLAB~\cite{Bliss0801} which built
upon MatlabMPI. Rather than requiring users to perform communication
between MATLAB processes explicitly, pMATLAB allows users to declare
distributed numerical arrays (or matrices) and an associated mapping
of these arrays to available processors in the style of High
Performance Fortran~(HPF)~\cite{KKZ07}. A map consists of a
grid specifying the partitioning of the array as well as a list of
processor IDs that define the processors that will hold the data.  Given
this data partitioning the program is automatically
parallelised. Overloaded MATLAB functions, which take distributed
arrays as arguments, automatically perform the required message
passing to coordinate computation on these arrays in parallel. Bliss
and Kepner report comparable performance to C MPI code for some
benchmarks, with a greatly reduced amount of code. A user
study they have performed indicates that there is not a steep learning curve
for converting a MATLAB application to a pMATLAB application and less
than 1\% of code requires modification. The HPF programming model limits 
the applications which can benefit from this type of parallelisation
to those which have regular data access patterns. 
Irregular access patterns on distributed matrices can result in
significant slow downs.

The team at MIT build upon pMATLAB again with
pMapper~\cite{Bliss08012007}.  pMapper provides fully automated
parallelisation of MATLAB code by generating the array distribution
maps for pMATLAB. A heuristic approach is taken to produce maps at
run-time. One feature of pMapper is that it is designed to
be independent of any single parallel architecture or parallel
library. Instead, when pMapper is installed on a system it performs an
initialisation phase which generates a performance model for the
system. This performance model contains timing information for
parallel functions with different input sizes. The performance model
is then used to generate maps specific to an architecture. pMapper is designed
specifically for signal processing applications and the benchmark
results are largely due to simulations.

The company MathWorks (the vendor of MATLAB) provides parallel
extensions for MATLAB in the form of a commercial Parallel Computing
Toolbox~\cite{sharma2009mlp}. The toolbox has been designed 
with a heavy focus on an explicit parallel programming model. They
achieve this by introducing several extensions to the MATLAB language including
parallel loops, distributed arrays and message passing functions.
The basic structure of the system uses several \emph{worker} processes
which can communicate with each other and the client MATLAB process.
The client formulates computations as a series of jobs which are submitted
to a scheduler, executed on workers and return a result. The toolbox
allows up to 8 workers to be running locally on a single machine and
can be scaled to multiple computers in a cluster with the use of MATLAB
Distributed Computing Server. The toolbox draws from several other
pieces of software such as MPI libraries and parallel math libraries.
It is worth mentioning that MathWorks conducted a survey and found that
``reusability of existing MATLAB code was cited as the most important
feature of any parallel computing toolset'', however they have introduced
several new constructs to the language which must be used in order
to achieve high levels of parallelism.

With reference to the goals of our framework, they are largely
unaddressed by message passing approaches to parallel MATLAB. Firstly,
existing systems are targeted toward distributed parallel
architectures. Although they can achieve parallelism on some modern
accelerator architectures for high-performance computing,
communication overheads can restrict even moderate utilisation of the
processing elements. In MatlabMPI, for example, inter-process
communication is achieved through the file system which is going to be
many orders of magnitude slower than specific communication means on
modern accelerator hardware. Secondly, as noted most message passing
systems require a large amount of intervention from the programmer in
order to achieve parallelism.

\subsubsection{Compilers}
FALCON~\cite{derose1996fmi,derose1996mft} is a programming environment
developed at the University of Illinois designed to support the
development of optimised numerical applications and libraries. The
input language is MATLAB which is translated to the target language,
Fortran 90, in three stages. The first stage, program analysis,
constructs an Abstract Syntax Tree (AST) and determines the type and
shape of variables.  As with any untyped language, in order to compile
the code, the types of variables must first be inferred. This can not
always be done statically, so FALCON utilises both static and dynamic
analysis, combined with user input to determine variable types. The
next stage uses a collection of transformation rules to restructure
the code in order to perform optimisations. This phase is fully
interactive with optimisations suggested by FALCON and selected by the
programmer. Finally, the code generation stage uses information
collected during the analysis phase to produce Fortran 90 code. This
code is annotated with compiler directives which allow automatic
parallelisation by Polaris~\cite{blume1996ppp}, a parallelising
Fortran compiler. Despite achieving speedups of up to 1000 times over
the MATLAB interpreter the work presented in FALCON focuses mainly on
producing high-performance sequential Fortran code. Parallelisation of
this code is left largely uninvestigated and as future work with no
results reported for benchmarks on multiprocessor
machines. Unfortunately this project is now seemingly dormant.

Work from FALCON continued with a MATLAB Just-In-Time (JIT) compiler
called MaJIC~\cite{Almasi2001}. In JIT compilation no static analysis
of the code is done. Instead, portions of the code are compiled at run
time in order to achieve better performance than that of purely
interpreted code. Although this project did not explore program
parallelisation in the compilation process it remains interesting since, as
far as the authors are aware, it is the only research project that
uses JIT compilation for MATLAB. JIT compilation is desirable for two
reasons. Firstly, it allows MATLAB to remain an interpreted (and
untyped) language which is important in facilitating rapid
prototyping.  Secondly, it means that optimisations can be applied
at run time that may not be apparent at compile time. In our work we
adopt JIT compilation techniques to efficiently
schedule matrix operations on the PowerXCell architecture.

Otter~\cite{quinn1998prp,quinn1998obg} is another MATLAB compiler
developed at Oregon State University. Unlike FALCON which translates
MATLAB scripts first to sequential Fortran code and then parallelises
that code, Otter translates MATLAB code directly to parallelised C
code with calls to MPI and existing parallel numerical libraries.  The
translation process used in Otter to generate C code is based on the translation process
developed for FALCON. However, extra compilation passes are used to introduce
parallelism. Calls to the Otter run-time library are inserted to provide a means
of
communication between processors or utilise existing numerical
libraries, such as ScaLAPACK when possible. Otter is
advantageous as it provides completely automatic parallelisation of
general code and is portable between all architectures supporting MPI. A
limitation, as with pMATLAB, is that it again achieves parallelism through
distributed arrays which require regular data access patterns to achieve
good performance. Also, it is bound to the usage of numerical libraries and
thus the matrix distributions associated with those libraries.  While
these compilers provide some automated parallelisation of code, they
are once again focused toward parallelism on distributed architectures.

MATCH~\cite{795917} is another MATLAB compiler however, unlike FALCON
and Otter, it is targeted toward heterogeneous parallel
architectures. These architectures consist of an interconnection of
various processing components such as embedded processors, digital
signal processors and commercial off-the-shelf components, each of
which can perform certain types of computation very
efficiently. Despite being developed nearly a decade ago, MATCH is relevant
in the context of our research because the
heterogeneous architectures presented resemble modern parallel
architectures in some ways. In particular these heterogeneous
architectures use a single general purpose processor (the
MicroSPARC-II in this case) to handle the control flow of the program
and specialised units perform computation-intensive portions of the
program more efficiently. This can be compared to the Cell
processor which has a PowerPC processor to handle control flow
and several Synergistic Processing Elements (SPEs) optimised for
computation. MATCH works by first producing an abstract syntax tree,
as is done in FALCON and Otter. The AST is then partitioned into
sub-trees whose nodes correspond to library
functions in MATLAB.  Each sub-tree is mapped to a given processing
resource depending on how efficiently that portion of the program can
be executed on the resource. This mapping can be produced
automatically by MATCH using timing information and a mixed integer
linear programming approach, however user guided mappings are also
possible. Code is then generated for each partition of the AST, 
depending on the architecture to which the partition is 
mapped. Built-in MATLAB functions, such as matrix
multiplies, are pre-compiled in architecture specific libraries.
 Distributed arrays are again used to generate
parallel code on target architectures for user defined functions.

\subsubsection{Backend Support}
Our framework falls into the category of backend support as it has
a computation engine that receives matrix operations from the Octave
interpreter, executes the operations and returns the result to the client.

Jacket~\cite{jacket1} is a MATLAB backend that runs on General
Purpose Graphic Processing Units (GPGPUs).  It is a commercial product
developed by Accelereyes and few details are available about its design. It is one of several systems that have emerged recently
for the acceleration of MATLAB code on GPGPUs~\cite{gpulib,gpumat}. 
In Jacket, the
programmer casts matrices into GPU matrices, which are
transferred to GPU memory. Operations on GPU matrices are
executed on the GPGPU by compiling code on-the-fly with the
NVIDIA/CUDA~\cite{nvidia2007compute} infrastructure.  Jacket also provides some syntax
extensions for executing for-loops with parallel semantics. Though
Jacket provides a high level of abstraction from the details of
parallelism, the programmer still requires knowledge of the underlying
GPGPU architecture to write efficient Jacket/MATLAB
code as they must understand the overheads involved in transferring
data between main memory and the device memory of the GPGPU. 
Furthermore, it appears that Jacket essentially provides MATLAB bindings to CUDA
functions and as such, instruction level parallelism is not exploited by the system.

Star-P~\cite{starpwhitepaper} is another commercial product which
began as a research project at MIT in 1998 and was commercialised in
2004. It involved the same authors who conducted the parallel MATLAB
survey discussed previously and incorporates ideas from several
existing parallel MATLAB systems. From a user's perspective, Star-P
behaves in a similar way as Jacket. Programmers denote a distributed
variable by tagging it with the characters $*p$, converting it to a
custom data type.
Star-P then provides overloaded versions of standard MATLAB
functions which operate on the custom data type in parallel.
Star-P targets a distributed computing
environment. The main purpose of the MATLAB client is to distribute
tasks to a cluster running the Star-P server, which will run
computations and deliver results.  The server software is a general
purpose computation engine which is not specific to MATLAB and
supports clients running many different languages, including Python. It
utilises existing math libraries to perform parallel computations on
each server.

\subsection{Task Scheduling\label{sec:The-Scheduling-Problem}}

In our work we attempt to improve utilisation of modern accelerator architectures by
scheduling matrix operations among parallel execution elements. 
We wish to schedule operations such that (1) the data dependencies between operations are satisfied, i.e.~an
operation can only begin execution after its operands have been computed, and (2)
the total time of execution (makespan) is minimised. This problem is well explored
in literature~\cite{344618}, with two main formulations --- the \emph{delay model}~\cite{hu1961psa} and the 
\emph{malleable tasks model}~\cite{turek}. 

\subsubsection{The Delay Model}\label{sec:delay}

In the delay model we are given a precedence graph
$G(V,E)$ as the input to the problem along with a number of available
processors $p$. $G$ is a directed acyclic graph (DAG) whose vertices 
$V$ represent the tasks and directed edges $E$
represent the dependencies between tasks. There is an arc from a node
$s\in V$ to a node $t\in V$ if task $t$ depends on task $s$. Vertices
are annotated with the cost of executing the task. An edge from $s$
to $t$ is annotated with the cost of communication between the tasks
$s$ and $t$.

The goal is to find a legal schedule that minimises the makespan.
A schedule consists of a start time for each task and the processor
to which it is assigned.
Note that a scheduled task must be executed completely on the processor
it has been assigned without interruption.

Our problem of scheduling matrix operations on SPEs resembles this
problem. Matrix operations become the vertices in the precedence
graph and the data dependencies between operations form the edges. The
costs of operations can be estimated by using a time model that is obtained by profiling
and regression analysis.

It has been proven that finding an optimal solution to the scheduling
problem with the delay model is NP-complete~\cite{garey1979cai}. 
As such, work has been done
on finding optimal, polynomial time solutions for simplified versions
of the problem. Hu~\cite{hu1961psa} addresses a variation of the problem
in which the task precedence graph is assumed to have a tree structure,
tasks are assumed to have unit execution costs and communication
costs are ignored. A simple, list scheduling algorithm is proposed
to find the optimal solution in polynomial time.
List scheduling algorithms are a common
approach to the scheduling problem that order tasks in a list according
to some heuristic. The heuristic typically ensures a topological ordering of
the task precedence graph. Tasks are then iteratively removed from
the list in order and assigned to the processor allowing the earliest
start time. Hu uses a heuristic in which tasks are
ordered by their distance from the root of the tree in the task precedence graph.

Coffman and Graham~\cite{coffman1972ost} also propose a polynomial
time, optimal algorithm with the simplifying assumptions that there
are only 2 processors, unit execution costs and no communication costs.
Papadimitriou and Yannakakis~\cite{papadimitriou1979sio} propose
a polynomial time, optimal algorithm with the simplifying assumption
that there are unit execution costs and the precedence graph is interval
ordered. In our work we must schedule operations that
have arbitrary precedence constraints over more than 2 processors. Hence,
the above mentioned algorithms cannot be used.

Recognising the difficulty of finding an optimal solution for the
general problem, many heuristic algorithms~\cite{344618} have arisen which are usually
variations of the list scheduling approach. Adam et al.~\cite{adam1974cls}
performed a simulated analysis of these algorithms and found that
a Highest Level First with Estimated Times (HLFET) heuristic gave
the best results for the chosen benchmarks. In this approach, tasks are ordered by the cost
of the longest path to an exit task (a sink node) in the precedence
graph. This value is known as the \emph{b-level} of the task. Tasks
with a larger b-level are scheduled first. Using this approach ensures
that tasks along the critical path of the graph will be scheduled
first. The critical path is the longest path through the precedence graph from
an entry node (source) to an exit node (sink) and is important as it represents the
minimum length of an optimal schedule. It was shown that this approach
can produce near-optimal schedules and has a run-time of $O(n^{2})$
for $n$ tasks. Graham~\cite{graham1966bcm} showed that the schedule
length, $SL$, generated using level-based, list-scheduling algorithms
such as HLFET was no more than twice the length of the optimal schedule,
$SL_{opt}$, such that $SL\leq(2-\frac{1}{p})SL_{opt}$ where $p$
is the number of processors.

When communication costs are allowed to be arbitrary, several other
algorithms have been proposed. A well known algorithm in this category
is the Insertion Scheduling Heuristic (ISH)~\cite{kruatrachue1988gsd}.
It works in a similar way to HLFET by using the b-level of
a task as its priority in a list scheduling algorithm. However, while
HLFET may leave gaps in the schedule, ISH seeks to fill these gaps
and thus reduce the schedule length. Every time a gap is introduced
in the schedule, ISH examines unscheduled tasks which are ready to
be scheduled and attempts to use them to legally fill the gap.
The time complexity of ISH is $O(n^{2})$. 

Another category of algorithms are proposed for scheduling tasks
with arbitrary precedence constraints which allow task duplication. 
These algorithms
recognise that by redundantly duplicating tasks, the time waiting
for a parent task to complete might be reduced which may reduce the
overall time of the schedule. The Duplication Scheduling Heuristic
(DSH)~\cite{kruatrachue1987duplication} is a representation of such algorithms. It works
by iteratively attempting to duplicate a tasks ancestors on a processor if it
allows the task to be scheduled at an earlier start time. 
The time complexity of the algorithm is $O(n^{4})$ which may be unsatisfactory
for applications with a limited time budget for scheduling.

More recently, polynomial time approximation algorithms have been proposed for
the general case of the delay model scheduling problem, with communication costs considered.
These yield a solution of bounded quality. One such algorithm~\cite{MSS96} 
gives an approximation guarantee of 
$2.33 - 1.33 m$ where $m$
is the number of edges in the task precedence graph. The algorithm is
based on rounding of a relaxed linear programming solution to find
a schedule with minimum makespan on an unbounded number of processors. A list
scheduling algorithm is then used to produce an optimal schedule for a finite
number of processors. Due to the use of linear programming, these algorithms
can again be impractical for time-critical applications.

\subsubsection{The Malleable Tasks Model}

The malleable tasks model is a more recent model than
the delay model and as such, there is less work on it. It is similar
to the delay model but with two main differences. Firstly, communication costs
are incorporated in the execution cost of each task. Secondly, tasks
in this model do not have to execute on a single processor but can
be divided up and executed over several processors with a reduced execution time.
As such vertices in the precedence graph are not merely labelled with
a single execution cost but instead a cost function, $c(k)$, which
is dependent on the number of processors, $k$, which the task is assigned
to. A schedule that satisfies the problem consists of a start time
for each task and a number of processors over which each task is to
execute.

We could apply the malleable task scheduling problem to our framework because
the tasks in our framework are matrix operations. These can be divided
into operations on sub-matrices which can be executed across several 
processing elements.

It has been proven that the scheduling problem for malleable tasks
with arbitrary precedence constraints is NP-hard with only 3 processors~\cite{leung}.
Even the problem of finding an optimal schedule for malleable tasks
\emph{without} precedence constraints has been shown to be NP-hard
with 5 or more processors~\cite{leung}. 

Despite the hardness of the problem there has been some recent work
on approximation algorithms. Lepère et al.~\cite{lepere} present
a polynomial time approximation algorithm for scheduling malleable tasks under
precedence constraints with an approximation ratio of $3+\sqrt{5}\approx5.23006$.
Their algorithm is developed by identifying the relationship between
the scheduling problem and the allotment problem, which is approximated
by rounding the solutions of a linear programming relaxation~\cite{skutella1998aar}.
This is combined with a list scheduling algorithm to provide the final approximation
algorithm. Jansen and Zhang~\cite{jansen} improve on the
work of Lepère et al., with an approximation ratio of $100/43+100(\sqrt{4349}-7)/2451\approx4.730598$. Existing approximation algorithms for the malleable tasks scheduling problem
may not be viable in time-critical situations. This is due to their use of linear programming 
which has polynomial time complexity, but high overheads in practice.

\section{System Overview} \label{sec:overview}

Our framework is a system extension for Octave. It takes the form of a shared library that is loaded by the Octave interpreter at runtime. The framework automatically parallelises matrix instructions for the Cell Broadband Engine and only requires minimal changes to existing Octave code when deployed. These changes consist of casting all matrix declarations to a new Octave data type called \psmatrix. Standard operators, such as \mytexttt{+}, \mytexttt{-} and \mytexttt{*}, as well as built-in functions, such as \mytexttt{sin} and \mytexttt{round}, have been overloaded to operate on the \psmatrix data 
type (see section~\ref{sec:impl}). 

The underlying idea of our framework is to execute several matrix instructions at the same time to optimally harness the computational power of the Cell Broadband Engine. However, the sequential execution semantics of matrix languages do not provide the notion of concurrent execution of matrix instructions, besides dividing matrix instructions into sub-operations which are distributed among parallel processing elements. To further increase the parallelism in Octave programs, we employ lazy evaluation of matrix instructions. Lazy evaluation delays the execution of matrix instructions until the result of an instruction is required. This concept is heavily used in functional programming languages and has numerous applications there, including avoiding unnecessary computations and error conditions, being able to operate on infinite data structures, and  defining control flow structures in the language itself~\cite{72554}.

\begin{figure}
\begin{center}
\includegraphics[scale=0.6]{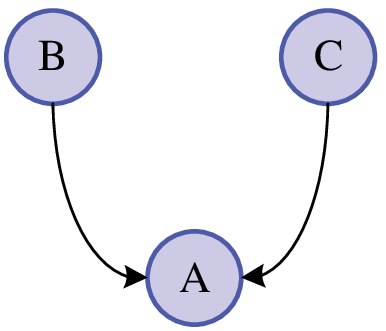}
\par\end{center}
\caption{A data dependence graph for the statement $A=B*C$. $A$ depends
on the values of matrices $B$ and $C$ so there is a directed edge in the graph from $B$ to
$A$ and from $C$ to $A$. \label{fig:overview-dep-graph}}
\end{figure}

Our framework uses lazy evaluation to collect a \emph{trace} of matrix instructions. The overloaded functions of the new data type facilitate the construction of the trace which is then analysed to determine the data dependencies between operations. The data dependencies in the trace loosen the strict sequential ordering of instructions to a partial ordering that allows independent matrix instructions to be executed in parallel. A
\emph{data dependence graph} $G(I,E)$ is constructed for the trace where $I$ is the set of nodes in the graph representing instructions in the trace, and $E$ is the set of data dependencies between pairs of matrix instructions. For example, the lazily evaluated statement $A=B*C$ imposes two directed edges $(B,A)$ and $(C,A)$ because the result $A$ of the matrix multiplication depends on the matrix operands $B$ and $C$, as shown in 
Figure~\ref{fig:overview-dep-graph}. The source nodes in this graph (i.e.~in-degree of 0)
are typically constant or computed matrices, whose value is already available.

Our framework constructs the data dependence graph on the fly when matrix instructions are lazily evaluated. Note that the constructed graph is acyclic even for loops. A matrix instruction that is executed multiple times inside a loop is represented by a set of nodes in the graph. For each execution instance of the matrix instruction there exists exactly one node in the graph.

A trace will continue to grow in length as the program
is executed until either (1) a statement is reached that requires the result of an unexecuted matrix operation in the trace and cannot be lazily evaluated, e.g., displaying the value of a matrix, or (2) the length of the trace has reached a certain threshold, i.e., it becomes opportunistic to execute the matrix operations in parallel. If either of these criteria are met, execution of the data dependence graph is triggered.

The first step in execution of the data dependence graph is \emph{lowering} the graph. The memory of the SPEs on the Cell architecture is at a premium, i.e., code and data share the same memory which is limited to 256kB. To be able to compute larger matrices, the framework decomposes matrix instructions into matrix instructions that operate on sub-matrices. This decomposition of the instructions not only enables the execution of matrix instructions on the parallel processing elements of the Cell but also exposes data parallelism in the matrix instructions. We refer to this process of decomposing the instructions of the data dependence graph into instructions that operate on sub-matrices as lowering (see section~\ref{sec:lowering}). 
The lowering process rewrites the original data dependence graph into a lowered data dependence graph, which has an increased number of operations and dependencies. 

The lowered data dependence graph is then scheduled among the parallel processing elements in the underlying architecture (see section~\ref{sec:scheduling}). Scheduling assigns
each parallel processing element a subset of the lowered operations which have a specified
order in which they are to be executed. The scheduling is performed in a way
that satisfies the data dependencies between operations and minimises the total execution
time (makespan) of the trace. Execution times of matrix operations are estimated using time-models
constructed from profiling data. This ensures that an accurate schedule is produced. Since the scheduling of operations happens at run-time it is also important that a schedule is produced quickly.

\begin{figure}
\begin{center}
\includegraphics[scale=0.6]{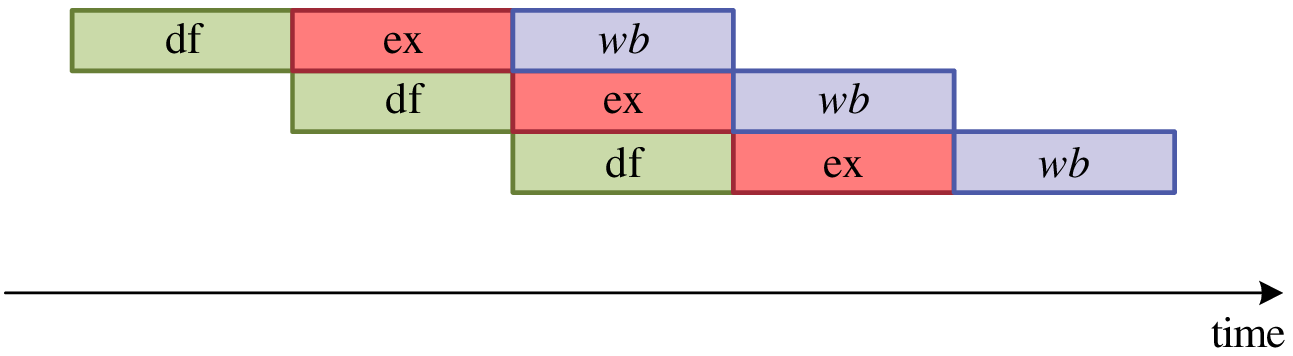}
\par\end{center}
\caption{Pipelined execution of matrix instructions. There are 3 stages in the pipeline: 
data fetch (df), execution (ex) and write back (wb), each of which can be overlapped. \label{fig:pipeline}}
\end{figure}

In the final step, the lowered matrix instructions are executed 
on the parallel processing elements according to the schedule. This component of the framework
is referred to as the \emph{computation engine}. The computation engine is abstracted from 
the details of the underlying architecture and instead viewed only as an asynchronous,  pipelined,
Multiple-Instruction/Multiple-Data (MIMD) architecture with a shared memory. The architecture executes matrix instructions concurrently in an asynchronous fashion.
To hide the communication between memory and the processing element, the architecture 
utilises a pipeline. The pipeline stages of a single matrix instruction are assumed
to be timely interleaved  as depicted in Figure~\ref{fig:pipeline}, which is an idealised scenario assuming that the durations of the stages have the same duration and there are no ``bubbles'' or gaps in the pipeline. We employ the
following pipeline stages in our computation engine:
\begin{enumerate}
\item \textbf{Data Fetch (df):} The operands of the matrix instruction are loaded
from main memory into the memory of the parallel processing element,
\item \textbf{Execute (ex):} The matrix instruction is executed on the parallel processing element,
\item \textbf{Write Back (wb):} the result of the matrix instruction is written back to main memory.
\end{enumerate}
In contrast to
a super-pipelined, super-scalar CPU~\cite{77493}, the matrix instructions are not assumed to be synchronised, i.e., there is no global clock that triggers a new step with a constant period. 

This abstraction from the details of the underlying parallel architecture allows the framework to be easily ported to many different architectures (such as GPGPUs or multi-core CPUs) by
customising the computation engine. In this work, we implement a computation engine for the Cell
Broadband Engine architecture (see section~\ref{sec:cell}). Each of the SPEs in the Cell processor acts as a processing element and executes a sequence of matrix instructions. We call the program that runs on the SPEs a \emph{Matrix Execution Unit} (MEU). The MEUs need to be synchronised globally. An \emph{execution control} mechanism guarantees that a matrix instruction on an MEU is only executed if its operands are already available in main memory. The execution control is run on the PowerPC Processing
Element (PPE) of the Cell. 

After the completion of the execution of lowered operations, the results are made available
for use by the Octave interpreter.

\begin{figure}
\begin{center}
\includegraphics[width=0.3\columnwidth]{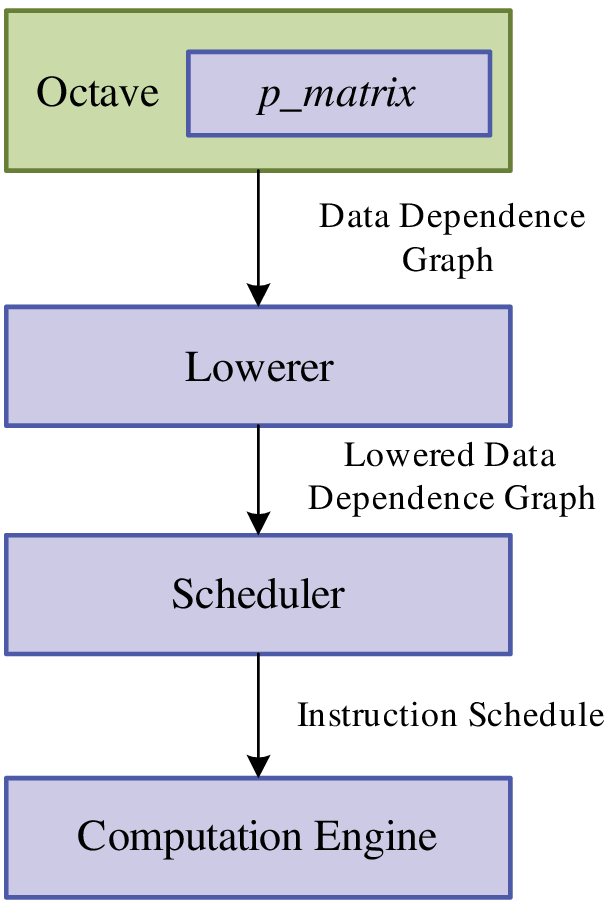}
\end{center}
\caption{The main software components of the framework. These are: (1) the
\psmatrix data type extension to the Octave interpreter, (2) the lowerer
for partitioning matrix operations, (3) the scheduler for scheduling matrix operations
among the parallel execution elements of the architecture, and (4) the computation
engine which executes the partitioned matrix operations.\label{fig:overview}}
\end{figure}

The software components of our framework are depicted in
Figure~\ref{fig:overview}. The first component is a data type
extension to the Octave interpreter called \psmatrix. The operators for the
new data type are overloaded to perform lazy evaluation and to obtain
the execution trace. From this, the data dependence graph of operations in the trace
is computed, on the fly.  The data dependence graph is passed on
to the lowerer which decomposes matrix instructions into
instructions that operate on sub-matrices. The scheduler computes the
schedule for the computation engine.  Finally, the matrix instructions
are executed on the computation engine according to the schedule. All four components
can be executed in parallel, i.e., the Octave interpreter, the
Lowerer, the Scheduler and the computation engine are executed in separate
execution threads allowing overlapped execution of all four
components.

\section{Motivating Example} \label{sec:motivating}

Assume we want to compute the value of $B=A^i$ using
an Octave script, where $A$ is a square matrix of dimensions $n\times n$ and  $i$ is a positive integer.
For the purpose of this example we let $A$ be a random $100\times 100$
matrix and $i=3$. A \naive Octave implementation for calculating matrix $B$ and 
displaying its value to the screen is given in Figure~\ref{alg:motivation}.  

To use our framework, the 
Octave programmer converts matrix declarations to the
custom \psmatrix Octave data type by wrapping them with the 
\psmatrix function. In our example program, the only matrix
declaration is for the matrix $A$ in line 1 of Figure~\ref{alg:motivation}.
The modified code that uses our framework is shown in Figure~\ref{alg:motivation-mod}.
Note that a user does not have to decide
which matrices should be converted to this new data type (all matrices can be safely converted),
and the need for this additional data type could be completely eliminated with alterations to the Octave interpreter.

\begin{figure}
\begin{minipage}[b]{0.49\textwidth}
\centering
\begin{lstlisting}[language=myoctave,frameround=fttt,frame=trBL]
A = rand(100);
i = 3;

B = A;
for k = 1:i-1
	B = B * A;
end

disp(B);
\end{lstlisting}
\caption{Octave script for the computation of $B=A^i$.\label{alg:motivation}}
\end{minipage}
\hspace{0.1cm}
\begin{minipage}[b]{0.49\textwidth}
\centering
\begin{lstlisting}[language=myoctave,frameround=fttt,frame=trBL]
A = p_matrix(rand(100)); (*@\label{line:declarea}@*)
i = 3;

B = A; (*@\label{line:assignb}@*)
for k = 1:i-1
	B = B * A; (*@\label{line:multiply}@*)
end

disp(B); (*@\label{line:dispb}@*)
\end{lstlisting}
\caption{Parallel data type modifications for $B=A^i$.\label{alg:motivation-mod}}
\end{minipage}
\end{figure}

When the script in Figure~\ref{alg:motivation-mod} is executed in the Octave interpreter, matrix operations involving
operands of the type \psmatrix are lazily evaluated by our framework. In the example program,
the only such matrix operation is the multiplication
\lstinline[language=myoctave]!B = B * A!, in line~\ref{line:multiply} of 
Figure~\ref{alg:motivation-mod}. The result of this operation (the matrix $B$) 
is never required
inside the for-loop. Hence, rather than eagerly executing this operation each
time it is reached inside the loop (as would be the case when executing in a
default installation of Octave), lazy evaluation defers execution of
the operation. Instead, the operation is recorded in an execution trace (shown in 
Figure~\ref{alg:motivation-unrolled}).
The trace is kept as an internal data structure and a data dependence
graph of the trace is constructed on the fly. The data dependence graph 
shows which operations depend on the results of 
other operations and determines a partial order in which operations must be executed to 
yield a correct result. This partial order enables the parallel execution of matrix 
operations in contrast to strict sequential execution of 
matrix languages.  

\begin{figure}
\begin{center}
\begin{eqnarray*}
B_{0} &=& A;\\
B_{1} &=& B_{0} * A;\\
B_{2} &=& B_{1} * A;
\end{eqnarray*}
\end{center}
\caption{The trace of lazily evaluated matrix operations.\label{alg:motivation-unrolled}}
\end{figure}

\begin{figure}
\centering
\subfigure[Data dependence graph after execution of line~\ref{line:declarea}.]
{\includegraphics[width=0.2\columnwidth]{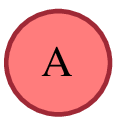} \label{fig:motivation-dep1}}
\hspace{3mm}
\subfigure[Data dependence graph after executing line 4.]{\includegraphics[width=0.2\columnwidth]{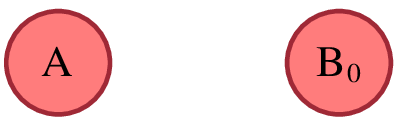} \label{fig:motivation-dep2}}
\hspace{3mm}
\subfigure[Data dependence graph after executing line 6 in the first iteration.]{\includegraphics[width=0.2\columnwidth]{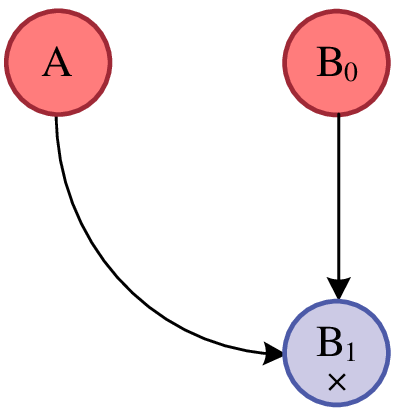} \label{fig:motivation-dep3}}
\hspace{3mm}
\subfigure[Data dependence graph after executing line 6 in the second iteration.]{\includegraphics[width=0.2\columnwidth]{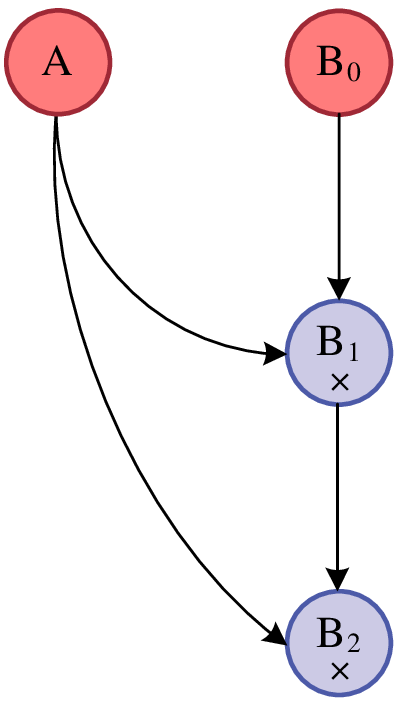} \label{fig:motivation-dep4}}
\caption{Showing the construction of the data-dependence graph as the program listed
in Figure~\ref{alg:motivation-unrolled} is executed. Red nodes represent constant
matrices and blue nodes represent matrix multiplication operations.}
\label{fig:motivation-dep}
\end{figure}

Let $B_{k}$ denote the value of $B$ in the $k$'th iteration of the for-loop.
Execution of the example program from Figure~\ref{alg:motivation-mod}, construction
of the trace (Figure~\ref{alg:motivation-unrolled}) and construction of the data 
dependence graph (Figure~\ref{fig:motivation-dep}) proceeds as follows:

\begin{enumerate}
\item Matrix~$A$ in line~\ref{line:declarea} is declared. The conversion
of the random matrix to type \psmatrix causes our framework to add the matrix to the data dependence
graph as a constant matrix (Figure~\ref{fig:motivation-dep1}). Note that constant matrices
(like $A$) are not true matrix operations as their result is already available and they
do not need to be executed. Instead they are added to the data dependence graph
to denote the dependence of another operation on that matrix.
\item
Matrix~$A$ is assigned to variable~$B_{0}$ in line~\ref{line:assignb}. Again this causes
a constant matrix to be added to the data dependence graph (Figure~\ref{fig:motivation-dep2}). 
Note that a deep copy of matrix~$A$ is not made upon assignment to $B_{0}$. For the sake of 
simplicity, we represent them separately in the data dependence graph.
\item In the first iteration of the for-loop, $B_{0}$ is multiplied by $A$ and the result 
stored in $B_{1}$. The result of this
operation is not yet required in the program. Hence, execution is deferred and a
multiplication operation
is added to the trace and data dependence graph (Figure~\ref{fig:motivation-dep3}). The operation
depends on 2 values --- matrix~$A$ and matrix~$B_{0}$. Arcs $(A, B_{1})$ and
$(B_{0}, B_{1})$ denote these dependencies in the dependence graph.
\item In the second iteration of the for-loop, $B_{1}$ is multiplied by $A$ and the result stored in $B_{2}$. Again, the result
of the multiplication is not required immediately, so another
multiplication operation is added to the trace and
data dependence graph (Figure~\ref{fig:motivation-dep4}). The operation
depends on 2 values --- matrix~$A$ and the result of the previous matrix
multiplication operation, $B_{1}$. Arcs $(A, B_{2})$ and $(B_{1}, B_{2})$ denote
these dependencies in the dependence graph.
\item The terminating condition of the for-loop is reached.
\end{enumerate}

The final statement of the program to be executed is given in line~9 of Figure~\ref{alg:motivation-mod}. It requests that the value of $B_{2}$ be printed to the screen. 
However, the value of $B_{2}$ has not yet
 been computed due to lazy evaluation. 
 This causes execution of the Octave program to be halted while the data
dependence graph from Figure~\ref{fig:motivation-dep4} is executed to obtain the required 
result.
Note that in this example, execution of the trace was forced by a required value. The
other cause of executing a trace is that the length of the trace becomes too large. If that
is the case then execution of matrix operations and execution of the Octave interpreter are performed concurrently.

The first step in execution of the data dependence graph is lowering the graph 
(see section~\ref{sec:lowering}).
Lowering partitions matrix operations on large matrices into operations
on sub-matrices. This is necessary because some matrices may be too large to
fit into the memory of parallel execution elements (e.g., with the Cell architecture), but
it has the beneficial side-effect of exposing data parallelism in matrix operations.
Assume that the parallel elements have enough memory to store operands of
dimensions $50\times 50$, however, the matrices in the example are of dimensions 
$100\times 100$.
We can partition the $100\times 100$ matrices into four $50\times 50$ blocks, and use block matrix
multiplication to perform the multiplication operations in our program. Block matrix
multiplications works in the same way as regular matrix multiplication, except that instead
of multiplying and adding the scalar elements of the two operands, we multiply and add the
partitioned sub-matrices of the operands. Figure~\ref{fig:motivation-block-matrix-mult}
shows how this partitioning occurs for the first multiplication operation, $B_{0}A$.

\begin{figure}[p]
\begin{center}
\includegraphics[width=\textwidth]{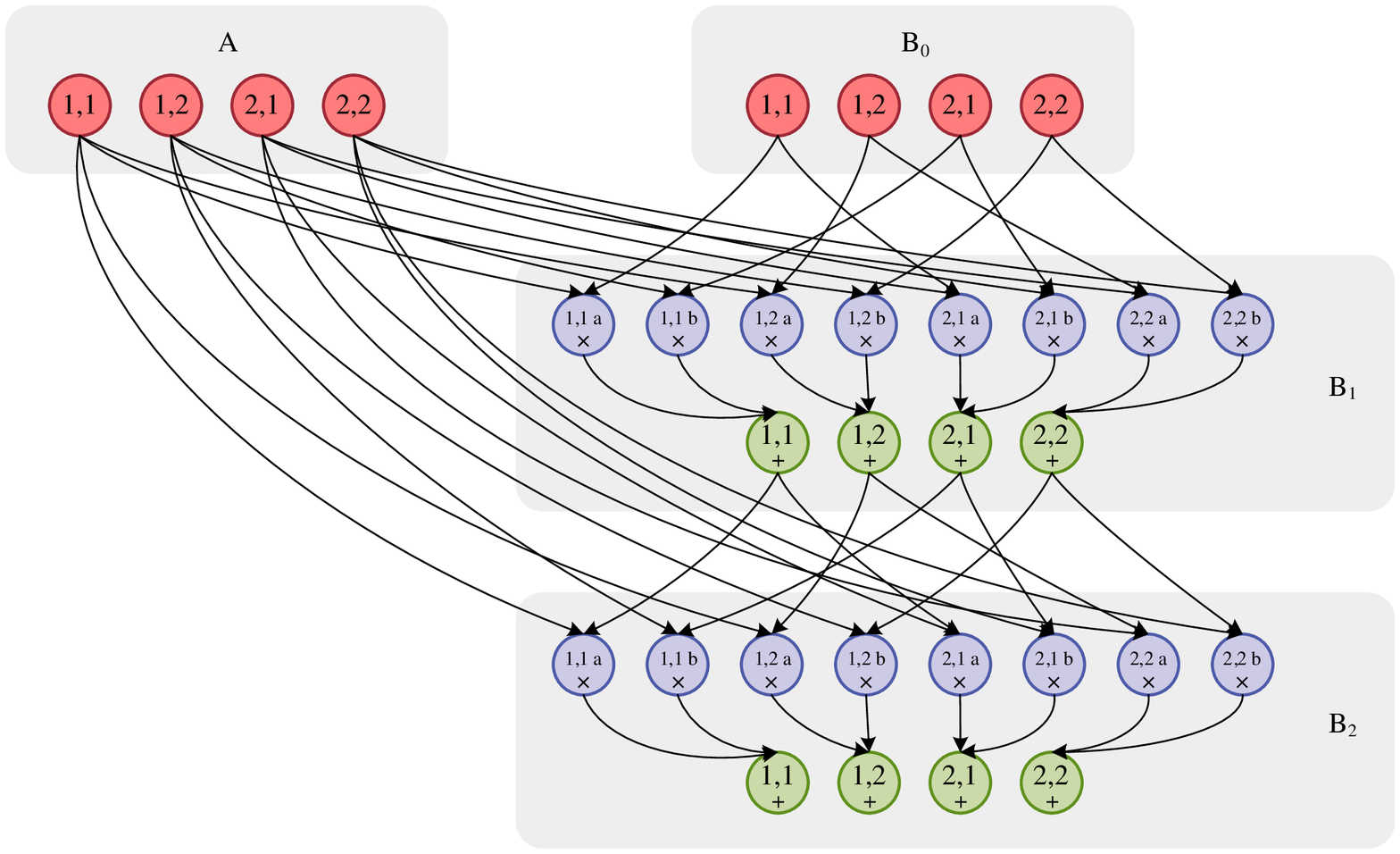}
\caption{The data dependence graph of lowered operations for the example program in
Figure~\ref{alg:motivation-unrolled}.  Red nodes represent constant matrices,
blue nodes represent matrix multiplication operations and green nodes represent
matrix addition operations. Each of the operations from the original data dependence
graph in Figure~\ref{fig:motivation-dep} corresponds to several operations in the lowered 
graph (which are grouped in the shaded areas of the diagram). A single matrix 
multiplication operation (e.g.~$B_{1}$) is lowered to a series of 
multiplications which are summed to produce the final result.
Partitioning of matrices is preserved across operations, leading to increased
opportunities for parallelism.}
\label{fig:motivation-lowered}
\end{center}
\end{figure}

Note that a single matrix operation in our original dependence graph will result in many lowered
operations after partitioning. These operations form a new data dependence graph, called
the lowered data dependence graph. The lowered data dependence graph for our example is
shown in Figure~\ref{fig:motivation-lowered}.

\begin{figure}[p]
\begin{center}

\begin{equation*}
\scriptsize
\begin{split}
B_{0}A &=
\left[%
\begin{array}{ccc:ccc}%
\MES{(b_{0})}{1}{1} & \cdots & \MES{(b_{0})}{1}{50} & \MES{(b_{0})}{1}{51} & \cdots & \MES{(b_{0})}{1}{100} \\
\vdots & \ddots & & & & \vdots\\
\MES{(b_{0})}{50}{1} & & & & & \MES{(b_{0})}{50}{100}\\
&&&&&\\
\hdashline
&&&&&\\
\MES{(b_{0})}{51}{1} & & & & & \MES{(b_{0})}{51}{100}\\
\vdots & & & & \ddots & \vdots\\
\MES{(b_{0})}{100}{1} & \cdots & \MES{(b_{0})}{100}{50} & \MES{(b_{0})}{100}{51} & \cdots & \MES{(b_{0})}{100}{100} \\
\end{array}
\right]%
\left[%
\begin{array}{ccc:ccc}%
\MES{a}{1}{1} & \cdots & \MES{a}{1}{50} & \MES{a}{1}{51} & \cdots & \MES{a}{1}{100} \\
\vdots & \ddots & & & & \vdots\\
\MES{a}{50}{1} & & & & & \MES{a}{50}{100}\\
&&&&&\\
\hdashline
&&&&&\\
\MES{a}{51}{1} & & & & & \MES{a}{51}{100}\\
\vdots & & & & \ddots & \vdots\\
\MES{a}{100}{1} & \cdots & \MES{a}{100}{50} & \MES{a}{100}{51} & \cdots & \MES{a}{100}{100} \\
\end{array}
\right] \\[2.5mm]
&=
\begin{bmatrix}
\MES{(B_{0})}{1}{1}&\MES{(B_{0})}{1}{2}\\
\MES{(B_{0})}{2}{1}&\MES{(B_{0})}{2}{2}\\
\end{bmatrix}
\begin{bmatrix}
\MES{A}{1}{1}&\MES{A}{1}{2}\\
\MES{A}{2}{1}&\MES{A}{2}{2}\\
\end{bmatrix} 
=
\begin{bmatrix}
\MES{(B_{0})}{1}{1}\MES{A}{1}{1} + \MES{(B_{0})}{1}{2}\MES{A}{2}{1} &
\MES{(B_{0})}{1}{1}\MES{A}{1}{2} + \MES{(B_{0})}{1}{2}\MES{A}{2}{2} \\
\MES{(B_{0})}{2}{1}\MES{A}{1}{1} + \MES{(B_{0})}{2}{2}\MES{A}{2}{1} &
\MES{(B_{0})}{2}{1}\MES{A}{1}{2} + \MES{(B_{0})}{2}{2}\MES{A}{2}{2} \\
\end{bmatrix}
\end{split}
\end{equation*}
\caption{Partitioning (lowering) for multiplication operation
 $B_{0}A$ from Figure~\ref{alg:motivation-unrolled}:
block-partitioning divides each $100\times 100$ operand
into four $50\times 50$ sub-matrices. Block matrix multiplication is used to multiply
sub-matrices. Each block of the result matrix is computed using two multiplications
and one add operation.}
\label{fig:motivation-block-matrix-mult}
\end{center}
\vspace{10mm}
\end{figure}

\begin{figure}[ht]
\begin{center}
\includegraphics[width=\textwidth, clip=true]{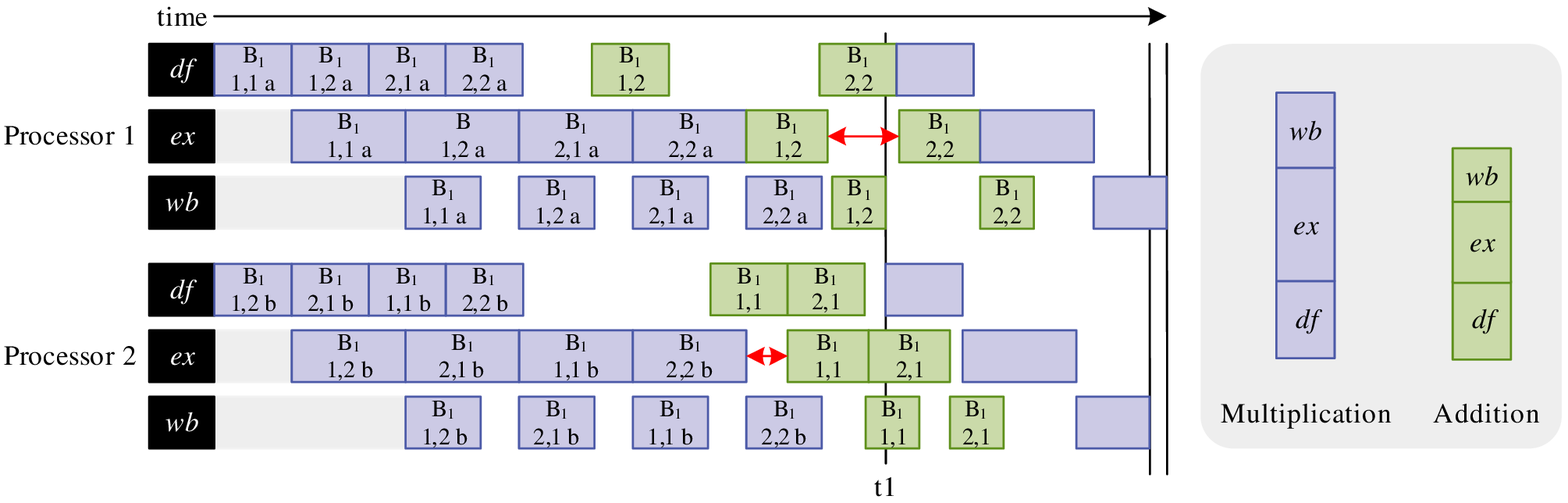}
\caption{A partial schedule of the operations from the lowered data dependence
graph from Figure~\ref{fig:motivation-lowered}. The sub-operations of operations
$B_{1}$ have all been scheduled. This schedule has been produced
by the heuristic scheduling algorithm described in subsection~\ref{sec:heuristic}.
The operation whose operands have been scheduled to complete execution at the earliest 
point in time will be scheduled
in the earliest available time-slot on any processor.
All three stages of the
pipeline must be considered for each processor --- data fetch (df), execution (ex) and write back (wb). These
stages can have different durations for different operations, such as multiplication and
addition, as shown on the right. 
 Red arrows show gaps in execution which we want to avoid to minimise the makespan. 
$t1$ marks the point in time where the write back of
operation~$(B_{1})_{1, 2}$ finishes. Thus, operation $(B_{2})_{1,2a}$ or
$(B_{2})_{2,2a}$ (which depend only on $(B_{1})_{1,2}$ and the constant matrix $A$)
can be scheduled to begin their data fetch stage any time after this point. 
The unlabelled blue rectangles on the end of the schedule show the result of scheduling
one of these multiplication operations on each of the processors. It can be seen
that the operation will complete earlier if scheduled on Processor 2, so the
heuristic will choose to schedule it on Processor 2. }
\label{fig:motivation-schedule}
\end{center}
\end{figure}

Once lowering is complete, operations are scheduled among the
execution elements of the underlying architecture (see section~\ref{sec:scheduling}).
The aim of scheduling is to assign operations
to processors in such a way that minimises the total execution time (makespan) of the operations
but also ensures that the data dependencies between operations are enforced:
an operation cannot execute before the results of all its operands are available.

Without specific reference to the details of the underlying architecture, it is viewed as
a pipelined architecture in which each processor can overlap computation of a matrix operation
with data transfers to and from main memory.  There are three pipeline stages
that are considered --- data fetch (df), execution (ex) and write back (wb).
The execution stage of an operation
can only begin after the completion of the data fetch of that operation and the
write back stage can only begin after the completion of the execution stage.
To produce an effective schedule, it
is necessary to estimate the time that each of these stages will take for each operation
that is to be executed.

A heuristic algorithm uses these execution time estimates to schedule the operations among
the execution elements. The heuristic
works by selecting the unscheduled operation whose operands have been scheduled to complete at the earliest point in time and
assigning that operation in the earliest available slot on a processor. A partial schedule
of lowered operations from the example is given in Figure~\ref{fig:motivation-schedule}. 
Note that the constant matrices (red nodes) are not scheduled as they do not 
need to be executed (their result is already available).
In this case, all of the sub-operations for operation $B_{1}$ have been scheduled
already. By examining the lowered data dependence graph from Figure~\ref{fig:motivation-lowered}, it can be seen that the next operations that are ready to be scheduled
(i.e.~whose operands have been scheduled already) are operations: $(B_{2})_{1,1a}$,
$(B_{2})_{1,1b}$, $(B_{2})_{1,2a}$, $(B_{2})_{1,2b}$, $(B_{2})_{2,1a}$,
$(B_{2})_{2,1b}$, $(B_{2})_{2,2a}$ and $(B_{2})_{2,2b}$.  Each of these operations 
depends on one of the
constant sub-matrices in $A$: $A_{1,1}$, $A_{1,2}$, $A_{2,1}$, $A_{2,2}$; as well as one 
of the addition sub-operations in $B_{1}$: $(B_{1})_{1,1}$,
$(B_{1})_{1,2}$, $(B_{1})_{2,1}$, $(B_{1})_{2,2}$. The addition operation
from $B_{1}$ that has been scheduled to complete at the earliest point in time is $(B_{1})_{1,2}$ and we
can consider all of the constant matrix operations from $A$ to have been scheduled at time 0.
Hence, our heuristic algorithm will choose one of the operations $(B_{2})_{1,2a}$ or
$(B_{2})_{2,2a}$ (which depend on $(B_{1})_{1,2}$) to be scheduled next. These
operations will complete earliest if scheduled on Processor 2 (as shown in Figure~\ref{fig:motivation-schedule})
so the heuristic would schedule the operation there. This process continues until
all operations have been scheduled.

Once the schedule is produced, the operations are executed on the underlying architecture
accordingly.  In the implementation of the framework for the Cell Broadband Engine (see
section~\ref{sec:cell}),
each of the SPEs acts as a processing element which is capable of executing
lowered matrix operations. Each SPE runs a specially designed, small virtual machine program called a Matrix Execution Unit which is optimised for computing matrix operations.
An execution control mechanism runs on the PPE of the Cell processor. It delivers
lowered operations to the MEUs according to the schedule. It also ensures that
an operation does not begin execution until all of its operands have finished being 
computed.

When all operations have completed execution, execution of the Octave program
resumes and our example program completes.

\section{Lowering}\label{sec:lowering}
Matrices can be partitioned into sub-matrices called blocks. 
A matrix can be partitioned across rows and columns as depicted by
the dashed lines in the following example:
\begin{equation*}
A=
\left[%
\begin{array}{cc:c}%
\MES{a}{1}{1}&\MES{a}{1}{2}&\MES{a}{1}{3}\\
\MES{a}{2}{1}&\MES{a}{2}{2}&\MES{a}{2}{3}\\
\hdashline
\MES{a}{3}{1}&\MES{a}{3}{2}&\MES{a}{3}{3}\\
\end{array}
\right]%
=
\begin{bmatrix}
\MES{A}{1}{1}&\MES{A}{1}{2}\\
\MES{A}{2}{1}&\MES{A}{2}{2}\\
\end{bmatrix}.
\end{equation*}
Matrix~$A$ above is thereby partitioned into a $2\times 2$ matrix whose entries \MES{A}{1}{1},
\MES{A}{1}{2}, \MES{A}{2}{1} and \MES{A}{2}{2} are blocks.

Operations on partitioned matrices can be performed by treating blocks as
numerical entries.
A single operation
on the original matrix will be converted into several operations on blocks. This
technique has been employed in the past in languages such High Performance Fortran~\cite{KKZ07}
for distributing matrix operations
among the nodes of a distributed computing environment.
 Matrix
partitioning is important for two reasons: (1) it enables us to perform matrix
operations on matrices that are too big to fit as a whole into the
local stores of processing elements, and (2) it exposes data parallelism that allows
many processing elements to work on the same matrix operation simultaneously.

In general, an
$n\times m$ matrix $A$ is partitioned into $p$ rows and $q$ columns of blocks. 
$\MES{A}{i}{j}$ is the block in the $i$'th row and $j$'th column and it has
$k_{i}$~rows and $l_{j}$~columns. If two matrices, $A$ and $B$ are said to have the
same partitioning, it implies that $p_{A}=p_B$, $q_{A}=q_B$, $\forall_{1\le i \le p_A}\!\!:\, k_{i_A}=k_{i_{B}}$ and $\forall_{1\le j \le q_A}\!\!:\, k_{j_A}=k_{j_{B}}$.
Similarly, if the columns of matrix $A$ are said to have the same partitioning as
the row of matrix $B$, it implies that $q_A=p_B$,
and $\forall_{1\le i \le q_A}\!\!:\, l_{i_A}=k_{i_B}$.
We use this notation throughout the remainder of this section.

Here we describe (1) the types of matrix operations which we focus on in our work
and how those operations are converted to operate on blocks, (2) the major concerns when developing a partitioning scheme, and (3)
a new partitioning scheme we developed called lowering.

\subsection{Block Matrix Operations} \label{sec:ops}

In our work we focus on three types of matrix operations: (1) unary element-wise operations, (2) binary element-wise operations, and (3) matrix multiplication. 
We choose to focus on this subset of matrix operations because they represent
 commonly used operations that fit the scope of the project. For a complete
 list of the block matrix operations we implemented in our framework,
 refer to Appendix~\ref{app:funcs}.

Unary element-wise operations have a single matrix operand. The result
of these operations is computed by applying a mathematical
function to every scalar element of the operand. The result matrix has
the same dimensions as the operand. Thus, these operations have the form:

\begin{equation*}
g(A) =
\begin{bmatrix}
f(\MES{a}{1}{1})&\hdots&f(\MES{a}{1}{m})\\
\vdots&&\vdots\\
f(\MES{a}{n}{1})&\hdots&f(\MES{a}{n}{m})
\end{bmatrix}
\end{equation*}

where $g$ is the operation, $f$ is the mathematical function applied by $g$, $A$ is the 
$n\times m$ operand and $a_{i,j}$ are its elements.

Examples of this type of operation
are matrix-scalar addition, rounding the elements of a matrix, finding the sine of each matrix element, and so on. In the case of matrix-scalar addition, the function $f$
simply adds a constant value $c$ to each of the matrix elements, i.e.~$f(a_{i,j})=a_{i,j}+c$.

These operations can be trivially modified to work on partitioned matrices because
each element of the result matrix depends only on the corresponding element
of the operand. Thus we can partition the matrix in any way we want and apply
the operation to each of the sub-matrices. Say we partition matrix $A$ into 3
rows and 2 columns of blocks:
\begin{equation*}
A =
\begin{bmatrix}
\MES{A}{1}{1}&\MES{A}{1}{2}\\
\MES{A}{2}{1}&\MES{A}{2}{2}\\
\MES{A}{3}{1}&\MES{A}{3}{2}\\
\end{bmatrix}
\end{equation*}
operation $g$ can be applied to all 6 sub-matrices separately:
\begin{equation*}
g(A) =
\begin{bmatrix}
g(\MES{A}{1}{1})&g(\MES{A}{1}{2})\\
g(\MES{A}{2}{1})&g(\MES{A}{2}{2})\\
g(\MES{A}{3}{1})&g(\MES{A}{3}{2})\\
\end{bmatrix}
\end{equation*}

Thus, the resulting matrix has the  same partitioning as the matrix operand.

Binary element-wise operations are similar to unary element-wise operations but
have two matrix operands. The result of the operation is computed by applying
a mathematical function to corresponding scalar elements of the two operands. 
Hence, it is a requirement of these operations that both matrix operands have the same
dimensions, and the resulting matrix will also have the same dimensions as the operands.
These operations have the form:

\begin{equation*}
g(A,B) =
\begin{bmatrix}
f(\MES{a}{1}{1}, \MES{b}{1}{1})&\hdots&f(\MES{a}{1}{m}, \MES{b}{1}{m})\\
\vdots&&\vdots\\
f(\MES{a}{n}{1}, \MES{b}{n}{1})&\hdots&f(\MES{a}{n}{m}, \MES{b}{n}{m})
\end{bmatrix}
\end{equation*}

where $g$ is the operation, $f$ is the mathematical function applied by $g$, $A$ is the 
first $n\times m$ operand with elements $a_{i,j}$ and $B$ is the 
second $n\times m$ operand with elements $b_{i,j}$.

Examples of this type of operation are matrix-matrix addition, matrix-matrix subtraction 
and element-wise multiplication. In the case of matrix-matrix addition, $f$ would
be defined as $f(a_{i,j}, b_{i,j})=a_{i,j} + b_{i,j}$.

These operations can also be trivially modified to work on partitioned matrices because
each element of the result matrix depends only on the corresponding elements
of the operands. In order for the operands $A$ and $B$ to be compatible with this operation they must have the same partitioning so that corresponding
blocks have the same dimensions. Say we partition operands $A$ and $B$ into 3
rows and 2 columns of blocks:
\begin{equation*}
A =
\begin{bmatrix}
\MES{A}{1}{1}&\MES{A}{1}{2}\\
\MES{A}{2}{1}&\MES{A}{2}{2}\\
\MES{A}{3}{1}&\MES{A}{3}{2}\\
\end{bmatrix}
B =
\begin{bmatrix}
\MES{B}{1}{1}&\MES{B}{1}{2}\\
\MES{B}{2}{1}&\MES{B}{2}{2}\\
\MES{B}{3}{1}&\MES{B}{3}{2}\\
\end{bmatrix}
\end{equation*}

operation $g$ can be applied to all 6 corresponding sub-matrices separately:
\begin{equation*}
g(A, B) =
\begin{bmatrix}
g(\MES{A}{1}{1}, \MES{B}{1}{1})&g(\MES{A}{1}{2}, \MES{B}{1}{2})\\
g(\MES{A}{2}{1}, \MES{B}{2}{1})&g(\MES{A}{2}{2}, \MES{B}{2}{2})\\
g(\MES{A}{3}{1}, \MES{B}{3}{1})&g(\MES{A}{3}{2}, \MES{B}{3}{3})\\
\end{bmatrix}
\end{equation*}

Thus, the resulting matrix has the same partitioning as $A$ and $B$.

The final type of operation we consider is matrix multiplication, which
is more complex than the previous two types of operation. Matrix
multiplication has two matrix operands, $A$ and $B$, which are required to have the same
inner dimensions. That is, if matrix $A$ is of dimension $n\times m$, $B$ must
be of dimensions $m\times k$.
Each element of the result matrix can then be defined by the following summation
\begin{equation*}
g(A,B)_{i,j} = (AB)_{i,j} = \displaystyle\sum_{r=1}^m a_{i,j}b_{i,j}
\end{equation*}
with a  dimensions of $n\times m$.

A technique called block matrix multiplication is used to allow multiplication
of partitioned matrices. This is the same technique employed in fast matrix multiplication
algorithms such as the Strassen algorithm~\cite{strassen1969gaussian} and the Coppersmith-Winograd algorithm~\cite{coppersmith1987matrix} (the fastest known algorithm). 
Block matrix multiplication works in the same way as regular matrix multiplication
but treats blocks of the partitioned
operands as scalar elements. 
In order for the operands $A$ and $B$ to be compatible for block matrix multiplication,
the column partitioning of matrix $A$ must match the row partitioning
of matrix $B$. 
If the operands are partitioned as follows:
\begin{equation*}
A =
\begin{bmatrix}
\MES{A}{1}{1}&\MES{A}{1}{2}\\
\MES{A}{2}{1}&\MES{A}{2}{2}\\
\MES{A}{3}{1}&\MES{A}{3}{2}\\
\end{bmatrix}
B =
\begin{bmatrix}
\MES{B}{1}{1} & \MES{B}{1}{2}\\
\MES{B}{2}{1} & \MES{B}{2}{2}\\
\end{bmatrix}
\end{equation*}

then block matrix multiplication would multiply these matrices as follows:
\begin{equation*}
\begin{split}
C = AB =
\begin{bmatrix}
\MES{A}{1}{1}\MES{B}{1}{1} + \MES{A}{1}{2}\MES{B}{2}{1} &
\MES{A}{1}{1}\MES{B}{1}{2} + \MES{A}{1}{2}\MES{B}{2}{2} \\
\MES{A}{2}{1}\MES{B}{1}{1} + \MES{A}{2}{2}\MES{B}{2}{1} &
\MES{A}{2}{1}\MES{B}{1}{2} + \MES{A}{2}{2}\MES{B}{2}{2} \\
\MES{A}{3}{1}\MES{B}{1}{1} + \MES{A}{3}{2}\MES{B}{2}{1} &
\MES{A}{3}{1}\MES{B}{1}{2} + \MES{A}{3}{2}\MES{B}{2}{2} \\
\end{bmatrix}
.
\end{split}
\end{equation*}
Instead of one matrix multiplication~$AB$ on the individual matrix
elements of matrices~$A$ and~$B$, we now have~$12$ multiplications
and~$6$ addition operations on the blocks of~$A$ and~$B$. In general,
each block of the result matrix is defined by 
\begin{equation}
C = \MES{(AB)}{i}{j}=\sum_{r=1}^{q_A}\MES{A}{i}{r}\:\MES{B}{r}{j}.
\label{eq:cij}
\end{equation}
The resulting matrix has the same row partitioning as matrix $A$ and
the same column partitioning as matrix $B$. 

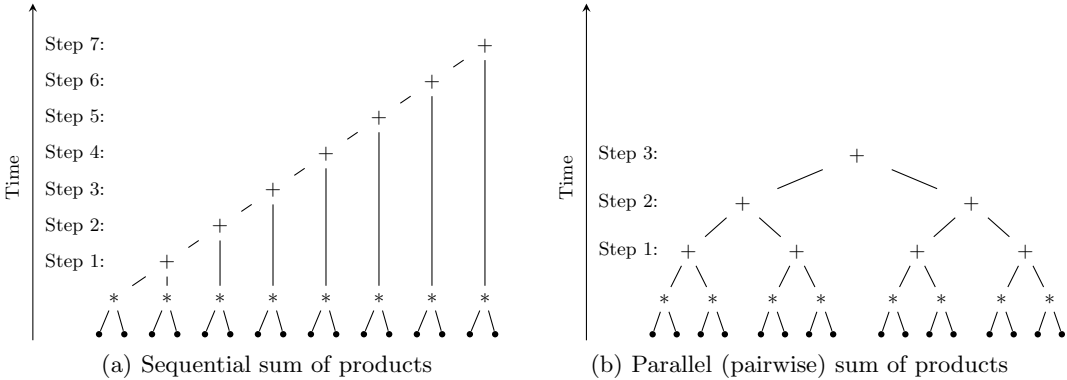
\begin{figure*}[htb]
\center
\tikzstyle{level 1}=[level distance=1.5cm, sibling distance=1.5cm]
\tikzstyle{level 2}=[level distance=1.5cm, sibling distance=1.5cm]
\tikzstyle{bag} = [text width=4em, text centered]
\tikzstyle{end} = [circle, minimum width=3pt,fill, inner sep=0pt]
\subfigure[Sequential sum of products]{
\begin{tikzpicture}[scale=0.8,transform shape]
\def\x{4.2mm}
\def\xx{4.6mm}
\def\xh{2.5mm}
\def\h{6mm}
\tikzstyle{end} = [circle, minimum width=3pt,fill, inner sep=0pt]
\node[end] (p0) at ($(0,0)+(0.0,0.0)$) {};
\node[end] (p1) at ($(p0)+(\x,0.0)$) {};
\node[end] (p2) at ($(p1)+(\xx,0.0)$) {};
\node[end] (p3) at ($(p2)+(\x,0.0)$) {};
\node[end] (p4) at ($(p3)+(\xx,0.0)$) {};
\node[end] (p5) at ($(p4)+(\x,0.0)$) {};
\node[end] (p6) at ($(p5)+(\xx,0.0)$) {};
\node[end] (p7) at ($(p6)+(\x,0.0)$) {};
\node[end] (p8) at ($(p7)+(\xx,0.0)$) {};
\node[end] (p9) at ($(p8)+(\x,0.0)$) {};
\node[end] (p10) at ($(p9)+(\xx,0.0)$) {};
\node[end] (p11) at ($(p10)+(\x,0.0)$) {};
\node[end] (p12) at ($(p11)+(\xx,0.0)$) {};
\node[end] (p13) at ($(p12)+(\x,0.0)$) {};
\node[end] (p14) at ($(p13)+(\xx,0.0)$) {};
\node[end] (p15) at ($(p14)+(\x,0.0)$) {};
\node[bag] (mul0) at ($(p0)+(\xh,\h)$) {$*$};
\node[bag] (mul1) at ($(p2)+(\xh,\h)$) {$*$};
\node[bag] (mul2) at ($(p4)+(\xh,\h)$) {$*$};
\node[bag] (mul3) at ($(p6)+(\xh,\h)$) {$*$};
\node[bag] (mul4) at ($(p8)+(\xh,\h)$) {$*$};
\node[bag] (mul5) at ($(p10)+(\xh,\h)$) {$*$};
\node[bag] (mul6) at ($(p12)+(\xh,\h)$) {$*$};
\node[bag] (mul7) at ($(p14)+(\xh,\h)$) {$*$};
\node[bag] (plus0) at ($(mul1)+(0,\h)$) {$+$};
\node[bag] (plus1) at ($(mul2)+2*(0,\h)$) {$+$};
\node[bag] (plus2) at ($(mul3)+3*(0,\h)$) {$+$};
\node[bag] (plus3) at ($(mul4)+4*(0,\h)$) {$+$};
\node[bag] (plus4) at ($(mul5)+5*(0,\h)$) {$+$};
\node[bag] (plus5) at ($(mul6)+6*(0,\h)$) {$+$};
\node[bag] (plus6) at ($(mul7)+7*(0,\h)$) {$+$};
\draw (p0) -- (mul0);
\draw (p1) -- (mul0);
\draw (p2) -- (mul1);
\draw (p3) -- (mul1);
\draw (p4) -- (mul2);
\draw (p5) -- (mul2);
\draw (p6) -- (mul3);
\draw (p7) -- (mul3);
\draw (p8) -- (mul4);
\draw (p9) -- (mul4);
\draw (p10) -- (mul5);
\draw (p11) -- (mul5);
\draw (p12) -- (mul6);
\draw (p13) -- (mul6);
\draw (p14) -- (mul7);
\draw (p15) -- (mul7);
\draw (mul0) -- (plus0);
\draw (mul1) -- (plus0);
\draw (plus0) -- (plus1);
\draw (mul2) -- (plus1);
\draw (plus1) -- (plus2);
\draw (mul3) -- (plus2);
\draw (plus2) -- (plus3);
\draw (mul4) -- (plus3);
\draw (plus3) -- (plus4);
\draw (mul5) -- (plus4);
\draw (plus4) -- (plus5);
\draw (mul6) -- (plus5);
\draw (plus5) -- (plus6);
\draw (mul7) -- (plus6);
\draw[->, >=stealth] ($(p0)-(11mm,1.0mm)$) -- ($(p0)-(11mm,-5.5cm)$) node[anchor=north, yshift=-2.4cm, xshift=-3.4mm] {\rotatebox{90}{\small Time}};
\node at ($(p0)-(4mm,0)+2*(0,\h)$) {\small Step 1:};
\node at ($(p0)-(4mm,0)+3*(0,\h)$) {\small Step 2:};
\node at ($(p0)-(4mm,0)+4*(0,\h)$) {\small Step 3:};
\node at ($(p0)-(4mm,0)+5*(0,\h)$) {\small Step 4:};
\node at ($(p0)-(4mm,0)+6*(0,\h)$) {\small Step 5:};
\node at ($(p0)-(4mm,0)+7*(0,\h)$) {\small Step 6:};
\node at ($(p0)-(4mm,0)+8*(0,\h)$) {\small Step 7:};
\end{tikzpicture}
}
\hskip-7mm%
\subfigure[Parallel (pairwise) sum of products]{
\tikzstyle{level 1}=[level distance=8mm, sibling distance=38mm]
\tikzstyle{level 2}=[level distance=8mm, sibling distance=18mm]
\tikzstyle{level 3}=[level distance=8mm, sibling distance=8mm]
\tikzstyle{level 4}=[level distance=5.8mm, sibling distance=4mm]
\tikzstyle{bag} = [text width=4em, text centered]
\tikzstyle{end} = [circle, minimum width=3pt,fill, inner sep=0pt]
\begin{tikzpicture}[grow=down, sloped, scale=0.8,transform shape]
\node[bag] {$+$}
child {
node[bag] {$+$}
child {
node[bag] (n1){$+$}
    child {
        node[bag] {$*$}
            child {
                node[end](n0){}{}
                edge from parent
            }
            child {
                node[end]{}{}
                edge from parent
            }
            edge from parent
    }
    child {
        node[bag] {$*$}
            child {
                node[end]{}{}
                edge from parent
            }
            child {
                node[end](n2){}{}
            }
            edge from parent
    }
}
child {
node[bag] {$+$}
    child {
        node[bag] {$*$}
            child {
                node[end]{}{}
                edge from parent
            }
            child {
                node[end]{}{}
                edge from parent
            }
            edge from parent
    }
    child {
        node[bag] {$*$}
            child {
                node[end]{}{}
                edge from parent
            }
            child {
                node[end]{}{}
            }
            edge from parent
    }
}
}
child {
node[bag] {$+$}
child {
node[bag] (n1){$+$}
    child {
        node[bag] {$*$}
            child {
                node[end]{}{}
                edge from parent
            }
            child {
                node[end]{}{}
                edge from parent
            }
            edge from parent
    }
    child {
        node[bag] {$*$}
            child {
                node[end]{}{}
                edge from parent
            }
            child {
                node[end]{}{}
            }
            edge from parent
    }
}
child {
node[bag] {$+$}
    child {
        node[bag] {$*$}
            child {
                node[end]{}{}
                edge from parent
            }
            child {
                node[end]{}{}
                edge from parent
            }
            edge from parent
    }
    child {
        node[bag] {$*$}
            child {
                node[end]{}{}
                edge from parent
            }
            child {
                node[end]{}{}
            }
            edge from parent
    }
}
}
;
\draw[->, >=stealth] ($(n0)-(11mm,1.0mm)$) -- ($(n0)-(11mm,-5.5cm)$) node[anchor=north, yshift=-2.4cm, xshift=-3.4mm] {\rotatebox{90}{\small Time}};
\def\h{8mm}
\node (s1) at ($(n0)-(4mm,0)+(0,14mm)$) {\small Step 1:};
\node at ($(s1)+(0,\h)$) {\small Step 2:};
\node at ($(s1)+2*(0,\h)$) {\small Step 3:};
\end{tikzpicture}
}
\caption{Sequential and parallel (pairwise) sum computations for matrix block
$\MES{C}{i}{j}=\sum_{r=1}^{q_A=8}\MES{A}{i}{r}\:\MES{B}{r}{j}$.
Sequential computation of the sum of products~$\MES{C}{i}{j}$ as depicted in subfigure~(a) requires $N-1$~steps
for $N$~operands. Provided we have enough parallel execution elements we can reorder the summations as depicted in subfigure~(b)
where we sum in pairs, reducing the computation to~$\lceil\log_2 N\rceil$ steps. Algorithm~\ref{alg:pairwisesum}
applies pairwise sums to increase parallelism with matrix multiplications on partitioned matrices.
\label{fig:sums}}
\end{figure*}

Block matrix multiplication can result in many addition operations
on blocks to compute each block of the result matrix (denoted by
the summation in Equation~\ref{eq:cij}). These additions can be performed
in any order because of the associative property of matrix addition.
We use pairwise summation to compute these additions in order maximise parallelism.
This allows several addition operations to be computed in parallel. For example,
if we had to add four matrix blocks together:
\begin{equation*}
Z = Y_1 + Y_2 + Y_3 + Y_4
\end{equation*}
we could do it in several ways. A \naive technique would be to compute
them in a sequential order as follows. The operations in the innermost parentheses are 
computed first:
\begin{equation*}
Z = (((Y_1 + Y_2) + Y_3) + Y_4)
\end{equation*}
However, this neglects parallelism available in the summation. If we instead, 
add the blocks as follows:
\begin{equation*}
Z = ((Y_1 + Y_2) + (Y_3 + Y_4))
\end{equation*}
Then, the additions $(Y_1 + Y_2)$ and  $(Y_3 + Y_4)$ are computable in parallel, 
which reduces the execution time of the matrix multiplication. 
Figure~\ref{fig:sums} illustrates the improvement in performance that can be
gained by using pairwise summation to add 8 matrix blocks together.

Algorithm~\ref{alg:pairwisesum} shows how we construct 
matrix instructions that compute the blocks of matrix~$C$
as stated in Equation~(\ref{eq:cij}) using pairwise summation. Therein
statement {\tt MatrixMult()} constructs a matrix multiplication operation from two
matrix block operands, and {\tt MatrixAdd()} constructs a matrix addition operation.

\begin{algorithm}[t]
\begin{lstlisting}[language=pseudocode]
(*@ \textbf{for all} $i\in\{1,\ldots,p_A\}$ @*)
(*@ \atab \textbf{for all} $j\in\{1,\ldots,q_B\}$ @*)
(*@ \atab \atab {\bf new} queue $q$ @*)
(*@ \atab \atab \textbf{for all} $r\in\{1,\ldots,q_A\}$ @*)
(*@ \atab \atab \atab \hskip1mm $q$.push\_back (\textbf{new} MatrixMult ($\MES{A}{i}{r}$, $\MES{B}{r}{j}$)) @*)
(*@ \atab \atab \textbf{while} $q$.size() $> 1$ @*)
(*@ \atab \atab \atab $s_1=$ $q$.pop\_front() @*)
(*@ \atab \atab \atab $s_s=$ $q$.pop\_front() @*)
(*@ \atab \atab \atab $q$.push\_back (\textbf{new} MatrixAdd ($s_1$, $s_2$)) @*)
\end{lstlisting}
\caption{Construction of instructions that compute the sum of products of matrix blocks~\MES{C}{i}{j} using pairwise summation.\label{alg:pairwisesum}
}
\end{algorithm}

There are known algorithms for more complex matrix operations on blocks, such
as matrix inverse~\cite{zhang2005schur} however we have not considered these in
our work so far.

\subsection{Partitioning Schemes}
A partitioning scheme determines how a matrix is split into blocks. Developing
an automatic partitioning scheme can be a non-trivial exercise. We recognise 
four main concerns when developing a 
partitioning scheme for modern accelerator architectures (summarised in 
Figure~\ref{fig:partitioning-concerns}):

\begin{enumerate}
\item Operations should be divided into approximately equal sized portions such
that the load  on processing elements is evenly balanced. This can reduce the makespan
of execution, as illustrated in Figure~\ref{fig:even-vs-uneven}.

\begin{figure}[t]
\centering
\subfigure[In this case, the operand $A$ of the operation is partitioned
unevenly. This results in an uneven load on the processing elements and an increased
makespan.]
{\includegraphics[width=0.45\columnwidth]{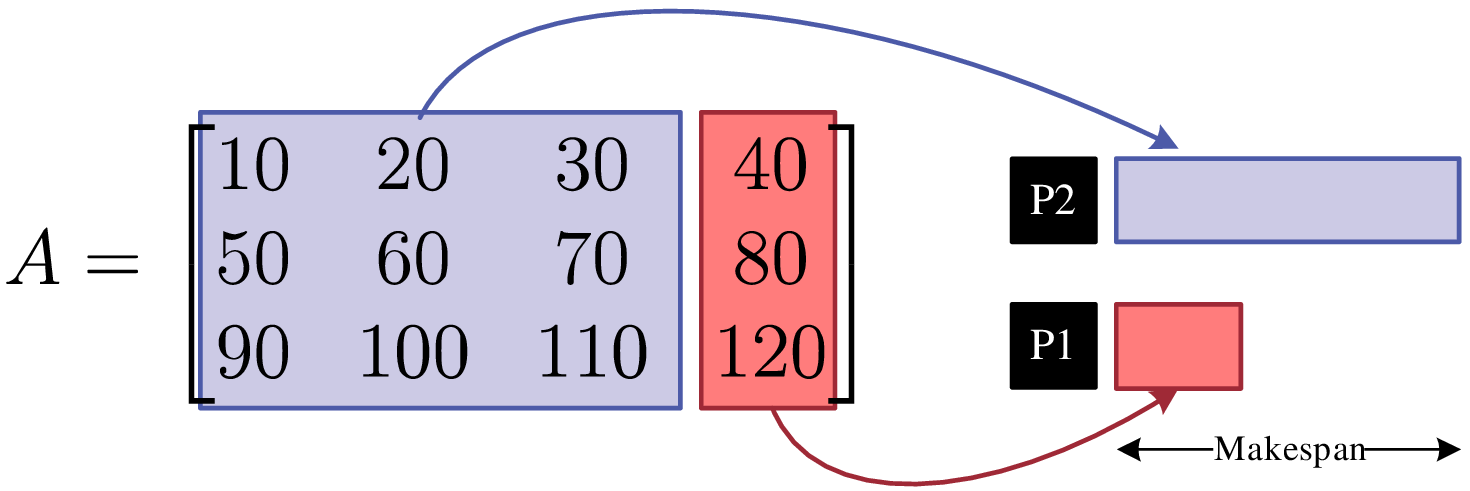} \label{fig:lowering-uneven}}
\hspace{3mm}
\subfigure[In this case, the operand $A$ of the operation is partitioned
evenly. This results in a balanced load on the processing elements and a reduced
makespan.]{\includegraphics[width=0.45\columnwidth]{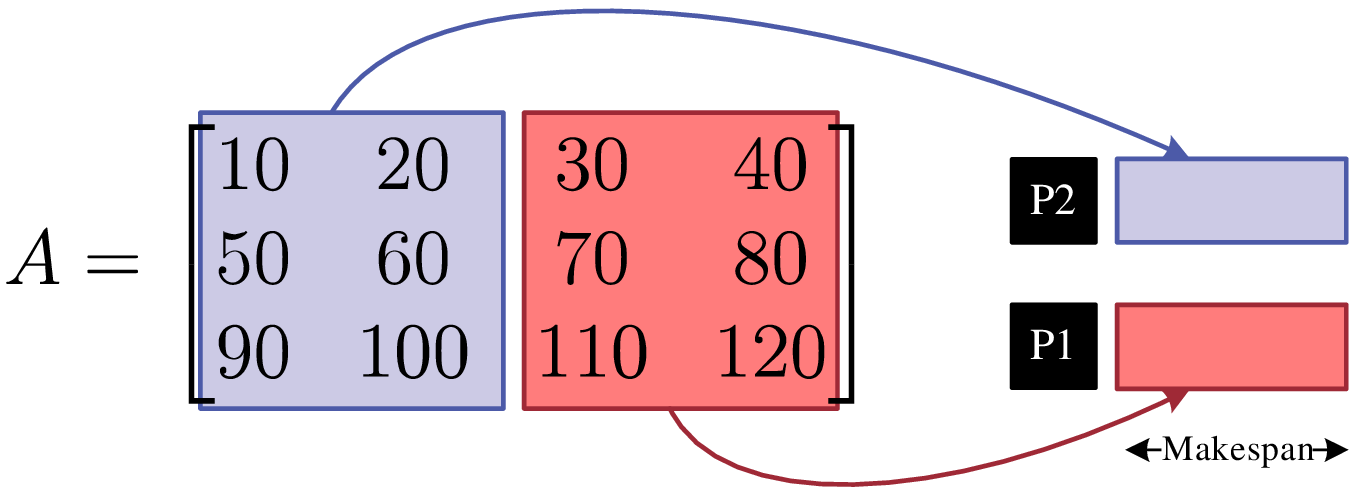} \label{fig:lowering-even}}

\caption{Partitioning matrices evenly improves the makespan. Assume we are given the
Octave code \lstinline[language=myoctave]!B = A / 2;!. To compute $B$,
$A$ is partitioned into two blocks which can be operated on by two different processing
elements. Two different partitionings of $A$ and their resulting schedules are given.
}
\label{fig:even-vs-uneven}
\end{figure}

\item The small memory available on the parallel execution elements should be maximally 
utilised. This criteria is specific to architectures like the Cell whose SPEs 
have a small 256kB local store (see section~\ref{sec:cell}). Peak performance is only obtained when this local store is fully utilised. 

Other architectures whose parallel elements have access to a 
large amount of memory have a different concern which is to determine the number
of blocks which a matrix should be partitioned into in order to obtain the best performance. In this case, there is
a tradeoff between the number of processors utilised to execute the operation and
the execution time of the operation. As a result, the partitioning of operations becomes 
related the scheduling of operations. This problem has been studied and is known
as the problem of scheduling malleable tasks under precedence constraints. A description
of the problem and proposed solutions is provided in the related work in section~\ref{sec:related-work}, however we do not address this concern in our scheme as we focus on the Cell architecture.

\item The number of synchronisation points in a trace should be minimised. Synchronisation
points arise in a trace when two operands of an operation have an incompatible partitioning. For example, the subtraction of two matrices $C=A-B$ requires
that both matrices have the same partitioning (as described in subsection~\ref{sec:ops}
of this section). Figure~\ref{fig:partition-bad} shows a situation where the
matrices $A$ and $B$ were the result of other matrix operations and do not
have the same partitioning. Thus, one of them must be re-partitioned in order
to make them compatible. In this case, matrix $B$ is re-partitioned into
the matrix $B^\prime$ so that
its blocks are the same dimensions as $A$. This allows the blocks of $C$
to then be computed in parallel. However, this re-partitioning has a
negative side-effect of reducing the amount of parallelism in a trace.
Each block in $B^\prime$ contains elements from every block of the matrix $B$.
Thus, in order to perform the re-partitioning, we must wait for all
of the blocks of $B$ to be computed. This can introduce gaps in the schedule
and result in an increased makespan, as shown in Figure~\ref{fig:partition-bad-schedule}.

In contrast, if $A$ and $B$ already have the same partitioning (as shown in 
Figure~\ref{fig:partition-good}) then no artificial synchronisation points are introduced.
In this case, each block of $C$ depends only on the corresponding blocks of $A$ and $B$.
For example, $C_1$ depends only on blocks $A_1$ and $B_1$. So we can begin computing
the result of $C_1$ as soon as $A_1$ and $B_1$ are available. We do not have
to wait until all of the blocks in $A$ or $B$ have been computed. This results
in increased parallelism in the trace and a reduced makespan, as shown in 
Figure~\ref{fig:partition-good-schedule}.

In order to avoid synchronisation points in a trace, we want to ensure that
the operands of a matrix operation have a compatible partitioning,
 even if the operand is a
result of another operation in the trace.

\begin{figure}
\centering
\subfigure[Operands $A$ and $B$ have a different partitioning. $B$ is re-partitioned.]
{\includegraphics[height=0.25\columnwidth]{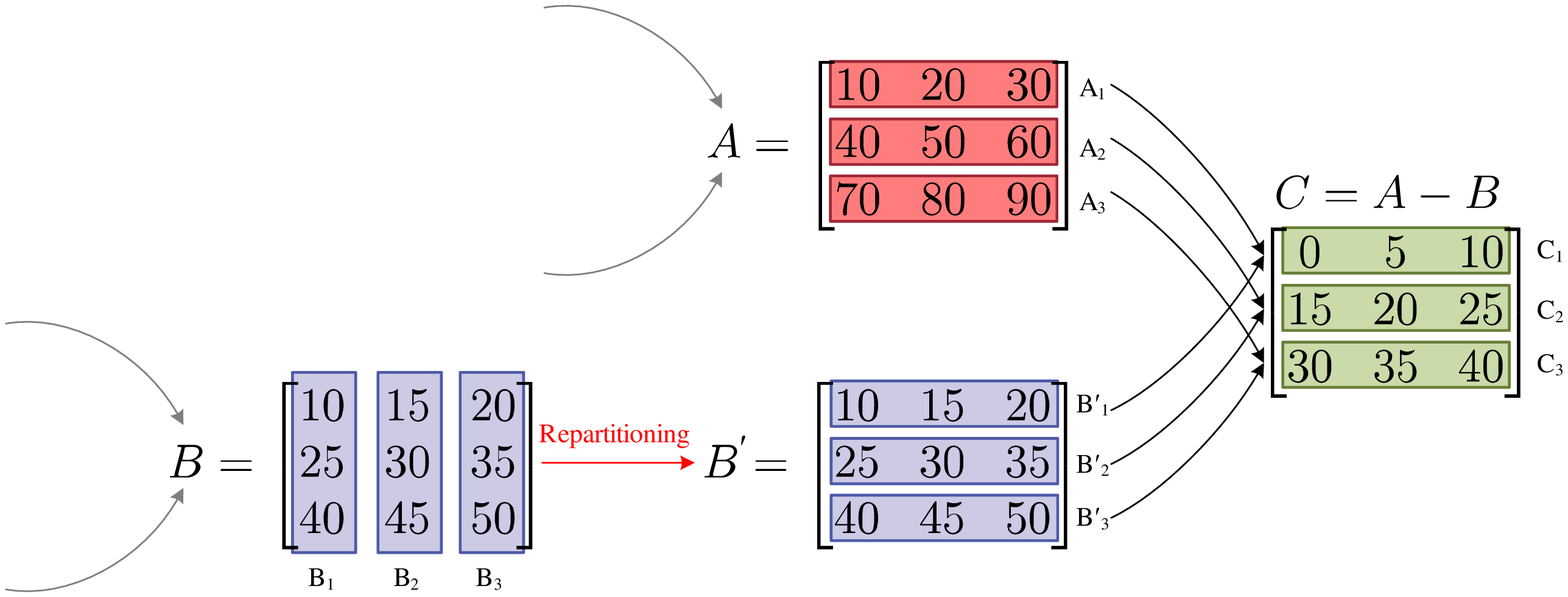} 
\label{fig:partition-bad}}
\subfigure[A barrier results from the re-partitioning of $B$, increasing the makespan
of the schedule.]
{\includegraphics[width=0.35\columnwidth]{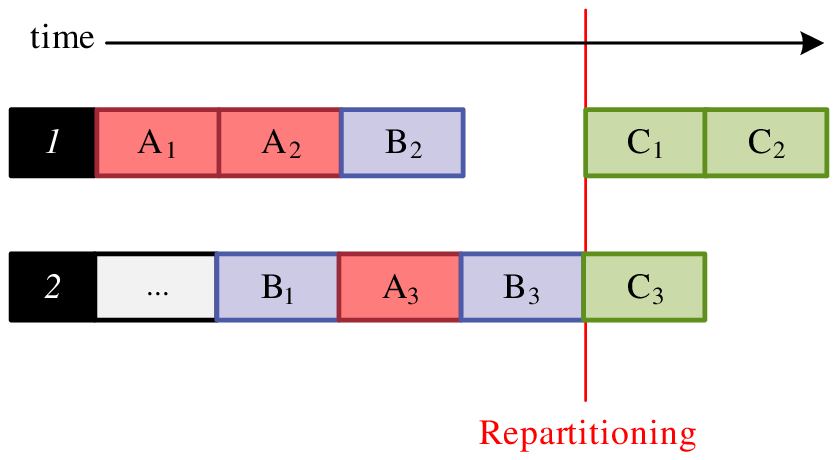} 
\label{fig:partition-bad-schedule}}

\subfigure[Operands $A$ and $B$ have the same partitioning.]
{\includegraphics[height=0.25\columnwidth]{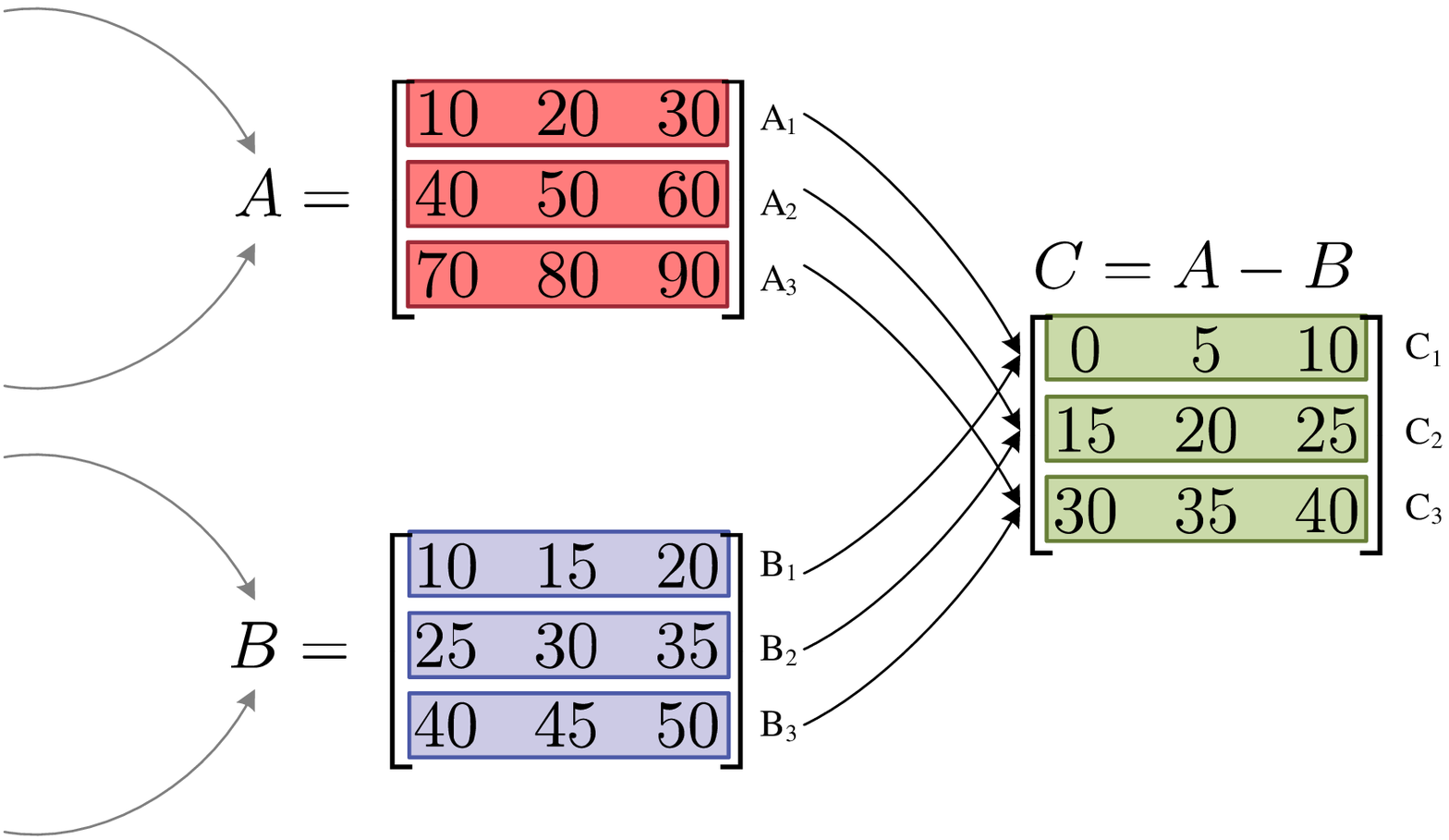} 
\label{fig:partition-good}} \\
\subfigure[No barrier is introduced and there is an improved
schedule due to increased parallelism.]
{\includegraphics[width=0.35\columnwidth]{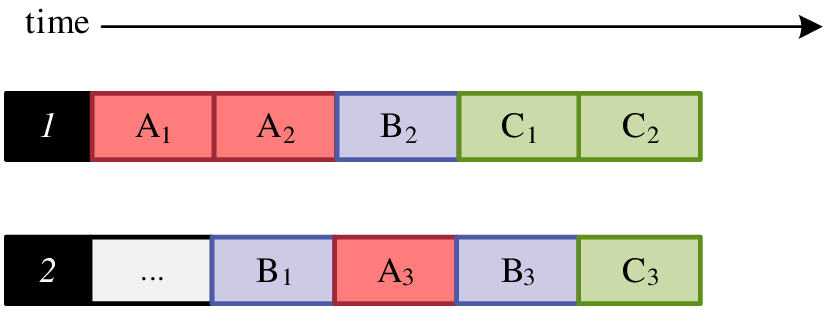}
\label{fig:partition-good-schedule}
}
\label{fig:partition-good-bad}
\caption{Re-partitioning matrices can degrade performance. 
A matrix subtraction $C=A-B$ is being computed.
Figure~\ref{fig:partition-bad} shows a situation where the operands $A$ and $B$
have a different partitioning and thus are incompatible. $B$ is re-partitioned
to $B^\prime$ to match $A$'s partitioning. This allows computation of $C$ to proceed in parallel. However, each block of $B^\prime$ depends on every block of $B$. This acts
as a barrier, forcing computation of the entire $B$ matrix before computation of
$C$ can begin. This reduces potential parallelism in the trace, as shown in Figure~\ref{fig:partition-bad-schedule}. In contrast, if $A$ and $B$ have the same partitioning
we can avoid re-partitioning and no barriers are introduced, resulting in an improved
makespan of the schedule, as shown in Figures~\ref{fig:partition-good} and \ref{fig:partition-good-schedule}.
}
\end{figure}

\item Partitioning should be efficient. Since partitioning occurs at run-time, it is
crucial that dividing operations does not incur a large amount of overhead. 


\end{enumerate}

\begin{figure}
\center
\fbox{
\begin{minipage}{0.6\textwidth}
\begin{enumerate}
\item Blocks should be of approximately equal size.
\item Blocks should be close to the buffer size, $S$.
\item Synchronisation points should be minimised.
\item Partitioning should be efficient.
\end{enumerate}
\end{minipage}
}
\caption{A summary of the four concerns which are important for a matrix partitioning
scheme.}
\label{fig:partitioning-concerns}
\end{figure}

These concerns can be in conflict with one another, making it difficult
to produce a good partitioning scheme. 
For example, a \naive solution 
is to consider each of the matrix operations in a trace 
separately and determine an individual
partitioning for its operands. We can optimise for concerns (1) and (2) of 
Figure~\ref{fig:partitioning-concerns}
by defining the partitioning problem as a least squares fitting. 
We want minimise the difference between the 
integer-valued block-sizes $k_{i}\times
l_{j}$ and the real-valued {\em ideal\/} blocks of
size~$\frac{nm}{pq}\approx\MSpace$ as follows:
{
\begin{eqnarray*}
\min f\!\!& =\sum_{i,j}\left(\frac{nm}{pq}-k_{i}l_{j}\right)^{2}\\
\mbox{s.t.} & 1\leq k_{i}l_{j}\leq \MSpace &i\in[1\tdots p],\:j\in[1 \tdots q]\\
 & \frac{nm}{pq}\approx \MSpace\\
 & k_{i}l_{j}\approx \MSpace\\
 & l_{i}\geq0 &i\in[1\tdots p]\\
 & k_{j}\geq0 &j\in[1\tdots q]\\
 & \sum_{i=1}^{p}l_{i}=n\\
 & \sum_{j=1}^{q}k_{j}=m.\end{eqnarray*}}

However, finding choices of $p$, $q$, $k_{i}$ and $l_{j}$ that optimally satisfy
these criteria for each operation neglects concern (3). That is, it produces
an optimal partitioning only for the operands of a single matrix operation meaning
that every operation may 
require operands with
a different partitioning. This would cause the need for re-partitioning of matrices
resulting in degraded performance. Furthermore, it could be computationally expensive to find these optimal values which neglects concern (4).
 
Another alternative would be to examine the data dependence graph, assuming that
the graph is a tree, and choose a partitioning for the leaf nodes in the graph. This partitioning could
then be propagated up the tree. This approach would help to satisfy concern (3)
by reducing the need for synchronisation points through the re-use of partitionings. However the problem with this approach is 
that the partitioning of inner nodes in the graph is determined by the partitioning of their 
descendants. This could lead to a partitioning for inner nodes which is unbalanced
or does not fully utilise the memory of the processing elements (neglecting concerns
(1) and (2)). In the worst case, this could lead to a partitioning where each block
of an inner node is a single element of a matrix, which would result in very poor
performance. Furthermore, in general the data dependence graph is a directed acyclic 
graph so propagation of partitionings could result in an operation whose partitioning
is determined by several different partitionings. In this case a barrier must be introduced
to resolve the conflict.

\subsection{A New Partitioning Scheme}
In this work, we introduce a new technique called \emph{lowering} for 
partitioning matrix operations in an execution trace. The main benefit
of our partitioning scheme is that all operands of an operation in a trace are guaranteed
to have a compatible partitioning with each other, without ever needing to re-partition
matrices. This maximises the amount of parallelism available in the trace.
Our partitioning scheme also finds a uniform partitioning of matrices, such that all
blocks in a matrix are of approximately the same size. 
Finally, since partitioning of a matrix relies solely on the dimensions of that matrix,
no other operations need to be examined in order to compute the partitioning (i.e.
partitioning is local to every matrix). This leads to an efficient algorithm
which examines each matrix in the execution trace only once.
Thus, we satisfy concerns (1), (3) and (4) from Figure~\ref{fig:partitioning-concerns}.
The only disadvantage of our partitioning scheme is that in certain cases it can lead to sub-matrices which are smaller than the capacity of the memory on the SPEs.
This can result in under utilisation of the SPEs. Thus there are situations
where we fail to address concern (2).

To be applicable to a wider
range of architectures, our partitioning scheme abstracts 
from the underlying hardware, but only to the extent that performance
is not sacrificed. We were able to reduce the dependencies of our
partitioning scheme to two quantities from the underlying Cell Broadband Engine 
architecture:
\begin{enumerate}
\item
The maximum number of matrix elements~\MSpace\ we can store in a
single matrix block. The local store of a Cell SPE provides 256kB of
memory to be shared by program code and data. We use a small
($\approx$26kB) kernel program on each SPE to execute matrix operations.  The
remaining~230kB of the local store is available for matrix blocks.
Because of Cell-related implementation techniques discussed in
subsection~\ref{sec:alignment}, we can devote up to 38kB to a single matrix
block, which amounts to $\MSpace=9216$~single precision floating point
matrix elements, or $\MSpace=4608$~matrix double precision floating point elements per 
block in.
\item
A divisor~\Div\ for the number of rows and columns of a block. 
For several Cell-related reasons discussed in subsection~\ref{sec:alignment}, 
the number of rows and columns in a block
must be a multiple of $\Div=4$ with single precision and a
multiple of $\Div=2$ with double precision. To guarantee that each block has 
has a multiple of~\Div\ rows and columns, it may be necessary to pad rows
and columns of a matrix with zero elements. An $n\times m$
matrix may require up to~$\Div-1$ additional rows and/or columns of zeroes, as
shown in Figure~\ref{subfig:memB}.
The number of elements of the resulting $n'\times m'$~matrix is
bound by $n\times m + (\Div-1)(n+m) \ge n'\times m'$.

\end{enumerate}

\def\S{\hskip3mm}
\def\LSKIP{-1.8mm}
\def\ASKIP{\hskip1.4mm}
\def\gvdots{\textcolor{gray}\vdots}
\def\gldots{\textcolor{gray}\cdots}
\begin{figure*}[ht]
\centering
\tikzset{node style ge/.style={draw=none, fill=none}}
\subfigure[Memory layout of an $n\times m$ matrix in row-major order: the leftmost column depicts the
starting address
of each row. An $n\times m$ matrix has a stride of $\ESpace\cdot m$ between adjacent rows of matrix elements
of size~\ESpace\ ($\ESpace =4\text{ bytes}$ for single precision values and $\ESpace=8\text{ bytes}$
for double precision values).\label{subfig:memA}]{
\begin{tikzpicture}
\node(tbl){\setlength{\arraycolsep}{0.7pt}
$\begin{array}{r@{\hskip4mm}cccc@{\:\:\cdots\:\:}cccc}
B&\MES{a}{1}{1}&\MES{a}{1}{2}&\MES{a}{1}{3}&\MES{a}{1}{4}&
\MES{a}{1}{m-2}&\MES{a}{1}{m-1}&\MES{a}{1}{m}\\
B+s m&\MES{a}{2}{1}&\MES{a}{2}{2}&\MES{a}{2}{3}&\MES{a}{2}{4}&
\MES{a}{2}{m-2}&\MES{a}{2}{m-1}&\MES{a}{2}{m}\\
B+2s m&\MES{a}{3}{1}&\MES{a}{3}{2}&\MES{a}{3}{3}&\MES{a}{3}{4}&
\MES{a}{3}{m-2}&\MES{a}{3}{m-1}&\MES{a}{3}{m}\\
\vdots\hskip6.1mm&\vdots&\vdots&\vdots&\vdots&\vdots&\vdots&\vdots\\
B+(n-1)s m&\MES{a}{n}{1}&\MES{a}{n}{2}&\MES{a}{n}{3}&\MES{a}{n}{4}&
\MES{a}{n}{m-2}&\MES{a}{n}{m-1}&\MES{a}{n}{m}\\
\end{array}$};
\end{tikzpicture}
}\\ \hskip3mm \subfigure[The above $n\times m$ matrix partitioned into
$p\times q$ blocks~$\MES{A}{i}{j}$.  Block~$\MES{A}{i}{j}$ has $k_i$
rows and $l_j$ columns.  Cell-specific implementation issues (see subsection~\ref{sec:alignment})
require the number of rows and columns of each block to be a multiple of
four with single precision and a multiple of two with double precision. 
This may require padding of outer-most blocks with zeroes. 
For a matrix operation~$A=B\cdot C$, padding matrix~$B$
with $p$ additional columns requires padding of matrix~$C$ with $p$
additional rows. To achieve maximum DMA transfer rates on the Cell,
transfer sizes must be multiples of 128~bytes, which may require
further padding.  The individual rows of a block are transferred to an
SPE using a DMA list transfer.  For block~\MES{A}{1}{1} we have $k_1$
list entries, consisting of the memory addresses $B$, $B+\ESpace
m'$,\ldots,$B+(k_1-1)\ESpace m'$.  Each list entry describes one row
of block~\MES{A}{1}{1}, as depicted by the dotted rectangles.
\label{subfig:memB}
]{
\tikzset{
  every node/.style={scale=0.65}
}
\begin{tikzpicture}
\tikzstyle{column 1}=[anchor = base east]
\tikzstyle{column 3}=[anchor = base west]
\tikzstyle{column 5}=[anchor = base east]
\tikzstyle{column 7}=[anchor = base west]
\tikzstyle{column 9}=[anchor = base east]
\tikzstyle{column 13}=[anchor = base west]
\tikzstyle{column 18}=[anchor = base east]
\definecolor{r1}{RGB}{255,255,210}
\matrix (A) [matrix of math nodes, nodes = {node style ge},
             row sep=-0.5 mm, column sep=-0.5 mm,
             ampersand replacement=\&,
]
{B\&\hskip3mm\&\MESg{a}{1}{1} \&\gldots\&\MESg{a}{1}{l_1}\&\S\&
    \MESg{a}{1}{l_1+1} \&\gldots\&\MESg{a}{1}{l_1+l_2}\&\S\&\gldots\&\S\&
    \MESg{a}{1}{m-l_q+1} \&\gldots\&\MESg{a}{1}{m}\&\MEg{0}{}{}\&\gldots\&\MEg{0}{}{}\\
 B+\ESpace m'\&\&\MESg{a}{2}{1} \&\makebox{}\&\MESg{a}{2}{l_1}\&\S\&
    \MESg{a}{2}{l_1+1} \&\gldots\&\MESg{a}{2}{l_1+l_2}\&\S\&\gldots\&\S\&
    \MESg{a}{2}{m-l_q+1} \&\gldots\&\MESg{a}{2}{m}\&\MEg{0}{}{}\&\gldots\&\MEg{0}{}{}\\[\LSKIP]
 \gvdots\ASKIP\&\&\gvdots\&\&\gvdots\&\S\&\gvdots\&\&\gvdots\&\S\&\&\&\gvdots\&\&\gvdots\&\gvdots\&\&\gvdots\\
 B+(k_1-1)\ESpace m'\&\&\MESg{a}{k_1}{1} \&\gldots\&\MESg{a}{k_1}{l_1}\&\S\&
    \MESg{a}{k_1}{l_1+1} \&\gldots\&\MESg{a}{k_1}{l_1+l_2}\&\S\&\gldots\&\S\&
    \MESg{a}{k_1}{m-l_q+1} \&\gldots\&\MESg{a}{k_1}{m}\&\MEg{0}{}{}\&\gldots\&\MEg{0}{}{}\\
 \gvdots\ASKIP\&\&\gvdots\&\&\gvdots\&\&\gvdots\&\&\gvdots\&\&\&\&\gvdots\&\&\gvdots\&\gvdots\&\&\gvdots\\
 \gvdots\ASKIP \&\hskip3mm\&\MESg{a}{n-k_p+1}{1} \&\gldots\&\MESg{a}{n-k_p+1}{l_1}\&\S\&
    \MESg{a}{n-k_p+1}{l_1+1} \&\gldots\&\MESg{a}{n-k_p+1}{l_1+l_2}\&\S\&\gldots\&\S\&
    \MESg{a}{n-k_p+1}{m-l_q+1} \&\gldots\&\MESg{a}{n-k_p+1}{m}\&\MEg{0}{}{}\&\gldots\&\MEg{0}{}{}\\
 \gvdots\ASKIP\&\&\gvdots\&\&\gvdots\&\&\gvdots\&\&\gvdots\&\&\&\&\gvdots\&\&\gvdots\&\gvdots\&\&\gvdots\\
 \gvdots\ASKIP\&\hskip3mm\&\MESg{a}{n}{1} \&\gldots\&\MESg{a}{n}{l_1}\&\S\&
    \MESg{a}{n}{l_1+1} \&\gldots\&\MESg{a}{n}{l_1+l_2}\&\S\&\gldots\&\S\&
    \MESg{a}{n}{m-l_q+1} \&\gldots\&\MESg{a}{n}{m}\&\MEg{0}{}{}\&\gldots\&\MEg{0}{}{}\\
 B+n\ESpace m'\&\&\MEg{0}{}{} \&\gldots\&\MEg{0}{}{}\&\S\&
    \MEg{0}{}{} \&\gldots\&\MEg{0}{}{}\&\S\&\gldots\&\S\&
    \MEg{0}{}{} \&\gldots\&\MEg{0}{}{}\&\MEg{0}{}{}\&\gldots\&\MEg{0}{}{}\\
 \gvdots\ASKIP\&\&\gvdots\&\&\gvdots\&\S\&\gvdots\&\&\gvdots\&\S\&\&\&\gvdots\&\&\gvdots\&\gvdots\&\&\gvdots\\
 B+(n'-1)\ESpace m'\&\&\MEg{0}{}{} \&\gldots\&\MEg{0}{}{}\&\S\&
    \MEg{0}{}{} \&\gldots\&\MEg{0}{}{}\&\S\&\gldots\&\S\&
    \MEg{0}{}{} \&\gldots\&\MEg{0}{}{}\&\MEg{0}{}{}\&\gldots\&\MEg{0}{}{}\\
};
\begin{pgfonlayer}{background}
  \draw[rounded corners,top color=r1,bottom color=r1,
        draw=white] ($(A-1-3.north west)$)
        rectangle ($(A-4-5.south east)$);
  \draw[rounded corners,top color=r1,bottom color=r1,
        draw=white] ($(A-1-7.north west)$)
        rectangle ($(A-4-9.south east)$);
  \draw[rounded corners,top color=r1,bottom color=r1,
        draw=white] ($(A-1-13.north west)$)
        rectangle ($(A-4-18.south east)$);
  \draw[rounded corners,top color=r1,bottom color=r1,
        draw=white] ($(A-6-3.north west)$)
        rectangle ($(A-11-5.south east)$);
  \draw[rounded corners,top color=r1,bottom color=r1,
        draw=white] ($(A-6-7.north west)$)
        rectangle ($(A-11-9.south east)$);
  \draw[rounded corners,top color=r1,bottom color=r1,
        draw=white] ($(A-6-13.north west)$)
        rectangle ($(A-11-18.south east)$);
  \draw[rounded corners, dotted, thick,
        draw=red] ($(A-1-3.north west)$)
        rectangle ($(A-1-5.south east)+(0,0.1)$);
  \draw[rounded corners, dotted, thick,
        draw=red] ($(A-2-3.north west)+(0,-0.06)$)
        rectangle ($(A-2-5.south east)+(0,0.1)$);
  \draw[rounded corners, dotted, thick,
        draw=red] ($(A-4-3.north west)+(0,-0.02)$)
        rectangle ($(A-4-5.south east)+(0,0.1)$);
\end{pgfonlayer}
\begin{pgfonlayer}{foreground}
  \draw node[rounded corners, rectangle, fill=r1,opacity=0.9] at ($(A-2-4)+(0,-0.3)$) {\large $\MES{A}{1}{1}$};
  \draw node[rounded corners, rectangle, fill=r1,opacity=0.9] at ($(A-2-8)+(0,-0.3)$) {\large $\MES{A}{1}{2}$};
  \draw node[rounded corners, rectangle, fill=r1,opacity=0.9] at ($(A-2-14)+(0,-0.3)$) {\large $\MES{A}{1}{q}$};
  \draw ($(A-8-4)+(0,0.20)$)  node[above] {\large $\MES{A}{p}{1}$};
  \draw ($(A-8-8)+(0,0.20)$)  node[above] {\large $\MES{A}{p}{2}$};
  \draw ($(A-8-14)+(0,0.20)$)  node[above] {\large $\MES{A}{p}{q}$};
\end{pgfonlayer}
\end{tikzpicture}
\tikzset{
  every node/.style={scale=1}
}

}
\caption{Memory layout of unpartitioned and partitioned padded matrices
 in row-major order.
\label{fig:matrixlayout}}
\end{figure*}

Figure~\ref{subfig:memB} shows how we partition an $n\times m$
matrix~$A$ into $p$~rows and $q$~columns of blocks. Block~$\MES{A}{i}{j}$ of this
partition has $k_{i}$~rows and $l_{j}$~columns.  The underlying idea of our 
partitioning scheme is that we limit the shape of a matrix block to a maximum of $\sqrt{S}$ rows and columns. When the
divisor \Div\ is also taken into account, each sub-matrix is limited further
to~$\Div\lfloor\frac{\sqrt{\MSpace}}{\Div}\rfloor$ rows and columns.
Making this assumption about the maximum number of rows and columns in each block allows
us to determine the number of rows of blocks $p$ and columns
of blocks $q$ which are required for a matrix:

\begin{eqnarray*}
p=\left\lceil
\frac{\left\lceil\frac{n}{\Div}\right\rceil}{\bigl\lfloor\frac{\sqrt{\MSpace}}{\Div}\bigr\rfloor}
\right\rceil,\text{ and }
q=\left\lceil
\frac{\left\lceil\frac{m}{\Div}\right\rceil}{\bigl\lfloor\frac{\sqrt{\MSpace}}{\Div}\bigr\rfloor}
\right\rceil.
\end{eqnarray*}
Therein~$\lceil\frac{n}{\Div}\rceil$ is the number of
groups-of-$\Div$-rows of a partition, and
$\lfloor\frac{\sqrt{\MSpace}}{\Div}\rfloor$ is the number of
groups-of-$\Div$-rows that fit in a single matrix block. Dividing the
former by the latter yields the number of rows~$p$ of blocks in a
partitioning. Likewise for the number of columns~$q$. 

Once we have determined the number of rows of blocks $p$, we divide the rows
of the original matrix into those blocks. This ensures that each block
in the matrix has approximately the same number of rows. Doing the same
for the columns results in each block of the matrix having approximately the
same number of elements, leading to a balanced partitioning (satisfying concern (1)). 
The number of rows~$k_{i}$ and columns~$l_{j}$ of a block are given by
\begin{eqnarray*} k_{i} = \Div\left(\left\lfloor
\frac{\lceil\frac{n}{\Div}\rceil}{p}\right\rfloor +r_{i}\right)\text{
and } l_{j} = \Div\left(\left\lfloor
\frac{\lceil\frac{m}{\Div}\rceil}{q}\right\rfloor
+s_{j}\right),\end{eqnarray*} 
When the number of rows in a matrix does not divide evenly into $p$, there will be left-over rows. These are distributed among the existing
row partitions. If a left-over row is included in a block, it is denoted by 
the inclusion of a non-zero $r_{i}$ term in the calculation of $k_{i}$. Likewise
for left-over columns. $r_{i}$ and $s_{j}$ are chosen such that $\sum_{i=1}^{p}r_{i}=\lceil\frac{n}{\Div}\rceil\bmod p$ where $r_{i}\in\{0,1\}$, and
$\sum_{j=1}^{q}s_{j}=\lceil\frac{m}{\Div}\rceil\bmod q$ where $s_{j}\in\{0,1\}$.
For consistency across
matrices we set~$r_{i}=1$ and $s_{j}=1$ for the smallest indices~$i$
and~$j$ of blocks~\MES{A}{i}{j} of a matrix partition.

Note that the number of rows in a block, $k_{i}$ depends on the divisor \Div ,
the number of rows in the original matrix $n$, the number of rows of blocks $p$
and its position $i$ in the partitioning (due to the left-over rows). $p$ depends
on $n$, \Div\ and $S$, the buffer size. Assuming \Div\ and $S$ are constants, 
this means that
$k_{i}$ depends only on $i$ and $n$ and likewise for the columns. Thus,
our partitioning scheme ensures that if two matrices $A$ and $B$
have the same number of rows and columns, they will be partitioned
in the same way. It also guarantees that if the number of columns in $A$
is equal to the number of rows $B$ then the column partitioning of $A$
will match the row partitioning of $B$.

With our partitioning scheme, we claim that matrices never have to be re-partitioned.
This implies that only the constant matrices in a trace are partitioned.
Matrices that result from other operations must be guaranteed to be compatible
in any operation in which they are used. Thus,
in our example in Figure~\ref{fig:partition-good-bad}, $A$ and $B$ must be guaranteed
to have the same partitioning, even if they are the results of other operations
in the trace. This is further complicated by the fact that $A$ and $B$ might be
the operands of more than one operation.

To show this, we first show that if the operands of a matrix operation have been partitioned according to our partitioning scheme then the operands are guaranteed to be compatible. The requirements for compatibility are described in 
subsection~\ref{sec:ops} of this section. We must do this for each type of operation we consider in our framework:
\begin{enumerate}
\item \textbf{Unary Element-wise Operations} Trivially compatible because these operations
do not have any requirement on the partitioning of the single operand.
\item \textbf{Binary Element-wise Operations} These operations have two
operands $A$ and $B$ which are required to have the same partitioning to be
compatible. This is ensured for our scheme because $A$ and $B$ are of the same
dimensions, thus they will have the same partitioning.
\item \textbf{Matrix Multiplication} This operation has two operands, $A$ and $B$. The
column partitioning of $A$ must match the row partitioning of $B$ for these to 
be compatible. This is ensured for our scheme because the number of columns
in $A$ matches the number of rows in $B$.
\end{enumerate}

We now show that if the operands of an operation have been partitioned according to our partitioning scheme, then the result of the operation will also be partitioned according to our scheme. The partitioning of the results of operations is also described in 
subsection~\ref{sec:ops} of this section.
We must do this for each type of operation we consider in our framework:
\begin{enumerate}
\item \textbf{Unary Element-wise Operations:} The result matrix has the same
partitioning as the input matrix. Thus it will be partitioned according
to our scheme.
\item \textbf{Binary Element-wise Operations:} The result matrix has the same
partitioning as both input matrices. Thus it will be partitioned according
to our scheme.
\item \textbf{Matrix Multiplication:} The result matrix has same number of rows and row-partitioning as first operand. It also has same number of columns and column-partitioning as the second operand. Thus it will be partitioned according to our scheme.
\end{enumerate}

Now recall that the data dependence graph of operations is a directed
acyclic graph, whose source nodes are constant matrices.
Thus, every operation either depends on constant matrices, or on other operations
which in turn depend on constant matrices. Since the constant matrices are partitioned
with our scheme, the results of other operations will be partitioned according
to our scheme. This means that every matrix operand in the trace will be partitioned
according to our scheme and so operands will always be compatible with each other.
So re-partitioning is never needed, satisfying concern (3).

The only downside of this partitioning scheme is that for very small-sized 
matrices and vectors, the partitioning can lead to
under-utilisation of the buffer. For example, assume $\Div=1$ (for simplicity)
and we wish to partition a row vector
of dimensions $1\times m$ with a buffer size of $S$. The partitioning
occurs as follows:
\begin{enumerate}
\item The number of rows $p$ of sub-matrices and columns $q$ of sub-matrices is given by:
\begin{eqnarray*}
p & = & \left\lceil \frac{1}{\sqrt{\MSpace}}\right\rceil \\
 & = & 1\\
q & = & \left\lceil \frac{m}{\sqrt{\MSpace}}\right\rceil \end{eqnarray*}
\item If we ignore the spare rows $r_{i}$ and spare columns $s_{j}$,  the number of rows per sub-matrix $k_{i}$ and columns per sub-matrix $l_{j}$ is given by:
\begin{eqnarray*}
k_{i} & = & \left\lfloor \frac{1}{p}\right\rfloor \\
& \le & 1\\
l_{j} & = & \left\lfloor \frac{m}{q}\right\rfloor \\
& \le & \frac{m}{\left\lceil\frac{m}{ \sqrt{\MSpace}}\right\rceil}\\
& \le & \sqrt{\MSpace}\\
\end{eqnarray*}
\end{enumerate}

Thus, the number of elements in a block is $k_{i}l_{j} \le \sqrt{\MSpace}$. So only
a square-root of the buffer size is used at most. This results in poorer performance
for operations with these kinds of matrices as operands due to the increased
number of sub-operations and relative overhead to computation ratio. This could
be improved by using a different partitioning for these kinds of matrices,
however this would force barriers to be introduced when the partitioning changes.
We leave this consideration as future work.

The output of the lowering process is a lowered data dependence graph. This graph is
produced through a graph transformation of the original data dependence graph.
The transformation algorithm examines each operation in the original graph in a topological
order, partitions its matrix operands and produces several new operations which
operate on the blocks of the partitioned matrices. Data dependencies are updated
according to the partitioned operations. Hence the lowered graph typically
has an increased number of nodes (operations) and edges (dependencies) over 
the original graph.

\section{Scheduling} \label{sec:scheduling}

The task of the scheduler is to distribute matrix operations among
the processing elements of the parallel architecture.
The input of the scheduler is the lowered
dependence graph annotated with estimated time durations for each
pipeline stage of an operation. The objectives of the scheduler are:
\begin{enumerate}
\item to produce a feasible schedule, i.e., operands of a matrix
  operation must not be scheduled after the operation,
\item to generate the schedule for the MEUs as fast as possible, and
\item to minimise the \emph{makespan}, i.e., the wall-clock time
  needed to execute the parallel schedule on the MEUs.
\end{enumerate}
The second and third objectives are hard to achieve because scheduling
is an NP-hard problem~\cite{garey1979cai}. Recognising the
difficulty of finding an optimal solution for the general problem,
many heuristics have arisen, as described in subsection~\ref{sec:The-Scheduling-Problem}
of the Related Work.

Our scheduling problem is a variant of the static scheduling problem
with arbitrary task precedence constraints, arbitrary execution times for tasks, 
and uniform workers/processors, as described in subsection~\ref{sec:delay}. 
The data dependence graph in our framework resembles the task
precedence graph, the workers are the MEUs and the time durations are
estimated ahead of time for each lowered matrix instruction. Although
there exists approximation algorithms for this problem (see subsection~\ref{sec:The-Scheduling-Problem}) they are based on rounding of a relaxed linear programming 
solution with high practical run-times. Thus,
they are not viable for our framework which needs to incur a low overhead. 
Furthermore, existing algorithms do not take into account the notion of
a pipeline for each of the workers where execution of tasks and communication between
processors can be overlapped. 
Thus, using these algorithms would lead to imprecision in the modelling:
the communication overheads of matrix instructions would be ignored
and the pipeline might stall because of imprecise modelling of matrix
operations.

In the following we discuss (1) how to find a time model to accurately estimate
the durations of each pipeline stage of a lowered matrix operation, (2)
a mathematical program for computing the optimal schedule for the pipelined 
scheduling problem with task
precedence constraints, and (3) a new greedy algorithm for solving the scheduling
problem efficiently.

An accurate time model for the scheduling is crucial to find an
effective schedule. The mathematical program for finding schedules is
not practical but for small problem sizes it gives us a yardstick to
compare how good our greedy algorithm performs in comparison to the
optimal solution.  Because the problem is intractable, we cannot hope
for computing schedules with the mathematical program for problems
with medium to larger size.
Results of the accuracy of time models and the performance of the heuristic
scheduling algorithm are provided in section~\ref{sec:experiment}.

\subsection{Time Model} \label{sec:time-model-results}

We compute a time model for each type of matrix instruction in our
framework, and we use multi-variate polynomial functions for the time
model that depend on the number of rows and columns of the input and
output matrices. The coefficients of the multi-variate polynomial are
computed using profiling and \emph{Ordinary Least Square} (OLS)
method.

For scheduling we require an accurate time model that estimates
the durations of the three phases of a matrix instruction before
executing it. Recall from section~\ref{sec:overview} that we model three pipeline
stages for a matrix instruction executing on a processing element:
\begin{enumerate}
\item \textbf{Data Fetch (df):} The operands of the matrix instruction are loaded
from main memory into the memory of the parallel processing element,
\item \textbf{Execute (ex):} The matrix instruction is executed on the parallel processing element,
\item \textbf{Write Back (wb):} the result of the matrix instruction is written back to main memory.
\end{enumerate}

For computing the time model we instrument the virtual
machine that runs on the SPEs (called MEUs). As a side-effect of execution we
obtain profiling information. The instrumented version of the MEUs is only executed for profiling purposes to obtain the execution
times for each pipeline stage of an instruction. Furthermore, we have
a carefully crafted input program that executes each type of matrix
instruction for every possible input problem size several times. Note that
it is feasible to profile every input problem size
since the matrices are limited by the SPE's small data and program memory of 256kB.

The Cell Broadband Engine offers hardware
counters~\cite{ibmcellhandbook} for performing the measurements with very
high precision. 
The execution time measurements on the SPEs have little variation due to
the lack of data caches. However, we  observed a higher volatility in the
measurements for the data fetch and write back stages
which is not surprising since the ring-bus
of the Cell that connects SPEs, memory and the PPE, is shared.

The coefficients of the multivariate polynomial are chosen such
that the deviation of the polynomial function from the measured time durations
is minimised. Assume we have
$l$ time measurements obtained by profiling one type of matrix operation (such
as matrix addition). For the $k$-th measurement we are given a vector $\vec{n}_k$ describing the input problem size and the durations $\Delta_\textit{df}(k)$, $\Delta_\textit{ex}(k)$, and
$\Delta_\textit{df}(k)$ for each phase of the pipeline. The input
problem size vector typically describes the dimensions of the operands
of the instruction. For example a
matrix addition has two elements in the input problem size vector
describing the number of rows and columns of the two matrices which
are to be added together.

We seek a multivariate polynomial for each pipeline phase of
each type of matrix instruction:
\begin{equation}
t_{\xi} (\vec{n}) = \sum_k a_k g_k(\vec{n})
\end{equation}
where $g_k(\vec{n})$ is a multivariate term of the multivariate
polynomial and $a_0, \ldots, a_n$ are the coefficients such that the
error
\begin{equation}
R_{\xi}=\sum_{k=1}^l (t_{\xi}(\vec{n}_k )- \Delta_{\xi}(k))^2
\end{equation}
becomes minimal (where $\xi$ is either \textit{df}, \textit{ex} or \textit{wb}).

In the following table we give the multivariate polynomials used for estimating 
the execution times of some types of matrix instructions, 
the data fetch, and write back duration. Note that the data fetch and write back
phases do not depend on the type of operation being executed.
\begin{center}
\begin{tabular}{l|l}
$t_{\xi}(\vec{n})$ & Operation \\
\hline
$a_0 + a_1 n_1 + a_2 n_2$ & data fetch and write back \\
$a_0 + a_1 n_1 n_2$ & scalar instr. execution \\
$a_0 + a_2 n_1 + a_1 n_1 n_2 n_3$ & matrix multipl.~execution\\
\end{tabular}
\end{center}
The variables $n_1$, $n_2$ and $n_3$ represent the rows and
columns of the input matrices which form the problem size vector $\vec{n}$. 
For matrix multiplications we do not
need to specify the number of rows of the second matrix because
it is equal to the number of columns of the first matrix. Furthermore,
we add an additional term $a_2 n_1$ to account for the overhead of the
inner-most loop in the matrix calculation to obtain a better fit of
the profile data. We used standard methods to obtain the coefficients~\cite{Kreyszig}.

For the remainder of the section, we use $t_\xi (i)$  to denote the function
$t_\xi(\vec{n})$ where $\vec{n}$ is the problem size vector of instruction $i$.

\subsection{Mathematical Program} \label{sec:ilp}

We develop an integer linear program that computes an optimal schedule
for a given problem instance that consists of the set of matrix instructions
$I=\{1,\ldots,n\}$, their data dependencies $E\subseteq I\times I$ where $(i,j)\in E$
denotes that $j$ depends on $i$, and the time parameters
$t_{\textit{df}}(i)$, $t_{\textit{ex}}(i)$,  and $t_{\textit{wb}}(i)$ for instructions $i \in I$.
For the model we introduce the following variables,
\begin{eqnarray*}
&x_{ij}\in\{0,1\} & i,j \in I \\
&t_{i}\in\mathbb{R}^{+}  & i \in I\\
&z\in\mathbb{R}^{+}
 \end{eqnarray*}
where $z$ is the makespan of the schedule, $t_{i}$ is the start
time of an instruction and $x_{ij}$ are elements of an adjacency
matrix for the \emph{schedule graph} which is a directed rooted
graph which forms the schedule of operations executed by each processing element.

\begin{figure}[b!]
\begin{center}
\begin{tikzpicture}
  \tikzstyle{vertex}=[circle,thick,draw=blue!75,fill=blue!20,minimum size=17pt,inner sep=0pt]
  \tikzstyle{hotvertex}=[circle,thick,draw=red!75,fill=red!20,minimum size=17pt,inner sep=0pt]
  \tikzstyle{edge} = [->,thick, line join=miter, auto]
  \tikzstyle{dotted line} = [dotted, very thick, >=stealth', line join=miter]
  \foreach \name/\x/\y/\text in {s/2/3/s,
                                 v1/0/1.5/v_1,
                                 v2/1/1.5/v_3,
                                 v3/2/1.5/,
                                 v4/4/1.5/,
                                 v5/0/0/v_2,
                                 v6/1/0/v_4,
                                 v7/2/0/,
                                 v8/4/0/}
    \node[vertex] (G-\name) at (\x,\y) {$\text$};
  \foreach \from/\to in {
      s/v1/,
      s/v2/,
      s/v3/,
      s/v4/,
      v1/v5/,
      v2/v6/,
      v3/v7/,
      v4/v8/  }
  \path[edge] (G-\from)  edge   (G-\to);

\draw [style=dotted] (2.5,1.5) -- (3.5,1.5);
\draw [style=dotted] (2.5,0) -- (3.5,0);
\draw [style=dotted] (0,-0.5) -- (0,-1);
\draw [style=dotted] (1,-0.5) -- (1,-1);
\draw [style=dotted] (2,-0.5) -- (2,-1);
\draw [style=dotted] (4,-0.5) -- (4,-1);
\end{tikzpicture}

$\underbrace{\hspace{0.30\columnwidth}}_p$
\end{center}
\caption{Schedule Graph: the start node $s$ has at most $p$ outgoing edges
and no incoming edges; every other node in the schedule graph has at
most one outgoing edge and exactly one incoming edge; self-loops are not
permitted. $s$ is a artificial instruction which is not executed. A line of nodes in the schedule graph represents the sequence
of instructions that are executed by one of the processing elements. For example,
instructions $v_1$ and $v_2$ are executed by the first processing element, in that
order. $v_3$ and $v_4$ are executed by the second processing element, and so on.}\label{fig:stream-graph}
\end{figure}
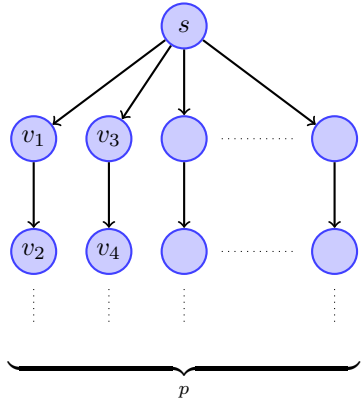

The structure of a schedule graph is depicted in Figure~\ref{fig:stream-graph}. 
It is defined to have a dedicated start node $s$, with at most $p$ outgoing
edges, and no incoming edges. The time parameters of the start node
are all zero and the start node is not a matrix instruction that is
executed on a processing element. The remaining nodes in the schedule graph
represent matrix instructions that will be executed.
Each of these nodes has exactly one predecessor node and at most one successor node.
The successor and predecessor node must not be the node itself, i.e.,
in the graph we do not allow self-loops. The successors nodes of the
start nodes (e.g.~$v_1$ and $v_3$ from Figure~\ref{fig:stream-graph}) represent the first nodes that will be executed by 
each processing element.  Their successor nodes ($v_2$ and $v_4$) are the second
instructions that will be executed, and so on.

We have at most $p$ outgoing edges for the start node
$s$ and hence at most $p$ processing elements. 
The linear constraints that ensure that elements $x_{ij}$ of the
adjacency matrix form a schedule graph, are given below:
\begin{eqnarray*}
 &\sum_{i=1}^{n}x_{1i}\leq p\\
 &x_{i1}=0  &i \in I\\
 &{\displaystyle \sum_{j=1}^{n}x_{ij}}\leq1  & i \in I\setminus\{1\} \\
 &{\displaystyle \sum_{j=1}^{n}x_{ji}}=1  & i \in I\setminus\{1\}\\
& x_{ii}=0 & i \in I\setminus\{1\} \\
 \end{eqnarray*}

For the task precedence relation $E$  we introduce the time constraints
\begin{equation*}
 t_{i}+t_{\textit{df}}(i)+t_{\textit{ex}}(i)+t_{\textit{wb}}(i)\leq t_{j} \quad (i,j) \in E\\
\end{equation*}
that forces an instruction $j\in I$ to be scheduled after its operands $i : (i,j) \in E$
are completed. The time $t_i$  is a global time for all matrix execution units, i.e., operands
may be scheduled on different matrix execution units and all matrix execution units have the
same wall-clock time.

For two subsequent instructions on a processing element, the durations of the three pipeline stages (data fetch, execute, and write back) cannot overlap.  To ensure this,
a time constraint is introduced for each pipeline stage:
\begin{eqnarray*}
 &{\displaystyle \sum_{i=1}^{n}\left(t_{i}+t_{\textit{df}}(i)\right)x_{ij} } \leq t_{j} &
  j \in I\\
&{\displaystyle \sum_{i=1}^{n}\left(t_{i}+t_{\textit{df-ex}}(i)\right)x_{ij} } \leq t_{j}+t_{\textit{df}}(j) &
  j \in I\\
 &{\displaystyle \sum_{i=1}^{n}\left(t_{i}+t_{\textit{df-ex-wb}}(i)\right)x_{ij} } \leq t_{j}+t_{\textit{df-ex}}(j) &
 j \in I
 \end{eqnarray*}
 where $t_{\textit{df-ex}}(i)=t_{\textit{df}}(i)+t_{\textit{ex}}(i)$ and $t_{\textit{df-ex-wb}}(i)=t_{\textit{df}}(i)+t_{\textit{ex}}(i)+t_{\textit{wb}}(i)$ for all $i\in I$.  The left-hand side of the constraints is a summation of the
time duration multiplied with $x_{ij}$ over all predecessors.
 However, in a schedule graph there exists at most one predecessor modelled by the constraints for the schedule graph.

 The expansion of above constraints have a non-linear term $t_{i}x_{ij}$ which we
 linearise by introducing a new variable $y_{ij}$. This new variable replaces the quadratic
 term $t_{i}x_{ij}$, and we add linear constraints to the program that force the equivalence.
 The linear constraints are developed by using standard techniques~\cite{DBLP:journals/orl/AdamsF05}, i.e.,
\begin{eqnarray*}
 & y_{ij}\in\mathbb{R}^{+} & i,j \in I \\
 & y_{ij}\leq Ux_{ij} &i,j \in I\\
 & t_{i}+Ux_{ij}-U\leq y_{ij}\leq t_{i} & i,j \in I
 \end{eqnarray*}
where $U$ is the sum of all time parameters, which is an upper bound for $y_{ji}$ and $t_{ji}$.

The set of constraints for the makespan can only be greater than the completion time of all instructions in $I$, i.e.,
\begin{equation}
  t_{i}+t_{\textit{df-ex-wb}}(i)\leq z \quad   i \in I
\end{equation}
and the objective function of the mathematical program is to minimise the makespan $z$. The
integer linear program is given at a glance in Appendix~\ref{app:integer-linear-program}
and its implementation as an AMPL~\cite{fourer1987ampl} script is given in Appendix~\ref{app:ampl}.

\subsection{Greedy Algorithm} \label{sec:heuristic}

The integer linear programming formulation (described
previously) of the task-precedence scheduling
problem for an asynchronous super-pipelined, super-scalar matrix engine is intractable.
To overcome this problem, we devise a simple heuristic as shown in
Algorithm~\ref{alg:heuristic} which exhibits a worst-case runtime of $\bigo(n\log n + m)$ where
 $n$ is the number instructions and $m$ is the number of dependencies in the
 task precedence graph. In the algorithm, the sets $d^{-}(i)$ and $d^{+}(i)$ denote the predecessor and successor
sets of an instruction $i\in I$ in the task precedence graph.

The algorithm is based on a list scheduling approach (see subsection~\ref{sec:delay} of
the related work). The list scheduling algorithm uses a heuristic to order
tasks in a list and then greedily distribute them among processors. 
The underlying idea of the heuristic is to
find a topological order for the acyclic data dependence graph $G(I,E)$.
This total order $(i_1,\ldots,i_n)$ for the instructions $I$ has the property
that for each data dependence  $(i,j)\in E$  instruction $j$ is listed after instruction $i$ in
the total order. The topological order ensures that instructions never deadlock
in its execution, i.e., the operands are either available or the instructions
have to wait a finite amount of time for their operands to become available. The topological order
is computed with the counter $c_i$ for an instruction $i \in I$. Initially, the counters
are set to the number of incoming edges, i.e., $|d^-(i)|$. Only instructions with counter values
of zero can be scheduled, and are maintained in a priority queue (min-heap). The priority is determined by the earliest point in time when an instruction can be scheduled. After scheduling an instruction $i$ on one of the processing elements, the counters of its successors $j \in d^+(i)$ in the task precedence graph are decremented by one. The successors are scheduled (put into the queue) as soon as their operands have been allocated to a processing element.

\begin{algorithm}[tbp]
\begin{lstlisting}[language=pseudocode]
(*@ \textbf{for all} $k\in\{1,\ldots,p\}$ @*)
(*@ \atab $s_{\textit{df}}(k) \leftarrow 0$ @*)
(*@ \atab $s_{\textit{ex}}(k) \leftarrow 0$ @*)
(*@ \atab $s_{\textit{wb}}(k) \leftarrow 0$ @*)
(*@ \textbf{for all} $i\in  I$ @*)
(*@ \atab $c_{i} \leftarrow |d^{-}(i)|$ @*)
(*@ \atab $e_{i} \leftarrow 0$ @*)
(*@ \atab \textbf{if} $|d^{-}(i)| = 0$ @*)
(*@ \atab \atab queue $i$ with cost $0$ @*)
(*@ \textbf{while} queue not empty @*)
(*@ \atab dequeue $i$ @*)
(*@ \atab $k \leftarrow \textit{minarg}_{r}\{h(i,r)\}$ @*)
(*@ \atab $t_i \leftarrow h(i,k)$ @*)
(*@ \atab add $i$ to stream $k$ @*)
(*@ \atab $s_{\textit{df}}(k) \leftarrow t_i +  t_\textit{df}(i)$ @*)
(*@ \atab $s_{\textit{ex}}(k) \leftarrow t_i + t_\textit{df-ex}(i) $ @*)
(*@ \atab $s_{\textit{wb}}(k) \leftarrow t_i + t_\textit{df-ex-wb}(i)$ @*)
(*@ \atab \textbf{for all} $j\in d^{+}(i)$ @*)
(*@ \atab \atab $c_{j} \leftarrow c_j - 1$ @*)
(*@ \atab \atab $e_{j} \leftarrow max(e_j,t_i + t_\textit{df-ex-wb}(i))$ @*)
(*@ \atab \atab \textbf{if} $c_{j}=0$ @*)
(*@ \atab \atab \atab \atab ~enqueue $j$ with cost $e_{j}$ @*)
\end{lstlisting}
\caption{Heuristic Scheduler\label{alg:heuristic}: earliest instruction in earliest stream.
}

\end{algorithm}

From the set of instructions whose operands have been scheduled (or do not
have operands) we choose the instruction that can be scheduled the earliest
on the time line. The earliest scheduling time $e_i$ of an instruction
$i \in I$ is determined by the completion time of the instruction's operands. When
an instruction is scheduled, the earliest scheduling time of its dependent instructions
is updated (cf.~line 20 in Algorithm~\ref{alg:heuristic}).

For each instruction we have exactly one enqueue and one dequeue from the priority
queue resulting in a worst-case complexity of $\bigo(n \log n)$. We also need to update
the earliest scheduling time $e_i$ of every instruction which has a worst-case complexity of $\bigo(m)$. Thus, the worst-case runtime complexity of the algorithm given in Figure~\ref{alg:heuristic} is $\bigo(n\log n + m)$. Note that we have
the underlying assumption that the number of processors $p$ is constant in this analysis.

\begin{figure}
\begin{center}
\includegraphics[width=0.6\columnwidth]{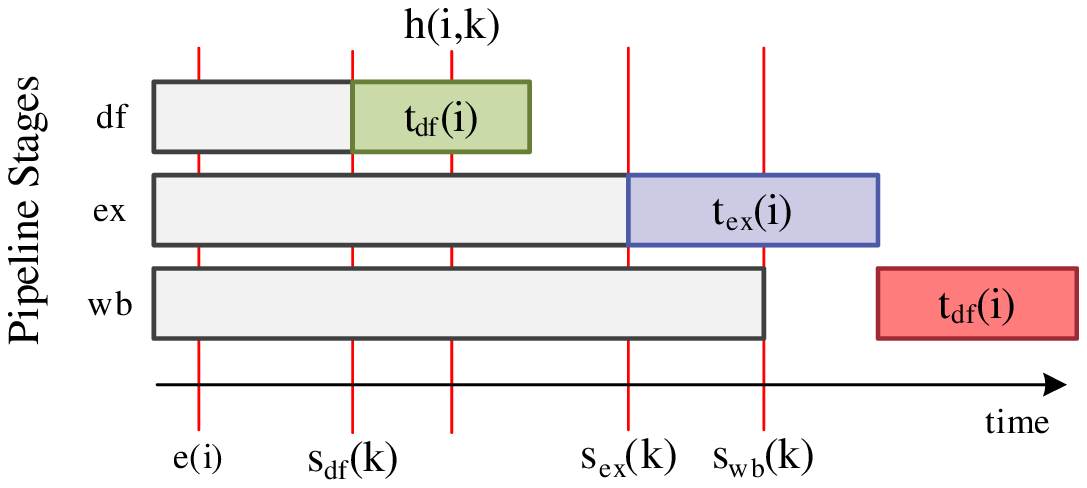}
\end{center}
\caption{Example input for slot function $h(i,k)$. $h(i,k)$ is the
earliest point in time when instruction $i$ can be scheduled. 
$s_\textit{df}(k)$,  $s_\textit{ex}(k)$, and $s_\textit{wb}(k)$ denote the last
completion times of each phase of the last instruction scheduled on a processing
element.  The start time of the data fetch of $i$ is not limited by the completion
time of its operands $e(i)$ but instead the completion of the data fetch of the previous
instruction $s_{df}(k)$. The execution phase is also limited by the completion of the 
execution of the previous instruction.  The start time of the write back of $i$ is limited by the completion of the execution of $i$. From the estimated start time
of the write back operation, we subtract $t_\textit{df}(i) + t_\textit{ex}(i)$
to yield $h(i,k)$ which is the earliest time at which $i$ can begin.
of the execution of $i$
pipeline.}\label{fig:slot-function}
\end{figure}

Once the earliest instruction is selected we choose to schedule the instruction on the
processing element
that can execute the instruction at the earliest point in time. For the selection of the processing element, we use a slot
function $h(i,k)$ which computes the earliest possible start of instruction $i$ on processing element $k$.
The slot function takes into account the completion times of the data fetch, execute, and
write back phase of the previous instruction. We keep track of the completion time
of the last scheduled instruction on processing element $k$ for each of the pipeline stages. These are denoted by $s_\textit{df}(k)$,  $s_\textit{ex}(k)$, and $s_\textit{wb}(k)$ for data fetch, execution and write back, respectively.
The slot function is given by
\begin{align*}
h(i,k)= &\max (
         \max
         (\max(e_i,s_\textit{df}(k))+t_\textit{df}(i),s_\textit{ex}(k))+\quad t_\textit{ex}(i),s_\textit{wb})
          - (t_\textit{df}(i) + t_\textit{ex}(i))
\end{align*}
and depicted in Figure~\ref{fig:slot-function}. For an instruction, $i$ being
scheduled on processing element $k$, the start
time of each of its pipeline stages could be limited by two things. The
first is the completion of $i$'s operands, i.e.~$i$ cannot begin data fetch until its operands are complete. The second is the completion of the previous instruction on processing element
$k$, i.e.~$i$ cannot begin its data fetch phase until the data fetch of the
previous instruction on $k$ is complete (and likewise for the execution and write
back phases).
The three max functions here are used to determine which of these is the limiting
factor in determining the start time of $i$.

In the example in Figure~\ref{fig:slot-function} the
earliest execution time $e(i)$ of instruction $i$ is before the
completion of the data fetch stage of the previous instruction
$s_\textit{df}$. Thus, the completion of the data fetch stage of the
previous instruction is taken as the earliest start time
for the data fetch of $i$. We add to this point in time the
estimated duration of data fetch for $i$ and compare it with the
completion of the execution stage of the previous instruction, $s_\textit{ex}$. This allows us to determine
the earliest point in time for commencing the execution stage
of $i$.  In the example, the completion of the execution stage of the
previous instruction takes longer and hence determines the earliest starting
time for the execution of $i$. We add to this point
in time the duration of the
execution phase of $i$ and compare it with the completion of the write back
phase of the previous instruction, $s_\textit{wb}$. This allows us to determine
at what time the operation can begin its write back. In this case, the write
back stage is limited by the completion of the execution phase of $i$. 
Finally, from the point in time that the write back of $i$ will finish, we
subtract the duration of data fetch
and execute stage to obtain the earliest point in time to slot instruction
$i$ into processing element $k$.

\section{Cell Computation Engine\label{sec:cell}}

\subsection{The Cell Broadband Engine} \label{sec:arch}
\begin{figure}
\begin{center}
\includegraphics[width=0.8\textwidth]{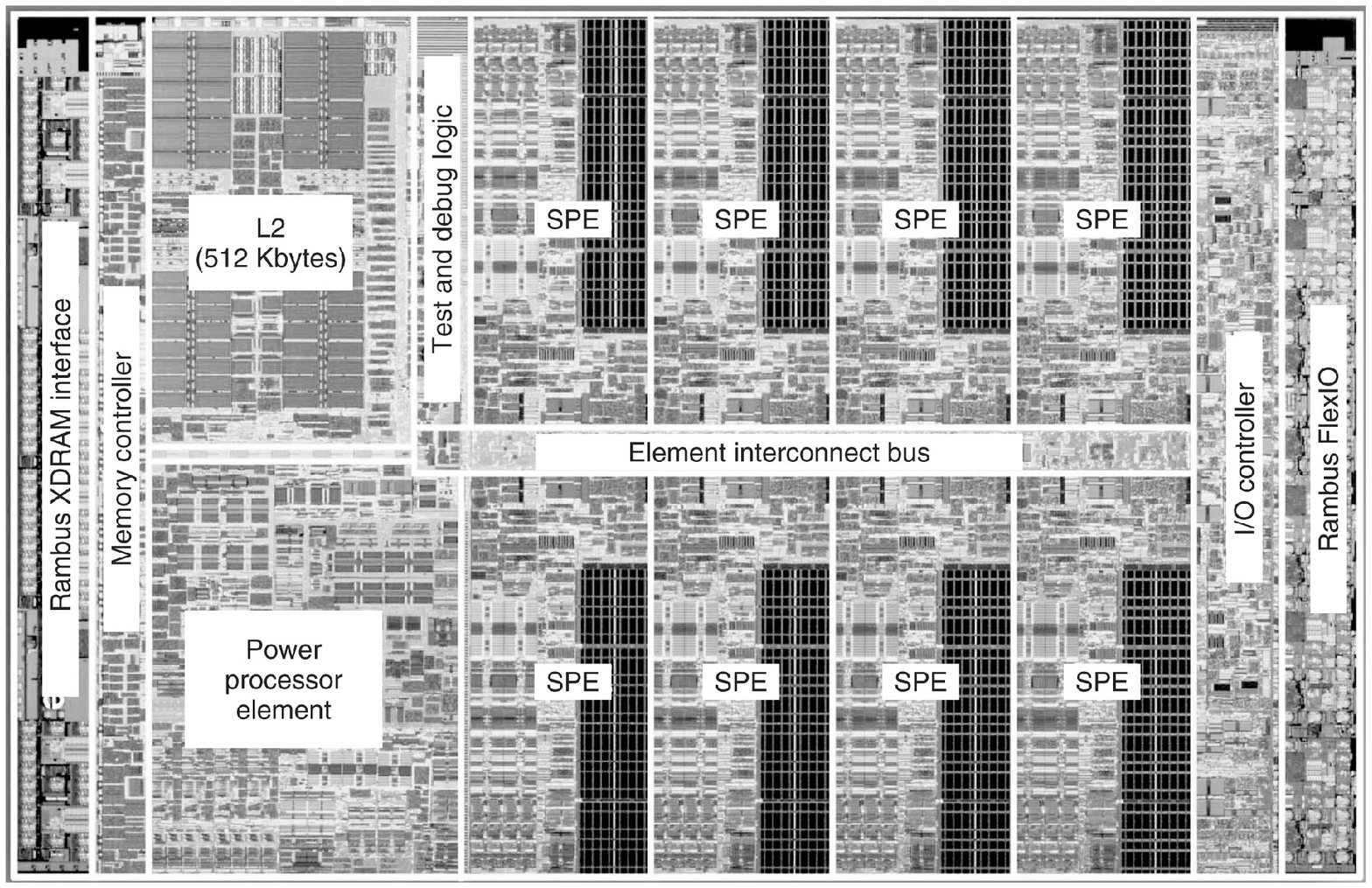}
\end{center}
\label{fig:cell-die}
\caption{A die photo of the Cell Broadband Engine~\protect\cite{1130803}. This shows the main components
of the processor: (1) the Power Processor Element (PPE), (2) the 8 Synergistic Processing
Elements (SPEs), and (3) the Element Interconnect Bus (EIB).}
\end{figure}

The Cell Broadband Engine architecture (Cell) is a heterogeneous multicore architecture that was jointly developed by
IBM, Sony and Toshiba~\cite{1130803}.
The structure of the Cell is shown in Figure~\ref{fig:cell-die}. It consists of a 64-bit PowerPC core (PPE),  eight SIMD cores called Synergistic Processing Elements (SPEs), a memory interface
controller and an I/O controller. The PPE and SPEs communicate through a high-speed Element Interconnect Bus (EIB).
At a 3.2~GHz clock rate, the theoretical peak performance for a single SPE with single-precision floating-point operations is $25.6$~GFLOPS, yielding
an overall performance of $204.8$~GFLOPS for 8~SPEs. For double-precision the theoretical peak performance for a
single SPE is ~$12.8$~GFLOPS, and~$102.4$~GFLOPS aggregate. The EIB supports a peak bandwidth of 204.8~GB/s for on-chip data
transfers between the PPE, SPEs, memory interface controller and I/O controller. The memory interface controller provides
25.2~GB/s peak bandwidth to main memory.

The PPE is the Cell's main processor, designated to run the operating
system, coordinate the SPEs and perform the control-intensive part of
applications. The PPE's memory hierarchy is similar to conventional
processors, with 32kB level-1 instruction and data caches and a
512kB level-2 cache.

SPEs are designed for high-performance
data-streaming and for data-intensive computations. Their memory
hierarchy consists of a 128x128-bit SIMD register file, 256kB of local
store memory and the off-chip main memory shared with the PPE. SPEs
can run SIMD operations at four different granularities: 16-way 8-bit
integers, 8-way 16-bit integers, four-way 32-bit integers, four-way
single-precision floating-point numbers, or two-way 64-bit double
precision floating point numbers. The 256kB local store of an SPE is
shared for code and data. Each SPE can only access the code and data
in its own local store. DMA transfers are used to move data between the
local store of an SPE and main memory, as well as between 
the local stores of different
SPEs. DMA transfers are asynchronous and enable SPEs to overlap
computation with communication. Unlike caches, SPE local stores must
be explicitly managed by software. 

Additionally, a mailbox mechanism is
provided for communication between the PPE and each of the SPEs. 
This acts as a queue which the PPE can enqueue data items onto and the target SPE
can dequeue values from. Each mailbox is capable of holding up to four 32 bit data items at 
any instant. If a PPE attempts to enqueue more than four data items, one data item will
be overwritten or execution will be blocked until an item is dequeued.

\subsection{Matrix Execution Units} \label{sec:meus}
As mentioned in section~\ref{sec:overview}, we implement the computation engine component
of our framework for the Cell architecture.
At the core of the Cell computation engine are the Matrix Execution Units (MEUs). An MEU is a small virtual machine which runs on an individual SPE and executes matrix operations.  There is one MEU running on each of the SPEs available in the Cell processor (typically 8). The fact that there are multiple MEUs capable of executing matrix instructions concurrently gives our framework a \emph{superscalar} property. Each MEU idles on its SPE, waiting to be notified by the PPE that there are matrix operations ready for execution. The PPE uses the mailbox mechanism available in the Cell processor to inform an MEU of the memory location of such a ready operation.

After receiving this message, the MEU uses Direct Memory Access (DMA) list commands to transfer the operands involved in the matrix operation from main memory to the local store of the SPE on which the MEU is running. Once the transfer is complete, the MEU executes the matrix operation and again uses DMA list commands to transfer the result of the operation back to main memory. This process repeats for each matrix operation that has been assigned to the MEU.

In order to achieve high performance on the Cell Broadband Engine Architecture, the latency of memory transfers between main memory and an SPE's local store must be hidden by overlapping them with computation. This technique is known as multi-buffering and is facilitated in the Cell processor by non-blocking DMA commands. When an SPE uses a DMA command to transfer data between its local store and main memory, the transfer is carried out by a separate processor known as the Memory Flow Controller (MFC) and execution can continue on the SPE. The SPE can then issue a further DMA command which will force completion of the transfer, when required.

The Cell Matrix Engine uses a form of multi-buffering known as triple-buffering to hide data transfer latencies. Three buffers are allocated on the local store of each SPE. At any point in time, exactly one of the following data items will be stored in each of the buffers:
\begin{enumerate}
\item The operands of the current operation being executed.
\item The result of the current operation being executed.
\item The result of the last operation that was executed as it is transferred out \emph{or} the operands of the next operation to be executed as they are transferred in.
\end{enumerate}

The use of this triple-buffering technique results in \emph{pipelined} execution of operations on each of the MEUs. These pipeline stages are taken into account by our scheduling algorithm and were
described in section~\ref{sec:scheduling}. However, here we describe them again in the context
of the Cell implementation. There are three pipeline stages:
\begin{enumerate}
\item \textbf{Data Fetch (df):} The operands of an operation are loaded from main memory into the 
local store of an SPE. We use the notation $df_{\textit{start}}(op, B_{\textit{i}})$ to denote
the initiation of a DMA transfer of operation $op$'s operands into buffer $B_{\textit{i}}$ on
an SPE.  This is a non-blocking call. We use the notation $df_{\textit{finish}}(op, B_{\textit{i}})$
to denote a blocking call that ensures the completion of the same transfer.
\item \textbf{Execute (ex):} The operation is executed on an SPE. We use the notation
$ex(op, B_{\textit{i}}, B_{\textit{j}})$ to denote the execution of operation $op$ whose
operands are contained in buffer $B_{\textit{i}}$ and whose result is to be placed in buffer
$B_{\textit{j}}$.
\item \textbf{Write Back (wb):} The result of the operation is transferred from the local store
of an SPE back to main memory. We use
the notation $wb(op, B_{\textit{j}})$ to denote a non-blocking call to initiate the transfer of
$op$'s result which is stored in buffer $B_{\textit{j}}$ back to main memory. Note that this
call also issues a barrier command which means that the write back transfer must be fully
complete before any further DMA commands to buffer $B_{\textit{j}}$ can begin. This prevents the chance of race conditions.
\end{enumerate}

\begin{figure}
\centering
\subfigure[Initially the execution pipeline is empty, and all buffers on the SPE
are empty. The first operation, $op_1$ is offered from the PPE and its operands
are fetched from main memory into buffer $B_{0}$. As there is no
operation that is ready to execute, the SPE initiates transfer of the next operations
($op_2$'s) operands into the second buffer, $B_{1}$. ]
{\includegraphics[width=0.65\columnwidth]{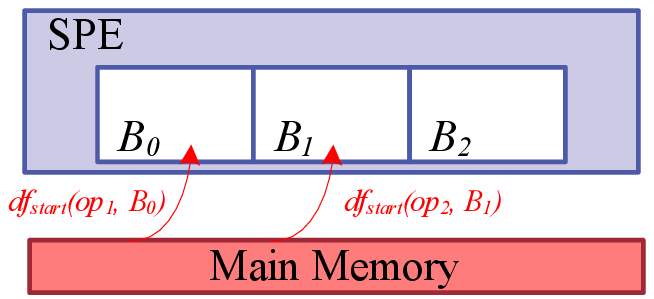} \label{fig:triple1}}
\subfigure[Execution on the SPE blocks until the completion of the data transfer of
$op_1$'s operands. When the transfer is completed, execution of $op_1$ begins and
the result is placed into buffer $B_{2}$. ]
{\includegraphics[width=0.65\columnwidth]{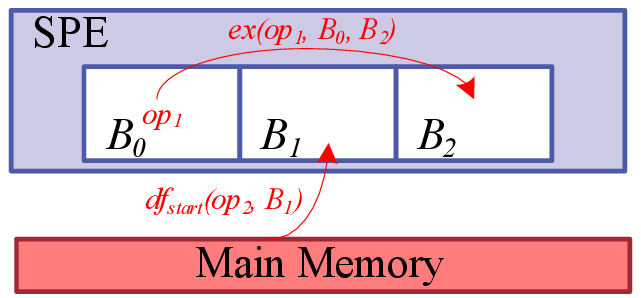} \label{fig:triple2}}
\subfigure[The transfer of the result of $op_1$ back from $B_{2}$ to main memory
is initiated. A DMA barrier command is placed after the write back operation and
the data fetch of the next operation ($op_3$) into $B_{3}$ is also requested.
However, this transfer does not begin until the write back of $op_1$ is complete due
to the barrier. Execution on the SPE is blocked until transfer of $op_2$ is complete and
then the execution of $op_2$ begins, with the result being placed in $B_{0}$.]
{\includegraphics[width=0.65\columnwidth]{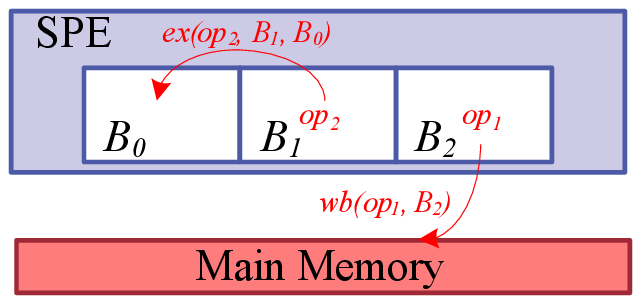} \label{fig:triple3}}
\subfigure[Once the write back of $op_1$ is complete, the data fetch of $op_3$ into
buffer $B_{2}$ begins. Execution of $op_2$ continues. We have now filled the execution
pipeline and are in the pipeline state as in Figure~\ref{fig:triple2}.]
{\includegraphics[width=0.65\columnwidth]{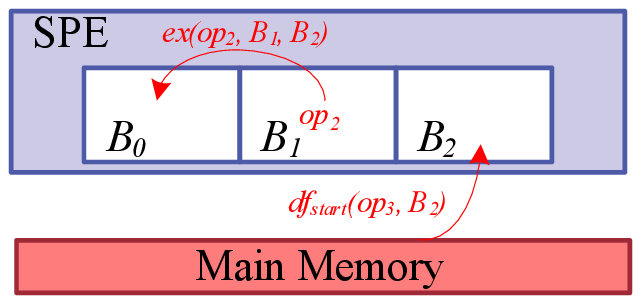} \label{fig:triple4}}
\caption{An illustration of the triple buffering process on the SPEs. Each SPE has
3 buffers labelled $B_{0}$, $B_{1}$ and $B_{2}$. Three matrix operations, $op_1$,
$op_2$ and $op_3$ are executed. \label{fig:triple}}
\end{figure}

Figure~\ref{fig:triple} describes the triple buffering process for executing three matrix
operations: $op_1$, $op_2$ and $op_3$.
Algorithm~\ref{alg:meu-loop} shows pseudocode of the execution loop
which runs on the MEU virtual machines and facilitates triple buffering. Therein, the variable
$in$ is the number of the buffer containing the operands of the current operation being executed,
$out$ is the number of the buffer containing the result of the current operation and
$next$ is the number of the buffer containing the operands of the next operation to be executed.

\begin{algorithm}
\begin{lstlisting}[language=pseudocode]
(*@ $i \leftarrow 0$ @*)
(*@ \textbf{loop} @*)
(*@ \atab $in \leftarrow i \bmod 3$ @*)
(*@ \atab $next \leftarrow (i + 1) \bmod 3$ @*)
(*@ \atab $out \leftarrow (i + 2) \bmod 3$ @*)
(*@ \atab $op_{\textit{next}} \leftarrow$ read mailbox  @*)
(*@ \atab \textbf{if} $op_{\textit{next}} =$ terminate @*)
(*@ \atab \atab exit @*)
(*@ \atab $df_{\textit{start}}(op_{\textit{next}}, B_{\textit{next}})$ @*)
(*@ \atab $df_{\textit{finish}}(op_{\textit{in}}, B_{\textit{in}})$ @*)
(*@ \atab $ex(op_{\textit{in}}, B_{\textit{in}}, B_{\textit{out}})$ @*)
(*@ \atab $wb(op_{\textit{in}}, B_{\textit{out}})$ @*)
(*@ \atab $i = i + 1$ @*)
\end{lstlisting}
\caption{Execution loop for the MEUs which facilitates triple buffering. \label{alg:meu-loop}}
\end{algorithm}


The matrix operations in the MEUs, such as matrix multiplication and matrix addition, must have
implementations that are highly optimised for the SPEs in order to obtain the best performance.
There are several considerations when optimising algorithms for the SPEs~\cite{ibmcellhandbook}. SIMD (Single Instruction,
Multiple Data) instructions must be used to exploit the 128-bit wide vector processing capabilities
of the SPEs. Branches must be eliminated to reduce the chance of costly branch misprediction and
loops should be unrolled to expose a maximum amount of instruction-level parallelism and
utilise both execution pipelines of the SPEs.

These factors make writing optimised operations for the SPEs a challenging and time-consuming task.
Where possible, we utilise the optimised libraries provided with the Cell SDK to perform operations
on the SPEs. For example, the Large Matrix Library~\cite{cellblasmanual} and SPE BLAS 
library~\cite{cellblasmanual} provide optimised matrix multiplication implementations which are used
by our framework.

Where existing library functions were not available, we implemented our own operations. While
not heavily optimised, these operations utilise the SIMD instructions through the vector
intrinsics provided in the Cell toolchain to yield moderate performance.

\subsection{Partitioning, Alignment and Padding of Matrices} 
\label{sec:alignment}

The small local stores of SPEs, coupled with the triple buffering technique described
in subsection~\ref{sec:meus} limit the size of matrices that can be operated on by the SPEs.
As mentioned in subsection~\ref{sec:arch}, the local store of an SPE provides 256kB of
memory which is shared for code and data. The MEU virtual machine program is approximately
26kB in size, leaving 230kB of space for matrices.  Triple buffering divides this
space further into 3 buffers of size $\approx$76kB, each of which should be capable of storing
the operands of a single matrix operation. We assume that each matrix operation can have up to two
operands, giving each matrix block a maximum size of $\approx$38kB. Given that single and double precision floating point numbers occupy 4 and
8 bytes of memory respectively, the maximum number of matrix elements that will fit in the buffer
of an SPE is $S=9728$ single precision elements or $S=4864$ double precision elements. 
However, the partitioning
scheme described in section~\ref{sec:lowering} results in blocks that are up to
~$\Div\lfloor\frac{\sqrt{\MSpace}}{\Div}\rfloor\times\Div\lfloor\frac{\sqrt{\MSpace}}{\Div}\rfloor$ in dimensions, where $\Div=4$ for single precision and $\Div=2$ for double precision (for
reasons described below). This results
in an effective buffer size for each operand of $S=9216$ for single and $S=4864$ for double
precision elements, as stated in section~\ref{sec:lowering}.

Each block of a matrix is transferred to the local store of an SPE 
using a DMA list transfer (a group of individual DMA commands). 
Each list entry transfers a single row of the matrix block as shown in Figure~\ref{subfig:memB}
of section~\ref{sec:lowering}.
In the Cell
architecture, DMA commands can only operate on memory sizes which are multiples of 16 bytes~\cite{ibmcellhandbook}.
The partitioning scheme described in section~\ref{sec:lowering} ensures that
the number of columns and rows in a block is always a multiple of \Div. 
Hence, we choose the divisor to be $\Div=4$ for single precision and $\Div=2$ for double precision which guarantees that each row of a block is a multiple of 16 bytes.

Optimal DMA performance
is seen when transfer sizes are multiples of 128 bytes~\cite{ibmcellhandbook}.
The maximum length of a row in a block (using our partitioning scheme) is 384
bytes. Thus blocks that reach this maximum will achieve optimal transfer performance.
A value for the divisor could be chosen such that row lengths are \emph{always} a multiple of
128 bytes ($\Div=32$ for single precision and $\Div=16$ for double
precision). Unfortunately, this would result in additional overhead in the extra padding of matrices. For example, if a single precision matrix had 4 columns, a value of $\Div=32$
would force 28 extra columns of padding to ensure that the length of each row is 32
elements.

Block dimensions which are a multiple of 16 bytes are also necessary in order
to easily utilise the SIMD operations on the SPEs which operate on 128 bit (16 byte) vectors.
For the same reason, the matrix libraries provided for the SPEs require matrices to be of
these dimensions.

A further requirement of DMA commands is that memory is aligned to at least a 16 byte 
boundary~\cite{ibmcellhandbook}.
The beginning of matrix arrays are aligned to 128 byte boundaries in main memory using the
\mytexttt{malloc\_align()} function call provided in the Cell SDK. Since
the partitioning scheme results in row lengths guaranteed to be a multiple of 16
bytes (as described above), this means that a partitioned block always begins on a 16 byte boundary in memory. 
Optimal DMA performance is seen when source and destination addresses are aligned to 128 byte boundaries (one cache-line)~\cite{ibmcellhandbook}, which is again
ensured for all blocks with an appropriate choice for the divisor \Div.

\subsection{Execution Control} \label{sec:execution-control}
\begin{figure}[htp]
\begin{center}
\includegraphics[width=0.9\columnwidth]{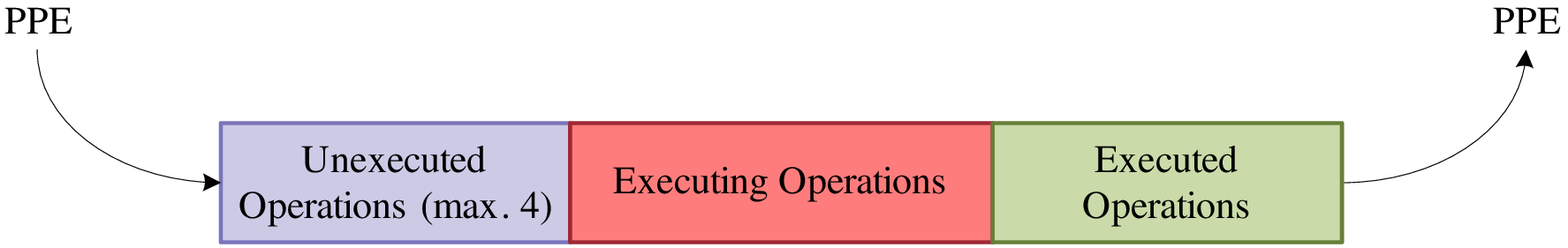}
\end{center}
\caption{Matrix execution queue. \label{fig:instructionstream}}
\end{figure}

A fast and efficient execution control mechanism must be used to deliver scheduled operations from the
PPE to the SPEs and track the completion of operations.  On the PPE, each SPE is modelled as an
\emph{execution queue}. The PPE enqueues operations on to the end 
of an execution queue if they are to be executed by the SPE represented by that
queue. Once an SPE has finished execution of the operation, the PPE can dequeue
the operation from the execution queue and register it as complete.
At any point in time, the execution queue contains three kinds of operations
(Figure~\ref{fig:instructionstream}):
\begin{enumerate}
\item \textbf{Unexecuted operations:} A number of operations that the PPE has offered to the SPE but
which the SPE has not yet acknowledged.
\item \textbf{Executing operations:} A number of operations that the SPE has acknowledged and started data transfer or execution of.
\item \textbf{Executed operations:} A number of operations which the SPE has finished executing and
whose result has been returned to main memory, but which the PPE has not yet acknowledged as complete.
\end{enumerate}

In order to enqueue an unexecuted operation, $op_{u}$, onto the execution queue, two criteria must be
met. Firstly, the number of unexecuted operations already in the execution queue must be less than
4. This is because the mailbox mechanism of each SPE which is used to deliver
operations from the PPE to SPEs can only hold up to four 32 bit
values. Secondly, all of $op_{u}$'s operands must be available in main memory. This means that if $op_{u}$
depends on the results of other operations, these must be fully computed by an SPE and returned to main
memory prior to $op_{u}$ being added to the execution queue. 

Ideally, the scheduler
would produce a schedule which includes the exact wall-clock time at which $op_{u}$
can begin executing.  However, indeterminisms in the
architecture can cause variations in the actual execution times of operations, which cannot
be anticipated by the scheduler. To ensure that all operands of an operation are available
prior to its execution, we introduce a \emph{guard count} value to each operation.  The guard
count is an integer value which represents the number of operands of an operation
which have not yet been computed. Before the execution of any operation, every operation's
guard count is equal to the total number of operands that the operation has.
For example, a matrix addition operation has 2 operands (the matrices to be added)
so its guard count is initially 2.
 When an executed
operation, $op_{e}$, has been acknowledged by the PPE as being complete, the guard count of all the
operations which depend on $op_{e}$ are decremented by 1. This means that it is safe to place an
unexecuted operation on the execution queue when its guard count is zero.

In order to signal the PPE to the completion of an operation, the SPE uses DMA commands to write
a counter value to a pre-determined location in main memory. The PPE polls this memory location
and when it changes, knows that an operation is complete and it is safe to dequeue the instruction
from the execution queue. This interprocessor communication technique
is chosen over other techniques because it provides better performance~\cite{ibmcellhandbook}. It is important that the PPE
is notified of completed operations as quickly as possible, as this allows dependant operations to begin execution sooner.

The execution control process consists of the PPE examining each of the execution queues in turn. It
enqueues as many unexecuted operations as possible on to the execution queue, in the order
that they are specified in the processors schedule. The SPE is notified of these operations through
the mailbox mechanism. The PPE then dequeues any complete operations from
the execution queue, decrementing the guard count of dependant operations by 1.
This process continues until all operations in the current schedule have completed
execution.

\section{Implementation Details and Tuning} \label{sec:impl}
The implementation of our framework has $\thickapprox$7000 lines
of C and C++ code. The code consists of two separately compiled programs: one written for the PPE
and one written for the SPEs. The SPE program consists of the implementation of the Matrix
Execution Units (MEUs) as described in section~\ref{sec:cell}. Although this component
is smaller (in lines of code) than the PPE program, 
it is crucial that it achieves high performance and does
not consume a significant portion of the small local store of the SPEs. Thus, it is written in C
code. The PPE program consists of the implementation of the remainder of the framework, i.e. 
the Octave extension, the lowerer, scheduler and execution control. It is written in C++ and 
performance is mostly not as critical as the SPE program. 

The development of the software was performed
on a conventional x86-based processor running Fedora Core 9 and the
latest version of the Cell SDK (3.1.0)~\cite{ibmcellhandbook}. The code for the Cell
can be compiled on this standard architecture by using cross-compilation
tools included in the Cell SDK. 
Testing was initially performed with the Sony Playstation 3 games console that
runs a Cell Broadband Engine processor. However, the Playstation 3 is limited
to 256MB of memory which impeded development and limited the size of test
cases that could be run. Later in the project we obtained access to a 
BladeCenter~QS22 server, as described in subsection~\ref{sec:setup}. Although
the QS22 has the same Cell-based processor as the Playstation 3, it offers
32GB of memory allowing large test case to be executed without causing memory swapping
and distorting time measurements.  

Here we describe implementation issues with regard to (1) debugging and testing
our framework, (2) details of the implementation
of the Octave extension, and (3) details of the performance tuning we performed.

\subsection{Debugging and Testing}
Debugging on the Cell processor is difficult for several reasons.
Firstly, concurrent programming on any architecture is challenging and the Cell
processor is no exception. Our framework has several threads of execution running simultaneously on the PPE, as well as each of the 8 SPEs running a separate
thread. There is no easy method for finding race conditions or dead locks between
these threads.
Secondly, standard
memory debugging tools such as Valgrind cannot be used on the Cell architecture
 because DMA memory transfers
between the SPEs and main memory interfere with their operation. This makes
debugging memory issues difficult, even when confined to the PPE.
Furthermore, the low-level programming model employed on the Cell processor makes programming a highly error-prone exercise. This is particularly true when
programming for the SPEs.

We used a few techniques to alleviate this issue. Firstly, we wrote our C++ framework
in a platform independent way. Only one component of the framework, the computation
engine, contained hardware-specific code (see the system overview in section~\ref{sec:overview}). We wrote versions of the computation engine
for the Cell processor, but also for a standard x86 architecture. The x86 implementation is sequential and
was not designed for performance but instead for debugging purposes. It contains
very simple, \naive implementations of matrix operations. This allowed us to
test our framework on a more conventional architecture before having to deal
with the idiosyncrasies of the Cell architecture. We were able to use standard debugging tools,
including Valgrind, on the x86 to eliminate errors in all components of our framework, besides
the Cell computation engine.
If we experienced anomalies when running the framework with the Cell architecture that were not
experience on the x86 architecture, we could narrow the bug to the Cell implementation of the computation engine. This dual-platform testing testing technique greatly 
reduced the time spent on debugging.

As a precursor to the integration of our framework with Octave, we developed a more simple
interface to our framework. This was in the form of a simple trace language which allowed
a user to to enter a trace of matrix operations to be executed.
The language does not have any control-flow mechanisms (like loops
or if-statements). An example of a script
for this language is shown in Figure~\ref{fig:trace-lang}. This script
simply defines two matrices and adds them together.
This interface allowed us to concentrate on the implementation of the back-end components
of our framework prior to the integration with Octave. However, this also served as another
debugging tool, allowing us to determine whether there was an issue with the Octave
extension or with a different component of the framework. An automatic test
generator was written to generate random test traces for this front end based
on a number of parameters such as trace length and matrix dimensions. This
further assisted us in debugging and testing our framework, particularly with larger
traces which would take too long to write by hand.

\begin{figure}[htb]
\centering
\begin{minipage}[t]{0.6\textwidth}
\begin{lstlisting}[language=myoctave,frameround=fttt,frame=trBL]
# define matrix A
A = [ 1,  2,  3,  4;
      5,  6,  7,  8;
      9,  10, 11, 12;
      13, 14, 15, 16];

# define matrix B
B = [ 1,  2,  3,  4;
      5,  6,  7,  8;
      9,  10, 11, 12;
      13, 14, 15, 16];

# add both matrices
C = MADD(A, B);
\end{lstlisting}
\caption{A script written for our simple trace language
interface. It declares two matrices $A$ and $B$ and adds
them together, storing the result in $C$.\label{fig:trace-lang}
}
\end{minipage}
\end{figure}

Debugging tools provided by the Cell SDK were also used to assist in
finding bugs in Cell specific components. We used the GNU
debugger (\mytexttt{gdb}) designed for the Cell processor that  
facilitates tracing the execution
of the program on both the PPE and SPEs simultaneously. 

Regression testing was employed throughout the project to reduce the chance
of introducing bugs and also to reduce debugging time. For every matrix operation
that was implemented in the framework, several tests were written. Initially
these tests were written for our simple trace language interface 
but later they were crafted as Octave scripts. Octave contains
several functions designed specifically for testing purposes, such
as the \mytexttt{assert} function. An example
of a simple Octave script for testing a matrix addition operation is included in 
Figure~\ref{fig:octave-test}.

\begin{figure}[htb]
\centering
\begin{minipage}[b]{0.6\textwidth}
\begin{lstlisting}[language=myoctave,frameround=fttt,frame=trBL]
% Matrix addition
x = p_matrix([1, 2; 3, 4]);
y = p_matrix([1, 2; 3, 4]);
z = x + y;
assert(!has_result(z));
assert(z, p_matrix([2, 4; 6, 8]));
assert(has_result(z));
\end{lstlisting}
\caption{An Octave test script for matrix addition. The \mytexttt{assert}
function is built into Octave and checks whether a condition is satisfied or that 
two results match.
The \mytexttt{has\_result} function is provided by our framework to determine
whether a result has been computed by lazy evaluation or whether its execution
is still outstanding.
\label{fig:octave-test}
}
\end{minipage}
\end{figure}

The debugging and testing techniques as  mentioned above, gave us a high level of confidence
in the quality and correctness of our implementation.

\subsection{Octave Extension}
GNU Octave~\cite{octave} is a scripting language and programming environment designed
for scientists and engineers. The programming environment consists of a command line
interface in which users execute scripts or individual statements interactively. As Octave is an open-source
project, its C++ source code is freely available for download and compilation. 

Octave has been designed in a way that facilitates the development of extensions
without the need to re-compile the Octave interpreter.
This is achieved by using shared libraries for extensions,
which can be loaded dynamically by the Octave interpreter at run-time.  
Extensions are called Oct-Files and can be written in C/C++ or FORTRAN. They are compiled by a 
utility program included with Octave called \mytexttt{mkoctfile}. This utility is a wrapper for 
the \mytexttt{gcc} compiler which ensures that the necessary compilation flags and linker options
are provided.

Octave permits extensions in the form of user-defined functions as well
as user-defined data types. An example of a simple user-defined function for printing a message to the interpreter console
is shown in Figure~\ref{fig:octave-hello}. The Oct-File
is pure C++ code however numerous macros are provided by 
the Octave system in order to simplify access to data types and
the declaration of functions.

\begin{figure}
\begin{minipage}[b]{\textwidth}
\begin{lstlisting}[language=c++,frameround=fttt,frame=trBL]
#include <octave/oct.h>

DEFUN_DLD (helloworld, args, nargout, ) {
	int nargin = args.length ();
	octave_stdout << "Hello World has " << nargin 
			<< " input arguments and "
			<< nargout << " output arguments.\n";
	return octave_value_list ();
}
\end{lstlisting}
\caption{
An example of a user-defined function written as an extension for Octave. Although
this is pure C++ code, macros and variables are provided by Octave
to assist with proper integration of the extension. Here, a function
called \mytexttt{helloworld} is defined which merely prints the number
of input and output arguments to the interpreter console. The \mytexttt{DEFUN\_DLD}
macro inserts the appropriate C++ code to allow \mytexttt{helloworld} 
to be called from inside of Octave.
\label{fig:octave-hello}
}
\end{minipage}
\end{figure}

\begin{figure}
\centering
\begin{minipage}[b]{0.6\textwidth}
\begin{lstlisting}[language=myoctave,frameround=fttt,frame=trBL]
A = [10 20 30; 40 50 70; 70 80 90];
A = p_matrix(A);
B = A + 5;
C = B;
disp(B(5));
\end{lstlisting}
\caption{Simple Octave script which uses our framework.
\label{fig:octave-simple}
}
\end{minipage}
\end{figure}

\begin{figure}[t]
\begin{center}
\includegraphics[width=0.6\textwidth]{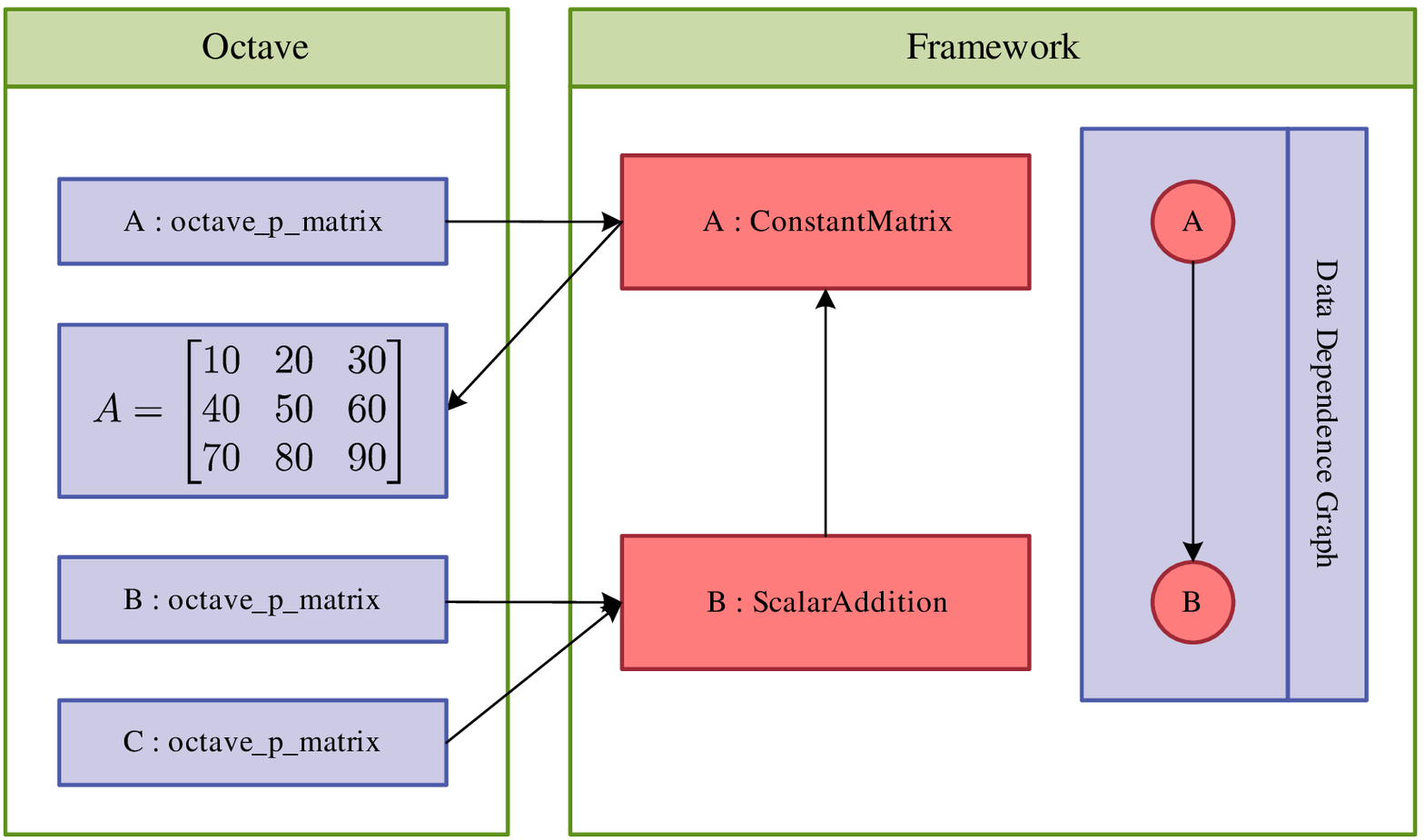}
\par\end{center}
\caption{The memory layout of C++ objects in Octave and our framework.
This shows the state of memory after executing line 4 of the Octave
script in Figure~\ref{fig:octave-simple}.
Every time a \psmatrix variable is created in Octave, an object of the
type \mytexttt{octave\_p\_matrix} is constructed. An object is also constructed
by our framework to represent that operation. When matrix $A$ is cast to the
\psmatrix type in line 2,
a \mytexttt{ConstantMatrix} object is created in our framework. This merely
has a pointer to the memory location of the matrix $A$. The operation
is also added to a data dependence graph on the fly. When the variable $B$
is declared in line 3, another \mytexttt{octave\_p\_matrix} object is constructed
by Octave. The operation to compute $B$ is not executed immediately. Instead,
a \mytexttt{ScalarAddition} object is created  by our framework representing
the addition of the scalar value 5 to the matrix $A$. This contains a pointer
to the matrix $A$, the operand of the operation. $B$ is also added to the
data dependence graph, with a dependency on $A$. In line 4, $B$ is assigned 
to $C$. Copy-on-write is used to prevent the execution of $B$ being forced
and also to prevent making an unneeded deep copy of $B$. Instead, $C$ has
a pointer to the same \mytexttt{ScalarAddition} object which computes its result.
\label{fig:octave-memory}
}
\end{figure}

We extended Octave by introducing a user-defined data type called \psmatrix. 
This is a custom matrix data type which can be used as a replacement
for the default Octave matrix data type. In order to cast standard Octave
matrices to the \psmatrix type, users call a function called \mytexttt{p\_matrix()}
with the standard matrix as an argument. For example, line 1 of the Octave code 
shown in Figure~\ref{fig:octave-simple} creates a $3\times 3$
 standard Octave matrix with the given elements. A user could then call 
the \mytexttt{p\_matrix()} function, as shown in line 2, to convert the matrix to our
custom data type. This is the only user intervention necessary
to use our framework. By replacing the standard Octave matrix data type with our own, 
no user intervention would be required but
this would necessitate modifying and re-compiling the Octave source code
which we wanted to avoid.

User defined data types in Octave are implemented as C++ classes which
inherit from the \\ \mytexttt{octave\_base\_value} class~\cite{octavetypes}. 
Operators
such as \mytexttt{+} and \mytexttt{*}, as well as built-in functions such as \mytexttt{sin} and \mytexttt{round} can be overloaded to behave as required for custom data types.

Our \psmatrix data type is implemented in a C++ class called \mytexttt{octave\_p\_matrix}.
For a complete list of the Octave operators and functions that were implemented for the \psmatrix
 data type refer to Appendix~\ref{app:funcs}.
When a variable of type \psmatrix is created in Octave, a corresponding
C++ \mytexttt{octave\_p\_matrix} object is generated. 
In line 2 of the example code in Figure~\ref{fig:octave-simple}, 
the \mytexttt{p\_matrix()} function is called to cast the matrix $A$ to our
custom data type. Thus, an \mytexttt{octave\_p\_matrix} object is generated by Octave.
Our framework also constructs a \mytexttt{ConstantMatrix} object to represent $A$.
This merely has a pointer to the memory location of the matrix $A$. The
\mytexttt{ConstantMatrix} object is added to the data dependence graph in our framework.
This is illustrated in Figure~\ref{fig:octave-memory}.

When a user
performs an operation on a \psmatrix variable, the overloaded operators
for the type \psmatrix defer execution of the operation.
For example, when a user executes line 3 of the Octave
code in Figure~\ref{fig:octave-simple} the value of $B$ is not computed immediately. Instead, we have overloaded the $+$ operator for
\psmatrix to create a new \mytexttt{ScalarAddition} C++ object in our framework to represents the addition of 5 to $B$. The \mytexttt{ScalarAddition} object contains a pointer to the 
\mytexttt{ConstantMatrix} to denote that $A$ is the operand of $B$, as shown in Figure~\ref{fig:octave-memory}.
This addition operation is also added to the data dependence graph in our framework, with
a data dependency on the matrix, $A$. 
Another \mytexttt{octave\_p\_matrix} object is generated by Octave to represent
the $B$ variable. This contains a pointer to the \mytexttt{ScalarAddition} object in
our framework, as depicted in Figure~\ref{fig:octave-memory}.

To improve performance of the framework, we 
employ copy-on-write semantics. If a matrix of the type
\psmatrix is assigned to another variable, no deep copy of that matrix is made.
For example, in line 4 of Figure~\ref{fig:octave-simple}, the variable $B$
is assigned to the variable $C$. Producing a deep copy of matrix $B$ here
would negatively impact performance in two ways. Firstly, we do not
need to make a copy of $B$ because $B$ and $C$ point to the exact same matrix value.
Thus, making a copy would introduce unnecessary overhead in memory copying operations.
Secondly, if we were to make a deep copy of $B$ here, it would require the the value
of $B$ be evaluated. Because of lazy evaluation, $B$ has not been computed at this point.
Forcing the computation of $B$ would unnecessarily shorten the length of the
trace of matrix operations and reduce the amount of parallelism to be exploited
by our framework. Instead of making a deep copy, the \mytexttt{octave\_p\_matrix} 
object representing $C$ is assigned a reference to the same \mytexttt{ScalarAddition}
operation as $B$ in our framework. This is shown in Figure~\ref{fig:octave-memory}.
If the matrices $B$ or $C$ are modified, for example
by changing the value of an element, then the operation is forced to be computed
and a deep copy of the matrix is made.

Our framework continues to accumulate matrix operations 
until the result of one of those operations is needed. Results of
operations are typically
needed when a user wants to print a matrix to the screen or access particular
elements of a matrix. Thus, we have overloaded the subscript operators on matrices,
as well as the \mytexttt{disp} function such that these trigger
the execution of outstanding operations in the framework. For example,
in the last line of code in Figure~\ref{fig:octave-simple}, the 
the 5'th element of $B$ is printed to the screen. A call to our overloaded subscript operator is made. This first determines whether the  
\mytexttt{ScalarAddition} operation to yield the value of $B$ has been computed yet.
In this case it has not so execution of the outstanding trace
of operations in the framework is initiated, and continues as described in 
section~\ref{sec:motivating}.
Once the result of $B$ has been computed, it is made
available to the Octave script and execution of the script can continue.
This technique gives our framework its lazy evaluation semantics.

Octave automatically handles garbage collection of its variables but this must
be tied to our framework so we can do the appropriate de-allocation of resources.
When a \psmatrix variable is garbage collected in Octave, the destructor of the corresponding 
\mytexttt{octave\_p\_matrix} object is called. This will de-allocate the matrix operation
object in our framework which is tied to that variable. In the example code in Figure~\ref{fig:octave-simple}  garbage collection of the variable $B$ will result in the de-allocation of the
\mytexttt{ScalarAddition} operation tied to $B$. However, we must be careful
not to garbage collect operations or matrices that are still needed.
If the matrix $A$ was garbage
collected prior to the evaluation of $B$ then our framework would not de-allocate $B$.
This could occur, for example, if the code \lstinline[language=myoctave]!clear A!
was called before line 5 of Figure~\ref{fig:octave-simple}. Our framework recognises
that the matrix $A$ is still needed to compute $B$. Thus, it will not be de-allocated
until execution of the next trace is complete.

\subsection{Performance Tuning}
It was necessary to optimise our framework to obtain the high performance reported
in the experimental subsection (see section~\ref{sec:experiment}).
In order to identify bottlenecks in the framework, we profiled the code using
the performance counters available in the Cell processor~\cite{ibmcellhandbook}. These are highly accurate
timers which operate at the time-base frequency of the architecture implementation
(26664325Hz for the BladeCenter QS22 implementation). 
All components of the framework were instrumented with profiling code in both
the PPE and SPE programs. A breakdown of time spent in each component of
the framework was obtained through use of this profiling and is shown in section~\ref{sec:experiment}.
 A compile time option
was used in our framework to enable or disable profiling as required. 

A large bottleneck we discovered through profiling was in the time
spent making virtual function calls. For example, matrix operation classes
in our framework are implemented with an inheritance hierarchy. There is a virtual
method implemented by each operation for obtaining the estimated data fetch,
execution and write back times for that operation. The heuristic scheduler
made many calls to this virtual function for each operation in order to compute
the schedule. We were able to reduce the number of calls to once for each operation
in a trace which resulted in an improvement in the time spent scheduling operations. 
This change alone led to a performance improvement of 20\% in the scheduling
process.

Using appropriate data structures had a significant impact on performance. Profiling
allowed us to determine that the scheduling of operations was initially dominating
the execution time of a trace for some benchmarks. That is, scheduling was taking
longer than the actual execution time. After investigating this, we found that
it was due to the use of an array for storing ready operations in the scheduler. 
This array has to be iterated over to find the earliest operation that could be scheduled
and this has to be done for every operation in the schedule (see section~\ref{sec:scheduling}). By using a min-heap priority queue instead of the array, we
were able to reduce the $O(n^2)$ complexity of this portion of the algorithm
to $O(n\log n)$. This had a large impact when there were many operations in the
trace and reduced the average time spent scheduling operations to
less than 15\% of the total execution time of the trace.

Another area where we made improvements was with memory fragmentation. Originally,
we just used the standard  \mytexttt{malloc} function to allocate memory 
for each matrix operation object in our framework. However, as can
be seen in section~\ref{sec:experiment}, benchmark traces could contain over
600000 matrix operations. This led to high fragmentation of memory and had
a negative impact on performance. We used memory pooling to pre-allocate large contiguous
blocks of memory at once. From these large blocks, matrix operation objects were
allocated. 

Memory pooling was also used to improve the performance of the allocation
of matrices. During the execution of matrix multiplication operations,
many temporary matrices are used to perform the summation of blocks
(described in section~\ref{sec:lowering}). Allocation of these
temporary matrices degraded performance. Instead of allocating each
matrix with a separate \mytexttt{malloc} call, we again used a pool
of contiguous memory. Not only did this improve the time spent
allocating matrices, but also deallocating them because returning
memory to the pool is a very cheap operation.

Finally, the execution control mechanism (described in section~\ref{sec:cell})
was another area where improvements were made. By instrumenting the SPEs with
profiling code, we were able to compute the idle time spent by each SPE.
By simplifying the execution control loop, which delivers operations to the 
SPEs, it was able to service them more quickly. This resulted in up to 
a 30\% decrease in the idle time of SPEs for some benchmarks.

\section{Experimental Results}\label{sec:experiment}
In the experimental evaluation of our framework we focused on the
performance improvements that users can expect from running Octave
application code with our framework running on the Cell processor.  
We compared our framework with three other typical system configurations:
\begin{enumerate}
\item A default installation of Octave on a contemporary Intel Core2 Quad processor.
\item A default installation of Octave on the Cell processor.
\item A default installation of MATLAB on an Intel Core2 Quad processor.
\end{enumerate} 

In addition to these comparisons, we also investigated our approach with respect to:
\begin{enumerate}
\item Its scalability to a larger number of SPEs.
\item The accuracy of our time models for estimating the execution time of matrix 
operations.
\item The quality of schedules derived by our heuristic scheduling algorithm.
\item The extent to which our framework was able to utilise the Cell SPEs.
\end{enumerate}

\subsection{Experimental Setup and Benchmark Suite} \label{sec:setup}

For the experiments, a BladeCenter~QS22 server was used to run our framework.
The details of its configuration are summarised in Table~\ref{tab:setup}.
The BladeCenter~QS22 is comprised of two IBM PowerXCell 8i processors (3.2GHz/ 1MB
L2) and has 32GB DDR2 memory. We used only one out of the two Cell processors available
on the BladeCenter due to non-uniform-memory-access (NUMA) issues which
complicated use of the second processor (though our framework could be configured
to work with an arbitrary number of SPEs). We
utilise all 8 of the SPEs as well as the PPE available in the Cell processor
(refer to section~\ref{sec:cell}).

Originally, we intended to perform the experiments on the Cell processor found
in the Playstation~3 games console. Despite both the the BladeCenter~QS22
and Playstation 3 being based on the same Cell Broadband Engine architecture 
with the same capabilities, the Playstation 3 has only $\thickapprox$200MB of
accessible RAM. Thus the BladeCenter~QS22 was used to alleviate this limitation
on the size of benchmark inputs.

\begin{table}[H]
\begin{center}
\begin{tabular}{|l|c|}
\hline
Blade model: & IBM BladeCenter QS22\tabularnewline
\hline
Number Cell processors: & 1 (8 SPEs, 1 PPE)\tabularnewline
\hline
Memory: & 32GB DDR2\tabularnewline
\hline
Linux distribution: & Red Hat EL Server 5.4 \tabularnewline
\hline
Linux kernel:& 2.6.18\tabularnewline
\hline
IBM Cell SDK:& 3.1.0\tabularnewline
\hline
GCC:& 4.1.1\tabularnewline
\hline
GCC optimisation flags:& O3\tabularnewline
\hline
Octave:& 3.0.3\tabularnewline
\hline
\end{tabular}
\caption{Experimental setup.\label{tab:setup}}
\end{center}
\end{table}

The following benchmarks were selected to evaluate the performance of our framework:

\begin{enumerate}
\item
\textbf{dft} Computation of the Discrete Fourier Transform (DFT) of a series of signals.
\item
\textbf{synth} A synthetically constructed benchmark with many small, independent matrix multiplications.
\item
\textbf{hill} Encryption and decryption using Hill ciphers.
\item
\textbf{hits} Computation of the Hyperlink-Induced Topic Search (HITS) algorithm for estimating the importance of a web-page.
\item
\textbf{kmeans} Computation of the k-means clustering of a 2D point set.
\item
\textbf{leontief} Computation of a Leontief input-output model, used to predict performance of economies.
\item
\textbf{markov} Computation of a Markov chain.
\item
\textbf{neural} Training of a single-layer neural network.
\item
\textbf{reachability} Computation of the reachability matrix of a graph.
\end{enumerate}

These benchmarks were chosen to represent commonly used kernel programs from scientific,
engineering and computer science domains.

\subsection{Speedups and Scalability} \label{sec:speedups}
We first compared our framework on the Cell processor with a default 
Octave installation on an Intel Core2 Quad Q9550~2.83GHz processor. We chose
the Intel Core2 Quad architecture for comparison because it represents
a contemporary and frequently used microprocessing architecture which
is more modern than the Cell processor.
The standard Octave installation on the Intel Core2 Quad
uses single-threaded ATLAS BLAS libraries for some matrix operations~\cite{atlas_sc98}.
These libraries utilise the SSE3 multimedia extensions of the processor for performance.

\begin{figure}
\begin{center}
\includegraphics[width=0.5\columnwidth]{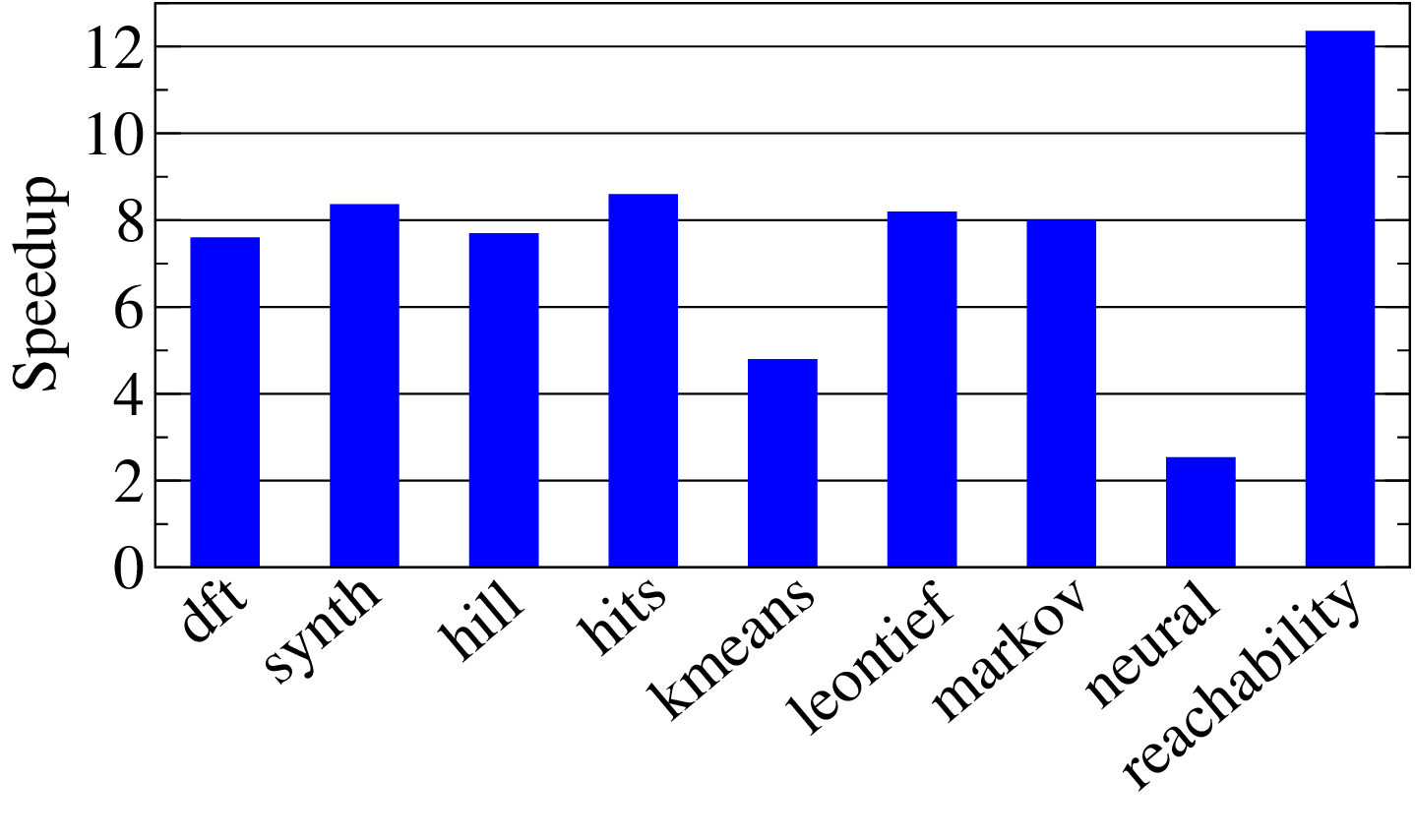}
\end{center}
\caption{Configuration 1: Speedup of our framework on the Cell processor over a default Octave installation on an Intel Core2 Quad processor for the 9 benchmark programs.}\label{fig:speedup}
\end{figure}

As depicted in Figure~\ref{fig:speedup}, our
framework achieves speedups of up to a factor of 12 times over standard
Octave on the Intel Core2 Quad.  For most benchmarks we achieve speedups of over 7.
The k-means clustering (``kmeans'') and neural network (``neural'')
benchmarks achieve lower speedups. These lower speedups can be attributed
to two factors. Firstly, the time models for some matrix operations in these
benchmarks may not be accurate. There is further evidence for this provided
in subsection~\ref{sec:exp-time-model}. Inaccurate time models would lead
to inaccurate estimation of the execution times of operations which could
result in degraded schedules. Fine-tuning of time-models could see improved
performance. The second reason  for lower speedups is that
these benchmarks contain many matrix operations on vectors. The partitioning
scheme employed in our framework results in sub-matrices which are very
small, as described in section~\ref{sec:lowering}. This results in under-utilisation
of the SPEs in the Cell processor and poorer performance. This could be improved
by modifying the partitioning scheme for vector operations.

The second configuration that we compared our framework to was a default
installation of Octave on the BladeCenter~QS22. That is, we compared the runtime
of the benchmarks BladeCenter server with and without our extension
switched on in Octave. It should be noted that the default Octave installation
utilises hardware-specific BLAS libraries which are provided with the
IBM Cell SDK. These libraries are highly optimised for the Cell
architecture~\cite{saxena2008optimization} and can utilise both Cell
processors available in the QS22 (a total of 16 SPEs).

\begin{figure}
\begin{center}
\includegraphics[width=0.5\columnwidth, clip=true]{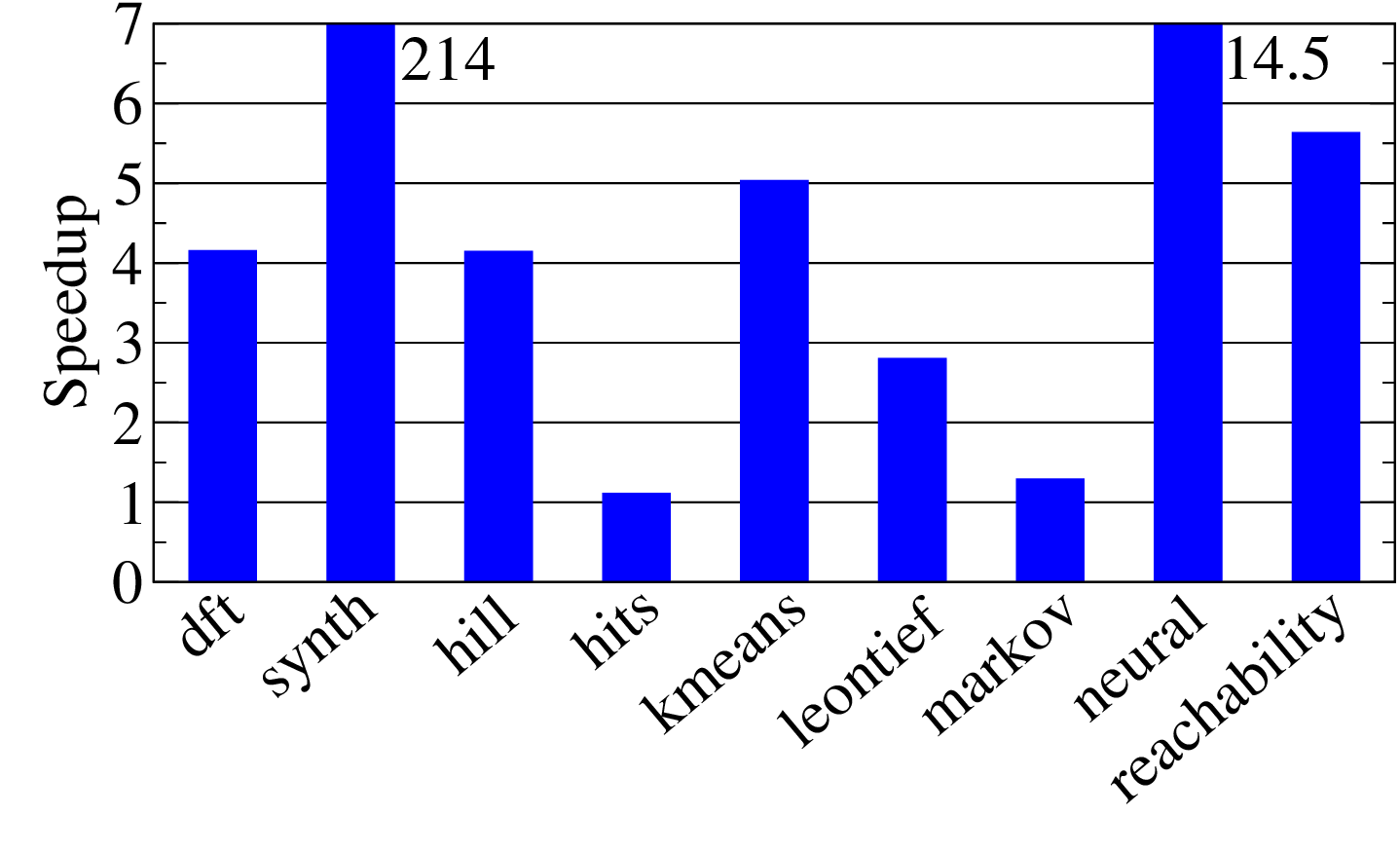}
\end{center}
\caption{Configuration 2: Speedup of our framework on the Cell processor over a default Octave installation on the BladeCenter QS22 for the 9 benchmark programs.}\label{fig:speedup-blas}
\end{figure}

With this configuration we achieved the speedups depicted in
Figure~\ref{fig:speedup-blas}. It can be seen that our framework achieves
a speedup with all benchmarks (though some are lower than others).
One may expect our framework to perform worse than the default
installation which uses highly optimised BLAS libraries. However, 
the BLAS libraries do not accelerate all Octave functions. For example,
computation of the $\sin$ function of elements in a matrix is not a standard BLAS
operation. As a result, the PPE is used to compute such functions, which can be relatively
slow. Our framework can utilise the SPEs for all operations that we have implemented and 
thus see an improvement in performance for these operations.

Furthermore, the Cell BLAS libraries exploit only the data
parallelism of matrix operations whereas our framework also exploits
instruction level parallelism. This is emphasised with the
``synth''~benchmark. This benchmark has a large amount of instruction
level parallelism because it contains many matrix
multiplications which are independent of each other. 
However, because each matrix multiplication operates
only on small matrices, there is no benefit in dividing an
operation among SPEs.  This causes BLAS to confine computation of the
operations to the PPE only. Each operation is computed sequentially 
by the default installation of Octave
leading to a very high runtime with this benchmark. In contrast,
our framework recognises that although it is not worth dividing up individual
operations among SPEs, the multiplications can be executed concurrently. This
allows our framework to obtain high utilisation of the Cell processor and
explains the speedup of over 200 times for this benchmark.

The seemingly erratic speedups across benchmarks in this configuration (i.e.~
speedups vary from 1.1 times to 214 times) can be explained with the same reasoning.
That is, the speedup of our framework depends heavily on how much instruction
level parallelism is available in a benchmark, as well as the proportion of
matrix operations in the benchmark which
can be executed by the optimised BLAS libraries in the default Octave installation.

In a variation of this configuration, we forced the default installation of
Octave to utilise only the PPE of the Cell processor (and not the SPEs). Our
framework achieved speedups of several hundred times in this case.

The third and final configuration that we compared our framework with was
a default installation of MATLAB running on an Intel Core2 Quad processor.
Since MATLAB is a commercial product, it is likely to be more optimised
than Octave, its open-source equivalent. However, we wanted to compare
our framework with MATLAB because it is more frequently used than Octave
and represents the current state of the art in high-level scientific computing.
MATLAB and Octave have very similar syntax and the benchmark
programs required no changes for them to run on MATLAB.

\begin{figure}
\begin{center}
\includegraphics[width=0.5\columnwidth, clip=true]{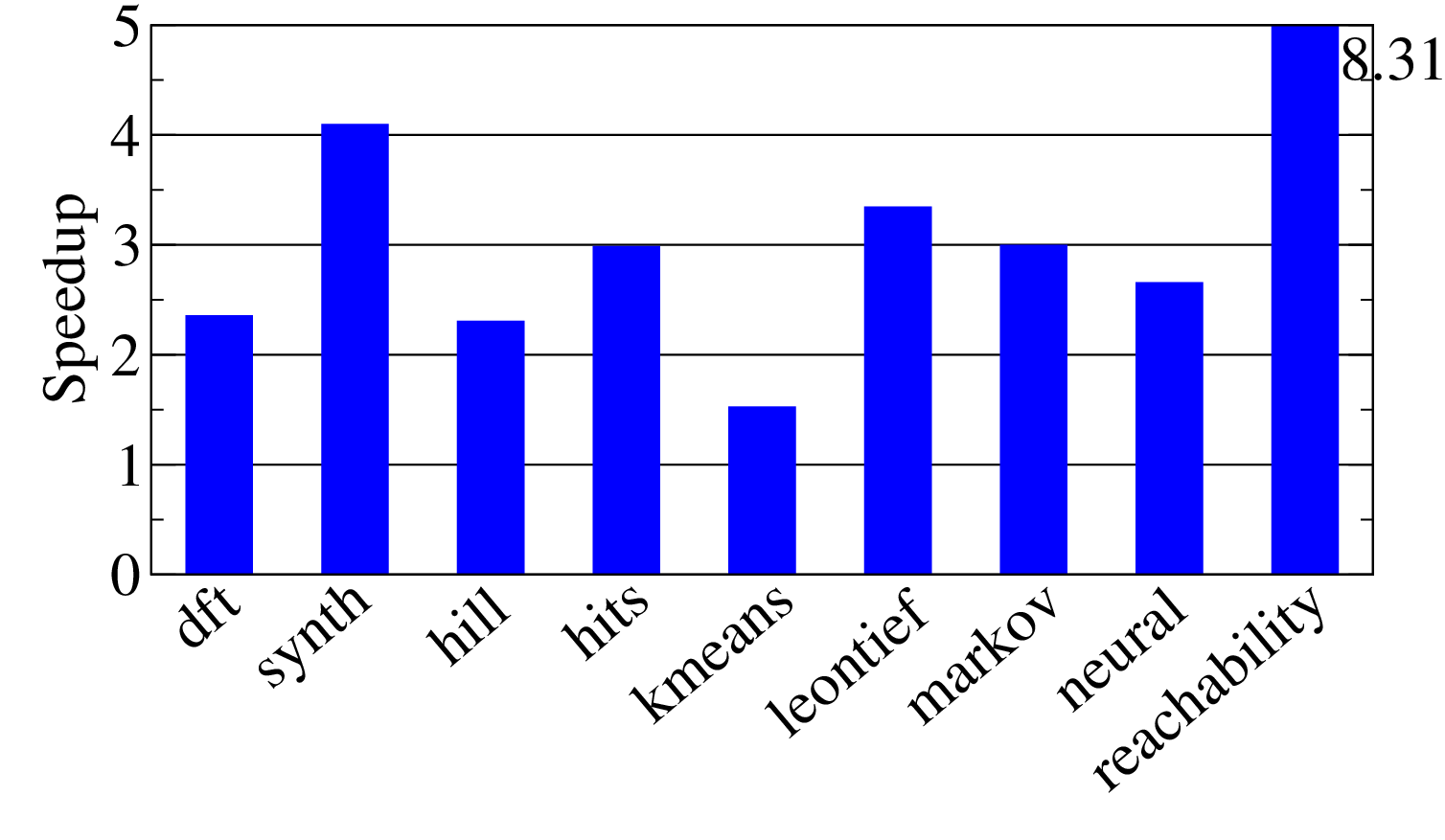}
\end{center}
\caption{Configuration 3: Speedup of our framework on the Cell processor over a default MATLAB installation on an Intel Core2Quad processor for the 9 benchmark programs.}\label{fig:speedup-matlab}
\end{figure}

In this experiment, we used the latest version of MATLAB (version 7.8) which
was found to utilise all four cores of the Intel Core2 Quad processor in
the execution of matrix operations. Again, we found we were able
to achieve speedups for all benchmarks and up to 8 times in the best case, 
as shown in Figure~\ref{fig:speedup-matlab}. However,
these speedups were lower than those from the first configuration which
confirmed our suspicions that MATLAB provides better performance than Octave.

Again, the speedups of our framework can be attributed to the exploitation
of instruction level parallelism and the scheduling of operations
among the parallel processing elements of the Cell architecture to
improve utilisation. Neither of these are currently performed by Octave or MATLAB
on any architecture.

\begin{figure}[tb]
\begin{center}
\subfigure[\label{fig:efficiency}]{\includegraphics[width=0.45\columnwidth]{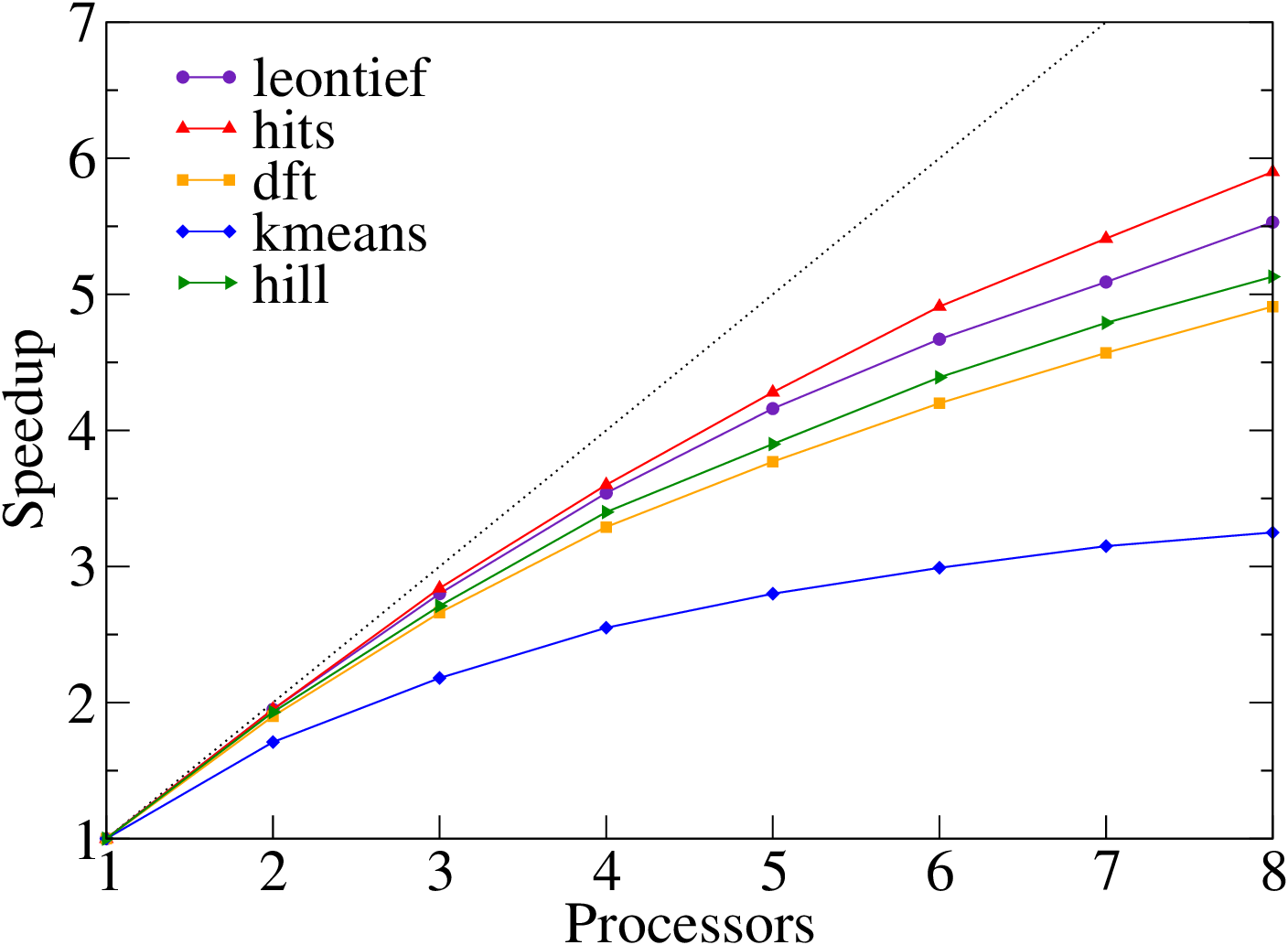}}
\subfigure[\label{fig:efficiency2}]{\includegraphics[width=0.45\columnwidth]{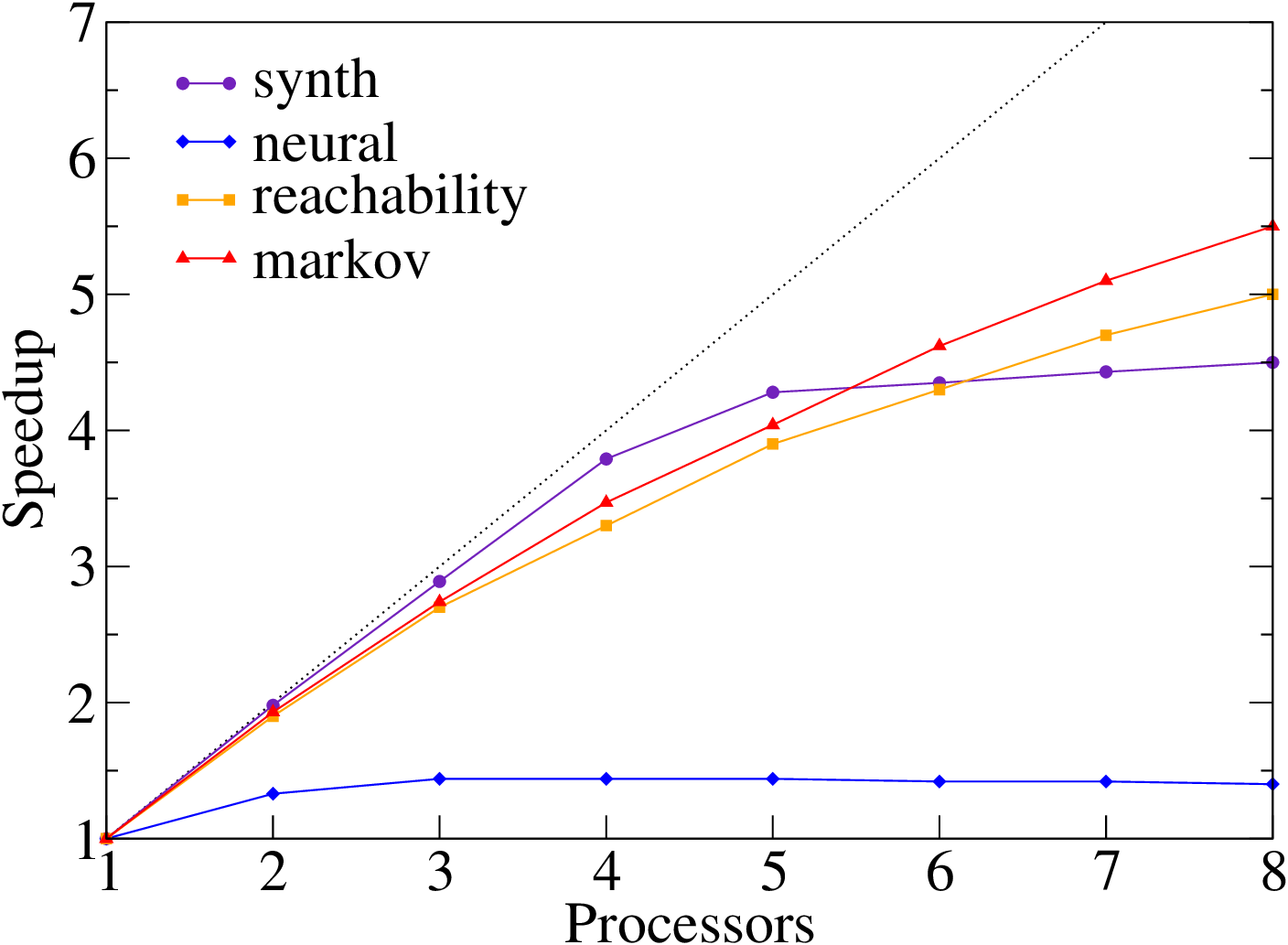}}
\end{center}
\caption{Performance scalability of the proposed framework on the IBM BladeCenter~QS22. This
shows the speedup achieved for each benchmark by varying the number of SPE cores utilised
from 1 to 8.
\label{fig:scalability}}
\end{figure}

In the next set of experiments, we investigated how the performance
of our framework scaled with the number of SPEs utilised
on the Cell processor. The results for each of our
benchmarks are shown in Figure~\ref{fig:scalability}.
As can be seen, performance deviates from linear speedup
depicted by the dotted line. This can be explained by considering
the proportion of sequential and parallelisable work for each benchmark,
as shown in
Table~\ref{tab:amdahl}.  These values were estimated
by measuring the time taken by each component of the framework
when utilising only 1 SPE. The proportion of time spent in lowering and scheduling
on the PPE could not be parallelised. However the proportion of time
spent executing matrix operations on the SPEs could be parallelised by
increasing the number of SPEs used.

When the amount of sequential work in each benchmark is taken in to
consideration, the theoretically achievable speedup with 8 processors 
can be computed using Amdahl's law~\cite{hill2008amdahl}.
This states that the maximum speedup achievable is given by $\frac{1}{(1-P)+\frac{P}{N}}$
where $P$ is the parallelisable portion of the program as a fraction and $N$ is the
number of processors used (8 in this case). These values for each benchmark, along
with the observed speedups are also shown in Table~\ref{tab:amdahl}. It
follows that most benchmarks do not deviate significantly from the
maximum theoretical speedup, indicating that to obtain
better performance it is necessary to further optimise the sequential
portion of the workload, i.e., the lowering and scheduling components.

Any deviations between the maximum theoretical speedups and the observed
speedups shown in Table~\ref{tab:amdahl} are due to
deficiencies that arise as the number of processors increases. For example, the Cell's EIB communications bus will have to cope with a higher load when the number of SPEs increases. The PPE will also face increasing workloads as it has to
service more SPEs. Further optimisations to the execution control mechanisms could help reduce this observed deviation from the maximum theoretical speedup by allowing SPEs to be serviced more quickly.


\begin{table}
\centering
\begin{tabular}{|c||c|c|c|c|}
\hline
Benchmark & Sequential (\%) & Parallelisable (\%)& Theoretical Speedup & Observed Speedup\tabularnewline
\hline
\hline
dft & 5.2 & 94.8 & 5.9 & 4.9\tabularnewline
\hline
synth & 7.8 & 92.2 & 5.2 & 4.5\tabularnewline
\hline
hill & 4.3 & 95.7 & 6.1 & 5.1\tabularnewline
\hline
hits & 3.7 & 96.3 & 6.3 & 5.9\tabularnewline
\hline
kmeans & 5 & 95 & 5.9 & 3.2\tabularnewline
\hline
leontief & 3.8 & 96.2 & 6.3 & 5.5\tabularnewline
\hline
markov & 3.5 & 96.5 & 6.4 & 5.5\tabularnewline
\hline
neural & 17.5 & 82.5 & 3.6 & 1.4\tabularnewline
\hline
reachability & 3.6 & 96.4 & 6.4 & 5.1\tabularnewline
\hline
\end{tabular}
\par
\vspace{5mm}
\caption{Sequential and parallelisable portions of benchmark programs are
measured by running the benchmarks on a single SPE. The sequential portion is
the time that is spent lowering and scheduling operations, allocating memory
and performing cleanup. The parallelisable portion is the time spent in the actual
execution of operations on the SPEs which can be reduced by utilising more SPEs.
The maximum theoretical speedups are computed for 8 SPEs using Amdahl's law.
This states that the maximum speedup achievable is given by $\frac{1}{(1-P)+\frac{P}{N}}$
where $P$ is the parallelisable portion of the program as a fraction and $N$ is the
number of processors used (8 in this case). The maximum
speedup that each benchmark was observed to achieve on our framework with 8 SPEs
is also shown for comparison.\label{tab:amdahl}}
\end{table}

\subsection{Time Model} \label{sec:exp-time-model}

For computing an effective schedule, it is important that the estimated execution
times of operations are accurate (see subsection~\ref{sec:exp-time-model} of the 
scheduling section). 
To verify the accuracy of our time models, we compared the estimated
makespan of a trace with the real, run-time makespan of the same trace.
The estimated makespan is calculated by the heuristic scheduler using
the time models of operations. The real makespan of the trace 
was measured by timing the execution phase of operations in the trace
using the hardware performance counters on the Cell processor. We did
this for many different data dependence graphs and computed the deviation between the two makespans.

\begin{figure}
\begin{center}
\includegraphics[width=0.5\columnwidth]{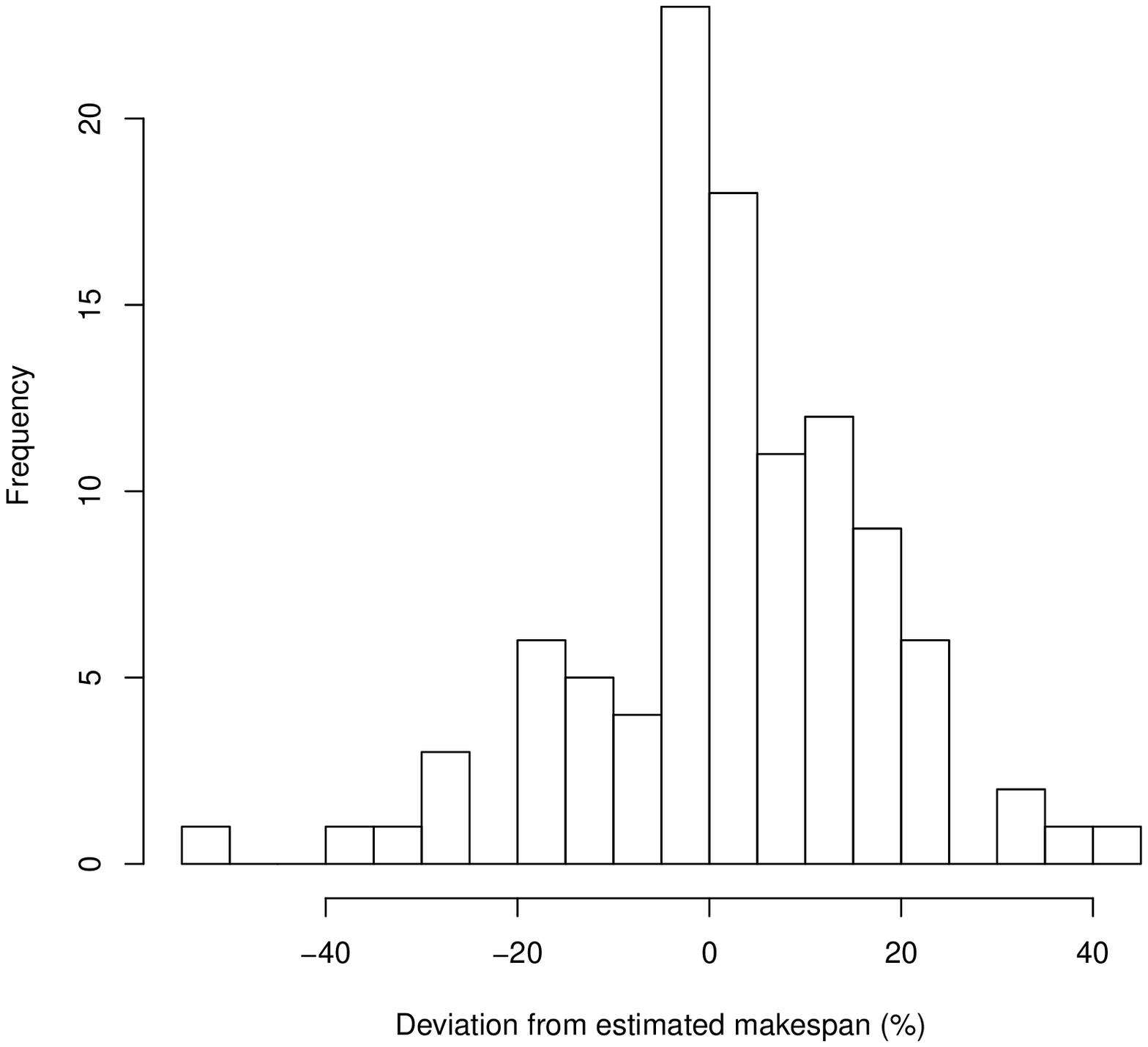}
\end{center}
\caption{The distribution of deviations between the estimated makespan
of the schedule produced by our heuristic algorithm and the observed makespan measured
using performance counters on the Cell. Deviations are computed for over 100 traces produced by
all benchmarks except ``kmeans'' and ``neural''. Most deviations are close to 0\% indicating
that our time models for matrix operations are accurate.}\label{fig:histogram}
\end{figure}

A histogram of these deviations is shown in Figure~\ref{fig:histogram}. The
majority of deviations are close to 0\%, indicating accurate time modelling. The
median of the deviations is 1.3\% and the skewness is -0.54, meaning that the
the distribution is slightly skewed towards the right. The skewness
of the distribution is caused by the overhead of the execution control mechanism
described in section~\ref{sec:cell}.  This mechanism is responsible for synchronising
the execution of matrix operations and the overheads involved are not taken
into account in the estimated makespan of a trace.  Thus the heuristic scheduler
underestimates the makespan accordingly.  

Note that Figure~\ref{fig:histogram} includes deviations 
for data dependence graphs generated by all benchmarks except
for benchmarks ``neural'' and ``kmeans''. We excluded those benchmarks because
there was found to be large deviations between the estimated
and actual makespans (up to 300\%). This provides evidence that the time
models for some operations in these benchmarks are inaccurate and need further 
fine tuning.

\subsection{Scheduling}
As a further experiment, we wanted to investigate the 
quality of our heuristic scheduler by comparing the makespan of the heuristic with the makespan
of the optimal solution. The number of operations and dependencies in 
each of the benchmark programs used to compute the speedups of our framework (see
subsection~\ref{sec:speedups}) are shown in Table~\ref{tab:stats}. The heuristic
scheduling algorithm was able to compute schedules for these problem sizes in
less than a second in all cases. However, these problem sizes were too 
large to compute an optimal solution using the mathematical program described in 
subsection~\ref{sec:ilp}. Thus, we had to reduce the input size of these benchmarks greatly, as listed 
in Table~\ref{tab:ilp}. CPLEX 10.0~\cite{cplex} was able to solve these smaller problem sizes 
within an hour.  

In all cases, the makespan of the heuristic schedule was within 1\% of the makespan of the optimal schedule. However, where the mathematical program took up to 40 minutes to produce
a solution, the time to compute the heuristic schedule was too small to be measured (for these problem sizes). This indicates 
that the heuristic is efficient and effective. 
Unfortunately, these small problem size
instances may not offer enough degrees of freedom to see a
large deviation between
the heuristic and the optimal solution.  We were not able to increase the problem
sizes without making the problems intractable to be solved by CPLEX.

\begin{table}[p]
\small
\begin{center}
\begin{tabular}{|c||c|c||c|c|}
\hline
 & \multicolumn{2}{c||}{Original} & \multicolumn{2}{c|}{Lowered}\tabularnewline
\hline
 & Nodes & Edges & Nodes & Edges\tabularnewline
\hline
\hline
dft & 9 & 9 & 72384 & 141284\tabularnewline
\hline
synth & 60003 & 120000 & 60003 & 120000\tabularnewline
\hline
hill & 9 & 8 & 73684 & 140608\tabularnewline
\hline
hits & 50 & 98 & 319950 & 639450\tabularnewline
\hline
kmeans & 34 & 23 & 137764 & 252743\tabularnewline
\hline
leontief & 204 & 402 & 129152 & 254720\tabularnewline
\hline
markov & 52 & 100 & 93400 & 186580\tabularnewline
\hline
neural & 10012 & 12489 & 231784 & 410151\tabularnewline
\hline
reachability & 960 & 1912 & 119600 & 239000\tabularnewline
\hline
\end{tabular}
\par\end{center}
\caption{Number of operations (nodes) and dependencies (edges) in the original
and lowered dependence graphs for each of the benchmarks used in the
experiments described in subsection~\ref{sec:speedups}.}\label{tab:stats}
\end{table}

\begin{table}[p]
\small
\begin{center}
\begin{tabular}{|c||c|c||c|}
\hline
 & Nodes & Edges & Time (mins:secs)\tabularnewline
\hline
\hline
dft & 34 & 44 & 0:02\tabularnewline
\hline
synth & 17 & 32 & 1:49\tabularnewline
\hline
hill & 28 & 32 & 0:28\tabularnewline
\hline
hits & 40 & 72 & 0:06\tabularnewline
\hline
kmeans & 79 & 64 & $<$ 0:01\tabularnewline
\hline
leontief & 46 & 88 & 0:06\tabularnewline
\hline
markov & 48 & 84 & 2:47\tabularnewline
\hline
neural & 38 & 35 & 0:04\tabularnewline
\hline
reachability & 53 & 104 & 41:19\tabularnewline
\hline
\end{tabular}
\par\end{center}

\caption{Number of operations (nodes) and dependencies (edges) in the largest problem sizes
that the integer linear programming solver could compute a solution for within 1 hour. Also included
is the time taken by the solver to produce a solution.}
\label{tab:ilp}
\end{table}

\begin{figure}
\centering
\includegraphics[width=0.5\columnwidth]{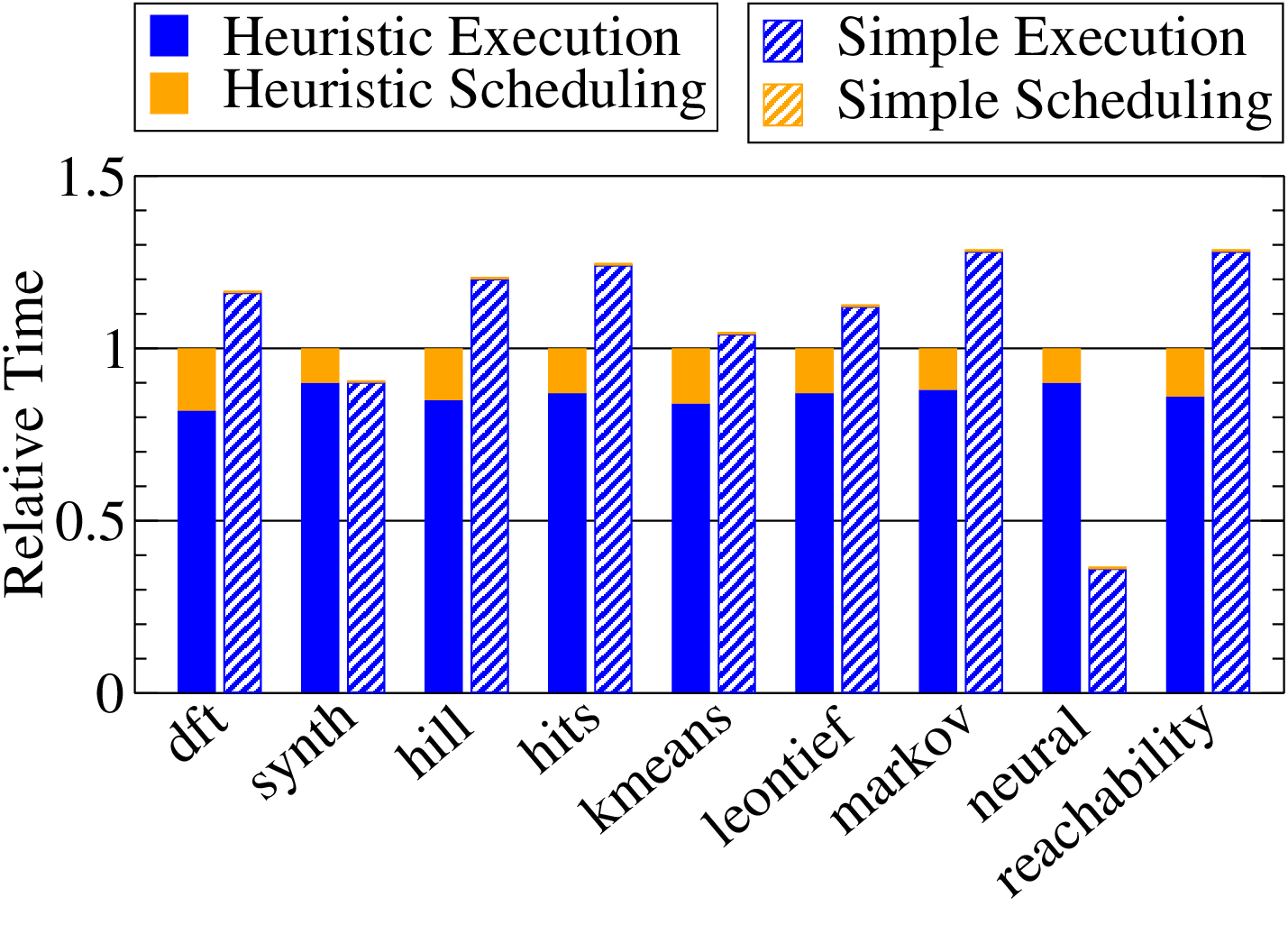}

\caption{Time taken to schedule and execute matrix operations using two scheduling
algorithms: our heuristic algorithm and a \naive approach which balances the number of operations
on each SPE. Note that although the time spent scheduling operations using our heuristic is much
longer than the time spent scheduling using the \naive approach, the total execution time is 
lower for most benchmarks.}
\label{fig:simple-vs-heuristic}
\end{figure}

Another method we used to evaluate the heuristic scheduling algorithm
was to compare it to other heuristic
scheduling techniques.  Figure~\ref{fig:simple-vs-heuristic} shows the time spent scheduling and executing matrix operations for two different scheduling techniques.
The first is our heuristic scheduling algorithm and the second is a very \naive scheduling
algorithm (simple scheduling) which merely distributes operations evenly and in a topological
order among the SPEs. This is done in a round-robin fashion such that each SPE has roughly an equal number of operations to execute.
It can be seen that the simple scheduling algorithm spends almost no time
scheduling operations. However, the time taken to execute the operations 
exceeds the sum of the scheduling and execution
times using our heuristic algorithm (for most benchmarks).
This shows that our scheduling algorithm has a worthwhile impact on the overall
execution time of a trace, even though it does incur noticeable overhead.
By reducing
the overheads of scheduling with further optimisations the benefits of our
scheduling algorithm would be even more substantial.
It also shows that, in general, scheduling is important in obtaining good performance
on modern accelerator architecture. Common, \naive approaches, such as the one
described, can lead to poor performance.

Two of the benchmarks execute more quickly with the \naive scheduling as shown 
in Figure~\ref{fig:simple-vs-heuristic}. The first is the
``synth'' benchmark. This is an expected result for this benchmark because every matrix
operation in the benchmark is independent and the same size. Hence, as long as operations
are distributed evenly among processors, an optimal schedule will be produced. So both
scheduling algorithms produce the same schedule, however the \naive algorithm does so
much more quickly and a corresponding drop in the execution time is observed.

The ``neural'' benchmark also runs more quickly with the \naive algorithm. This provides
further evidence to support the suggestion that some operations in the benchmark have a
time model that is inaccurate. This would cause a degraded schedule to be produced by our heuristic algorithm and increase the execution time of the operations as a result.

An online scheduling approach was also experimented with, in which a pool of operations
which were ready to be executed was kept. SPEs selected an operation from this pool
arbitrarily. Similar results to the \naive scheduling algorithm were observed.

\subsection{Utilisation of Parallel Execution Elements}

\begin{figure}
\centering
\includegraphics[width=0.5\columnwidth]{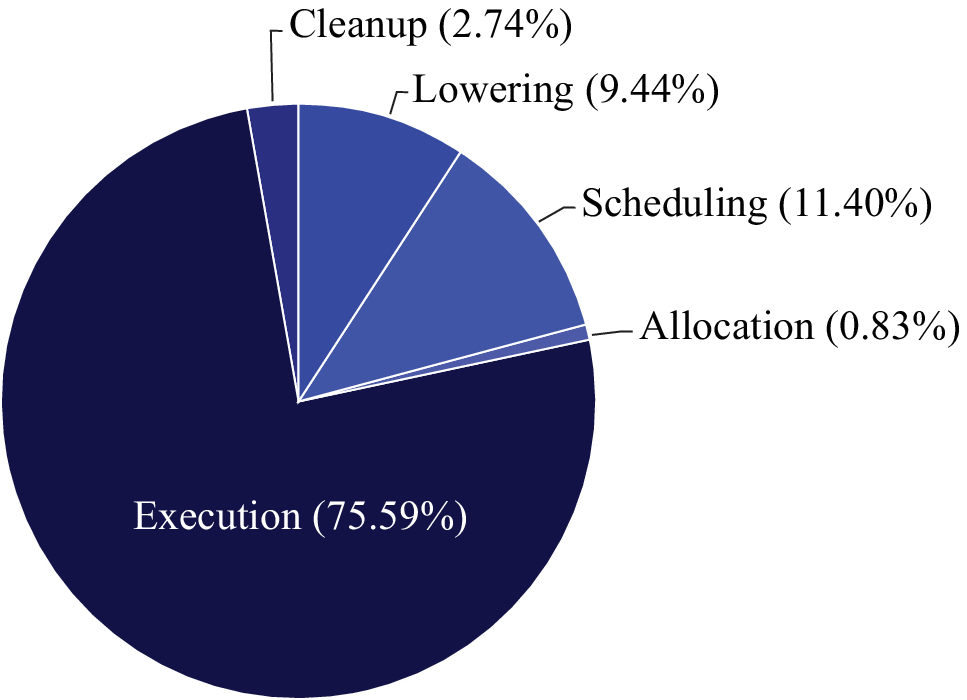}

\caption{Breakdown of execution time into phases, measured with the performance counters
available on the Cell. 8 SPEs were used during these measurements. 
The majority of the time executing a trace is spent executing the matrix operations.}\label{fig:breakdown-pie}
\end{figure}

In the remaining set of experiments we investigated the extent to which our
framework utilised the parallel execution elements (SPEs) of the Cell.
Figure~\ref{fig:breakdown-pie} shows a breakdown of execution time spent for the
different phases of our framework. These numbers are averaged across all benchmarks when run
with 8~SPEs. It follows that the majority of time (more than 75\%) is spent executing
matrix operations on Cell SPEs. The remaining 24.41\% is mainly spent on scheduling
(11.4\%) and lowering of matrices into blocks~(9.44\%). Only 0.83\% of time is spent
on matrix allocation, and 2.74\% on deallocation and other cleanup.

\begin{figure}
\centering
\includegraphics[width=0.5\columnwidth]{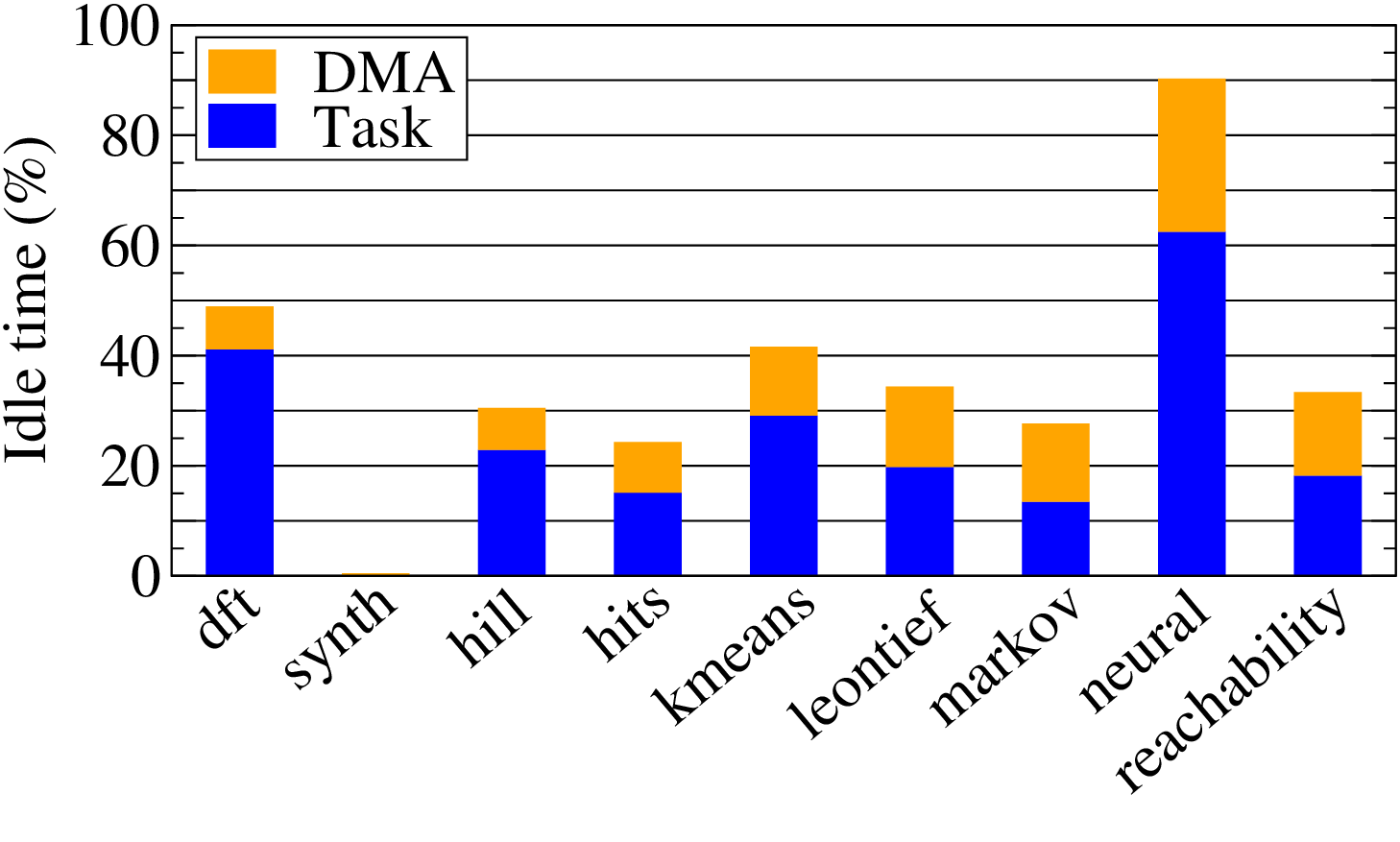}

\caption{The percentage of time that SPEs spent idle during each benchmark. 8~SPEs were
used for this experiment.  Idle time is broken into two components: ``DMA'' idle time is the
time spent waiting for the completion of DMA transfers and ``Task'' time is the time spent
waiting for an operation to be sent by the PPE. The ``synth'' benchmark exhibits
close to 0\% idle time and reaches close to the peak performance of
the Cell processor.}\label{fig:idle}
\end{figure}

In another experiment, we measured the time that the SPEs spent idle during the 
execution phase of the framework. The results of this experiment are shown in Figure~\ref{fig:idle}.
The idle time is composed of the time the SPEs spent waiting for tasks to be offered by the PPE (``Task'') and the time spent waiting for the completion of
DMA transfers (``DMA''). Again these numbers were determined for 8~SPEs.
It follows from Figure~\ref{fig:idle} that with the ``synth'' benchmark we approached the Cell's peak
performance, which is manifested in idle times close to zero. 

The remaining benchmarks incur idle times as a result of three main factors. Firstly, operations
such as matrix addition are memory-bound (rather than computationally bound). Although our triple buffering technique overlaps computation with communication, memory-bound operations can still induce DMA waiting times because memory transfers for these operations are more expensive than computation of the
result.

The second reason for idle times is that certain traces may contain a large number of data
dependencies which reduce the amount of instruction level parallelism for execution.
This can result in idle time spent waiting for the operands of an operation to become available.
Thirdly, if operations are computed too quickly by the SPEs, the execution control mechanism 
may not be able to service them fast enough, resulting in time spent waiting for a task to be
offered by the PPE. The impact of this could be reduced by aggregating several operations into
a single operation which takes a longer time to execute.

\begin{figure}
\centering
\includegraphics[width=0.5\columnwidth]{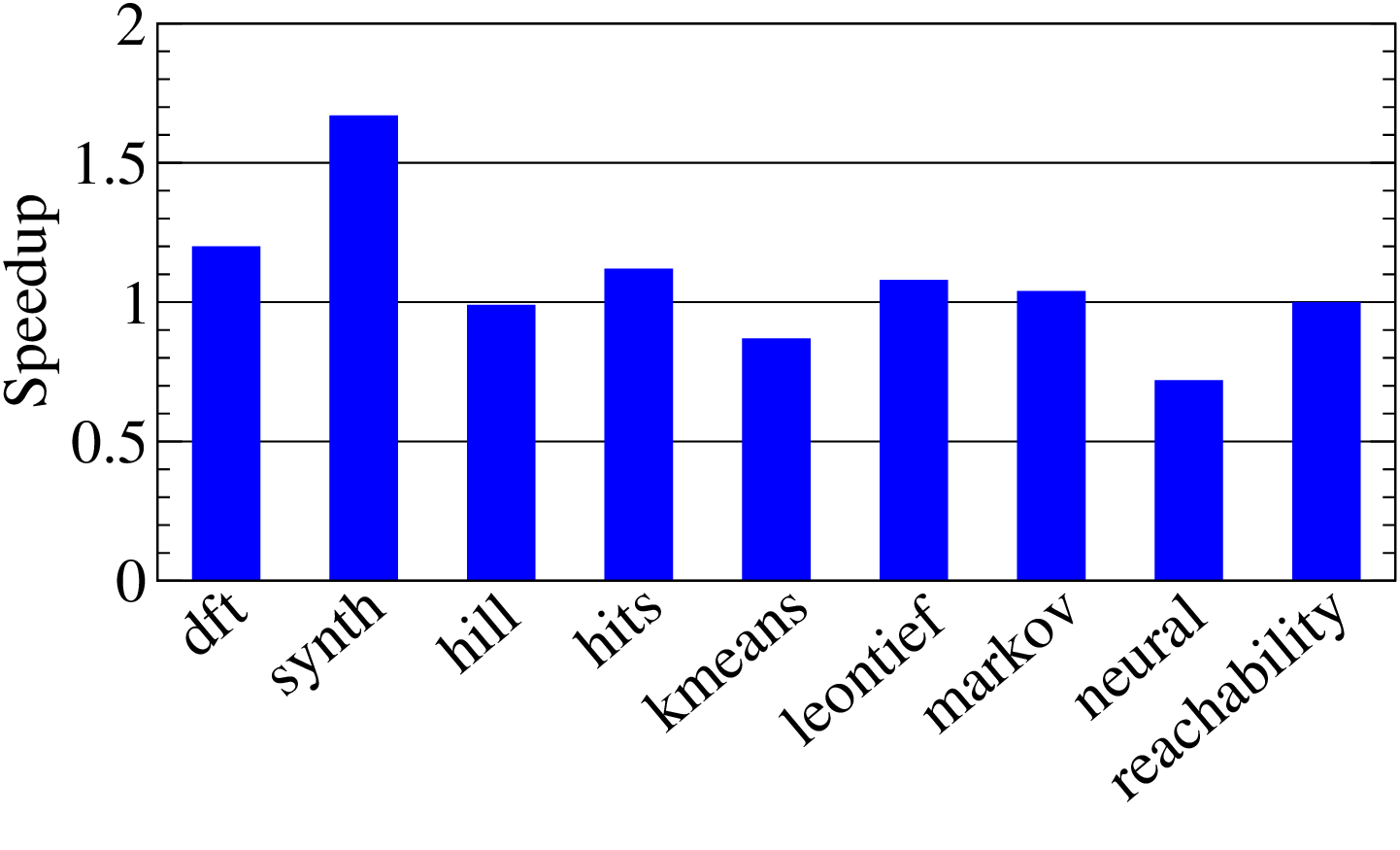}
\caption{Speedups achieved by overlapping the scheduling, lowering
and execution phases of a trace, versus sequential execution of those phases.}\label{fig:serial-vs-parallel}
\end{figure}

The final experiment we measured the impact of overlapping scheduling of operations, lowering of operations and execution of operations. As described in
section~\ref{sec:overview}, if the size of a trace reaches a maximum threshold
value, execution of the Octave interpreter will continue while the trace
is executed. Scheduling of the trace, lowering of the trace and operation execution are 
executed in separate threads. Thus it is possible for multiple traces to 
be in execution concurrently.
Figure~\ref{fig:serial-vs-parallel} shows the speedup in the execution time of each
benchmark when these separate threads are used versus each of the stages being executed in sequence.
Most benchmarks achieve a speedup when compared
with sequential execution.  Those that do not, such as ``kmeans'' and
``neural'' may not produce traces that are large enough to allow any extra parallelism to be exploited
and instead, the overhead of using several threads for execution causes a slowdown. Fine tuning of the
threshold at which execution of a trace is triggered may result in further speedups from using this 
technique.

\section{Conclusion} \label{sec:conclusion}

In this work we developed a new framework for fully utilising the performance
of modern accelerator architectures in the execution of matrix languages.
We provide an implementation of the framework for executing Octave
programs on the Cell Broadband Engine architecture. 

The framework is very easy to use, with only minor changes needed to existing
Octave code in order to take advantage of the parallel architecture. As opposed to
existing systems, which typically only take advantage of the data parallelism
of matrix operations, our framework additionally exploits instruction level parallelism,
pipeline parallelism and task parallelism to obtain better performance. 

Lazy evaluation is used to extract a trace of matrix operations from a program at run-time.
The data dependencies of operations in a trace are examined to elicit instruction level
parallelism. A novel partitioning scheme, called lowering, is used to divide operations on matrices in a way that maximises the parallelism available
in a trace. A new heuristic scheduling algorithm is then used to schedule operations
among the processing elements in a way that improves their utilisation and reduces
the execution time of a trace. In order to produce an accurate schedule, we employ time modelling of matrix operations to estimate their execution times.

We performed an extensive evaluation of our framework with positive results.
Octave benchmarks executing on our framework for the Cell Broadband Engine
architecture are up to a factor of 12 faster than execution on standard Octave on
more recent and expensive Intel Core2 Quad processors. Our framework also out-performed 
an out-of-the box installation of Octave running on the Cell processor, as well as an
installation of MATLAB
running on an Intel Core2 Quad processor.

We also evaluated a new heuristic scheduling algorithm by comparing the schedules
it produced with the optimal schedules produced by a mathematical program 
we developed. We found that the makespan of the optimal schedule
deviated no more than 1\% from the makespan of the schedule produced
by the heuristic algorithm, for all computable problem sizes.

Further evaluation of our framework revealed that partitioning and scheduling
of operations did not incur a significant amount of overhead and one of our
benchmark programs was able to reach the peak performance obtainable on the 
Cell Broadband Engine Architecture.

Though it is widely believed that automatic parallelisation techniques for imperative
programs are infeasible, we showed that automating the parallelisation
of sequential matrix language programs is achievable for modern accelerator architectures.
Speedups of several magnitudes are possible when various kinds of parallelism, including instruction level, data, 
pipeline and task parallelism, are exploited. We showed that the main contributors to these speedups
are instruction level and data parallelism that are obtained through novel lazy evaluation and lowering techniques. 

\bibliographystyle{acm}
\bibliography{tr.bib}


\appendix
\section{Integer Linear Programming Model}\label{app:integer-linear-program}
\begin{eqnarray*}
\min & z\\
\\
\mbox{s.t.} &\sum_{i=1}^{n}x_{1i}\leq p\\
 &x_{i1}=0  &i \in I\\
 &{\displaystyle \sum_{j=1}^{n}x_{ij}}\leq1  & i \in I\setminus\{1\} \\
 &{\displaystyle \sum_{j=1}^{n}x_{ji}}=1  & i \in I\setminus\{1\}\\
& x_{ii}=0 & i \in I\setminus\{1\} \\
 &{\displaystyle \sum_{i=1}^{n}\left(y_{ij}+t_{\textit{df}}(i) x_{ij} \right) } \leq t_{j} &
  j \in I\\
&{\displaystyle \sum_{i=1}^{n}\left(y_{ij}+t_{\textit{df-ex}}(i) x_{ij}\right) } \leq t_{j}+t_{\textit{df}}(j) &
  j \in I\\
 &{\displaystyle \sum_{i=1}^{n}\left(y_{ij}+t_{\textit{df-ex-wb}}(i) x_{ij} \right) } \leq t_{j}+t_{\textit{df-ex}}(j) &
 j \in I \\
 & y_{ij}\leq Ux_{ij} &i,j \in I\\
 & t_{i}+Ux_{ij}-U\leq y_{ij}\leq t_{i} & i,j \in I \\
 &  t_{i}+t_{\textit{df-ex-wb}}(i)\leq z  &   i \in I \\
&t_{i}\in\mathbb{R}^{+}  & i \in I\\
 &x_{ij}\in\{0,1\} & i,j \in I \\
 & y_{ij}\in\mathbb{R}^{+} & i,j \in I \\
&z\in\mathbb{R}^{+}
\end{eqnarray*}

\pagebreak 

\section{AMPL Script}\label{app:ampl}
\begin{lstlisting}[basicstyle=\tiny]
###############################################################################
## Input Sets and Parameters
##

# set of matrix instructions
set I;

# set of task precedences
set E within I cross I;

# time parameters for instructions

# time duration for data fetch stage 
param t_df{I};

# time duration for execute
param t_ex{I};

# time duration for execute
param t_wb{I};

# number of processors
param num_processors;

###############################################################################
## Set and Parameters for Model 
##

# synthetic start task 
param start symbolic;

# set of matrix instruction with start task
set IS := I union {start};

# set adjacency elements that are not 
set A := {(i,j) in (IS cross I): i <> j};

# upper bound for variables t
param U := sum{i in I} (t_df[i] + t_ex[i] + t_wb[i]);

# set of predecessors
set preds {u in I} := {(v,w) in E: w = u};

###############################################################################
## Variables 
##

# adjacency matrix of stream graph
var x{A}, binary;

# linearisation time y[i,j] = x[i,j] * t[i,j]
var y{A}, >=0;

# start time of instruction
var t{I}, >=0;

# makespan
var z, >=0;

###############################################################################
## Objective
##

# minimize makespan
minimize objective:
    z;

###############################################################################
## Constraint
##

# makespan is greater than or equal to the completion of all instructions
subject to makespan {i in I}:
    t[i] + t_df[i] + t_ex[i] + t_wb[i] <= z;

# instruction precedence constraint
subject to precedence {(i,j) in E}: 
    t[i] + t_df[i] + t_ex[i] + t_wb[i] <= t[j];
    
# stream succession constraint for each stage
subject to stream_df {j in I}: 
    sum{i in I:i <> j} (y[i,j] + t_df[i] * x[i,j]) <= t[j];
subject to stream_ex {j in I}: 
    sum{i in I:i <> j} (y[i,j] + (t_df[i] + t_ex[i]) * x[i,j]) <= t[j] + t_df[j]; 
subject to stream_wb {j in I}: 
    sum{i in I:i <> j} (y[i,j] + (t_df[i] + t_ex[i] + t_wb[i]) * x[i,j]) <= t[j] + t_df[j] + t_ex[j]; 

# stream graph processor constraint
subject to processors:
    sum{i in I} x[start,i]  <= num_processors;

# stream graph successor constraint
subject to successors {i in I}:
    sum{j in I: i <> j } x[i,j] <= 1; 

# stream graph predecessor constraint
subject to predecessors {j in I}:
    sum{i in IS: i <> j} x[i,j] = 1; 

###############################################################################
## Linearisation Constraints
##

# linearisation of quadratic term y[i,j] = x[i,j] * t[i]
# for all (i,j) in A. 
subject to linearize_t1 {(i,j) in (I cross I): i <> j}: 
    y[i,j] <= U * x[i,j];
subject to linearize_t2 {(i,j) in (I cross I): i <> j}: 
    t[i] - U  + U * x[i,j]  <= y[i,j];
subject to linearize_t3 {(i,j) in (I cross I): i <> j}: 
    y[i,j] <= t[i];
\end{lstlisting}

\pagebreak
\section{List of Implemented Octave Functions} \label{app:funcs}
The following Octave operators and functions were implemented to work on our custom
\psmatrix data type. These operations utilise our framework to execute in parallel.
\begin{itemize}
\item \lstinline!<p_matrix> + <p_matrix>!
\item \lstinline!<p_matrix> + <scalar>!
\item \lstinline!<p_matrix> - <p_matrix>!
\item \lstinline!<p_matrix> - <scalar>!
\item \lstinline!<p_matrix> * <p_matrix>!
\item \lstinline!<p_matrix> * <scalar>!
\item \lstinline!<p_matrix> .* <p_matrix>!
\item \lstinline!<p_matrix> ./ <p_matrix>!
\item \lstinline!<p_matrix> .^ <p_matrix>!
\item \lstinline!abs(<p_matrix>)!
\item \lstinline!mod(<p_matrix>, <scalar>)!
\item \lstinline!sin(<p_matrix>)!
\item \lstinline!cos(<p_matrix>)!
\item \lstinline!sign(<p_matrix>)!
\item \lstinline!round(<p_matrix>)!
\item \lstinline!<p_matrix> == <p_matrix>!
\item \lstinline$<p_matrix> != <p_matrix>$
\end{itemize}

\end{document}